\documentclass[12pt]{JHEP3}
\input epsf.tex
\epsfclipon

\usepackage{epsfig}
\usepackage{epstopdf}
\usepackage{epsf}
\input{epsf.sty}
\usepackage{graphicx,amsmath,amssymb}
\usepackage{epsfig,multicol}
\allowdisplaybreaks

\usepackage{bbm,bm,amsmath,amssymb}
\def\blfootnote{\xdef\@thefnmark{}\@footnotetext}

\long\def\symbolfootnote[#1]#2{\begingroup%
\def\thefootnote{\fnsymbol{footnote}}\footnote[#1]{#2}\endgroup}

\newcommand{\be}{\begin{eqnarray}}
\newcommand{\ee}{\end{eqnarray}}
\newcommand{\ben}{\begin{eqnarray*}}
\newcommand{\een}{\end{eqnarray*}}

\newcommand{\bcent}{\begin{center}}
\newcommand{\ecent}{\end{center}}
\newcommand{\benum}{\begin{enumerate}}
\newcommand{\eenum}{\end{enumerate}}
\newcommand{\bdesc}{\begin{description}}
\newcommand{\edesc}{\end{description}}
\newcommand{\bitem}{\begin{itemize}}
\newcommand{\eitem}{\end{itemize}}
\newcommand{\bquote}{\begin{quote}}
\newcommand{\equote}{\end{quote}}
\newcommand{\bhalfp}{\begin{minipage}{0.45\textwidth}}
\newcommand{\ehalfp}{\end{minipage}}
\newcommand{\bhead}{\begin{center}\bf \Large}
\newcommand{\ehead}{\end{center}\bigskip}

 \newcommand{\bfC}{{\bf C}}

\def\be{\begin{equation}}
\def\ee{\end{equation}}
\def\ba{\begin{eqnarray}}
\def\ea{\end{eqnarray}}

\newcommand{\roughly}[1]{\mathrel{\raise.3ex\hbox{$#1$\kern-0.85em
\lower1ex\hbox{$\sim$}}}}

\def\2pi{\left(2\pi\right)}

\def\beq{\begin{equation}}
\def\eeq{\end{equation}}
\def\bg{\begin{eqnarray}}
\def\nd{\end{eqnarray}}
\def\bea{\begin{eqnarray}}
\def\eea{\end{eqnarray}}

\def\D3{\overline{\mbox{D3}}}

\title{Toward the Gravity Dual of Heterotic Small Instantons}

\author{Fang Chen$^1$, Keshav Dasgupta$^1$, Paul Franche$^1$, Radu Tatar$^2$\\
\vskip.03in
${}^1$ Ernest Rutherford Physics Building, McGill University,\\
3600 University Street, Montr{\'e}al QC, Canada H3A 2T8\\
${}^2$ Division of Theoretical Physics,\\
Department of Mathematical Sciences,\\
The University of Liverpool, Liverpool, England, UK L69 3BX\\
{\tt fangchen, keshav, franchep@hep.physics.mcgill.ca, Radu.Tatar@liverpool.ac.uk}}
\date{October 2010}

\abstract{The question of what happens when the heterotic $SO(32)$ instanton becomes small was answered sometime back
by Witten. The heterotic theory develops an enhanced $Sp(2k)$ gauge symmetry for $k$ small instantons, besides
the allowed $SO(32)$ gauge symmetry. An interesting question now is to ask
what happens when we take the large $k$ limit.
In this paper we argue that in some special cases, where Gauss' law allows the large $k$ limit, the dynamics of the
large $k$ small instantons can be captured by a dual gravitational description. For the cases that we elaborate in
this paper, the gravity duals are non-K\"ahler manifolds although in general they could be non-geometric.
These small instantons are heterotic five-branes and the duality allows us to study the strongly
coupled field theories on these five-branes.
We review and elaborate on some of the recent observations pointing towards this duality, and argue that in
certain cases the gauge/gravity duality may be understood as small instanton transitions under which the instantons
smoothen out and consequently lose the $Sp(2k)$ gauge symmetry.
This may
explain how branes disappear on the dual side and are replaced by fluxes.
We analyse the torsion classes
before and after the transitions, and discuss briefly how the
ADHM sigma model and related vector bundles could be studied for these scenarios.}

\maketitle

\begin{document}

\section{Introduction}

\noindent In the full moduli space of string theory, the
heterotic theory \cite{gross} has always been an important corner where phenomenologically useful models are most
easily accessible. Part of its appeal lies in the existence of anomaly cancelling $SO(32)$ or $E_8 \times E_8$
vector bundle that is crucial for embedding standard model in string theory. The existence of a minimal
supersymmetric multiplet is also an additional benefit.

On the other hand type IIB theory has its own share of advantages. The non-abelian multiplet in this theory
come from non-perturbative branes such that exactly similar physics, as from the heterotic theory, can be
studied here using these branes. Additionally, type IIB theory has a full non-perturbative
completion: the so-called F-theory \cite{vafaF} where local and non-local branes participate to realise the
quantum corrections. In fact the F-theory completion of type IIB theory is directly related to the
heterotic theories. Thus various vacua of heterotic theories should be thought of as duals to the various
seven-brane configurations in F-theory compactified on Calabi-Yau spaces.

In recent times gauge/gravity dualities have been studied exclusively in type II theories and especially in type
IIB theory. The duality that we are most interested in IIB is the geometric transition \cite{vafaGT}, where the
strongly coupled far IR
dynamics of an ${\cal N} = 1$ pure SYM theory is studied in terms of a weakly coupled type IIB supergravity on
a deformed conifold with three-form fluxes. The far IR theory, on the other hand, is realised as type IIB
D5-branes wrapped on the two-cycle of a resolved conifold. Thus this duality is a geometric transition where,
under a conifold transition, the wrapped D5-branes {\it disappear} and are replaced by three-form fluxes on a
deformed conifold. The type IIA dual of this in terms of D4- and NS5-branes was understood in \cite{kyungho}.

Unfortunately similar dualities have not been addressed in details in the heterotic side. To our knowledge the
first attempt to address this issue was done in \cite{gtpaper2} (see also \cite{israel} for a more recent analysis).
The difficulty in the heterotic side lies in two things: understanding the vector bundles and solving the
Bianchi identity. For example one would be tempted to realise the geometric transition in the heterotic theory
by taking the S-dual of the original IIB transition i.e replacing the IIB D5-branes with the NS5-branes and
interpreting the NS5-branes as heterotic five-branes.
However it is
not {\it a-priori} clear whether the dual deformed conifold geometry would indeed solve the Bianchi identity.
Additionally, it is not clear how the vector bundles could be pulled across a conifold transition.

In our earlier papers \cite{gtpaper2, gtpaper1} we gave a local description of this 
duality\footnote{By {\it local} we mean 
that the sugra background is studied around a specific chosen point in the internal six-dimensional space. For 
example we choose a point 
($r_0, \langle\theta_i\rangle, \langle\phi_i\rangle, \langle\psi\rangle$) in 
\cite{gtpaper1, gtpaper2} which is away from the $r=0$ conifold point. 
 This is because the full global picture was
hard to construct, and any naive procedure always tends to lead to non-supersymmetric
solutions. In deriving the local metric, we took a simpler model where all
the spheres were replaced by tori with periodic coordinates ($x,
\theta_1$) and ($y,\theta_2$). The coordinate $z$ formed a
non-trivial $U(1)$ fibration over the $T^2$ base. Here ($r, x, y, z, \theta_1, \theta_2$) is the coordinate of a point 
away from 
($r_0, \langle\phi_1\rangle, \langle\phi_2\rangle, \langle\psi\rangle, \langle\theta_1\rangle, 
\langle\theta_2\rangle$). 
The replacement
of spheres by two tori was directly motivated from the
corresponding brane constructions of \cite{dasmukhi}, where
non-compact NS5 branes required the existence of tori instead of
spheres in the T-dual picture. On the other hand 
the term {\it global} means roughly adding back the curvature, warping, etc., replacing tori by
spheres, so that at the end of the day, we have a supersymmetric
solution to the equations of motion. In \cite{chen} we managed to provide the full global picture of geometric
transition. Note also that the only known global solution, i.e \cite{pandoz}, before our work
was unfortunately not supersymmetric (see
\cite{cvetic}, \cite{anke} for details) although it satisfied the type IIB EOMs.}. 
Our local analysis
reproduced a global configuration that was more general than a deformed conifold with fluxes. In this paper we will
show why this is true: the type IIB story cannot be directly dualised to the heterotic side. Bianchi identity
will in fact change the IIB solutions, and so the heterotic duals will not quite be the same as the IIB ones.

In fact the precise duality in the heterotic side can be presented succinctly in the language of small
instantons \cite{wittensmall}. The large $N$ limit of these small instantons can be captured by a dual gravitational
background which is generically non-geometric. The interesting thing about our analysis is the observation that the
geometric transition is a small instanton transition where the small instantons become ``smooth'' on the dual side
and therefore lose the $Sp(2N)$ gauge symmetry. This gives a possible explanation of the disappearance of the
branes in the dual side. In this paper we will only work with the $SO(32)$ heterotic theory and leave the
$E_8 \times E_8$ case for the sequel. The $E_8 \times E_8$ case presumably follows similar path as illustrated in
\cite{ganorhanany}. For earlier studies on small instanton transition, the reader may refer to \cite{ovrutsmall}.

The work in this paper is a direct followup of our last paper \cite{chen} where various supersymmetric
duals in type II and M-theory were presented in the geometric transition set-up. However in \cite{chen} formal
proofs for supersymmetry of the solutions,
using say torsion classes, were not presented. In this paper we will rectify these short-comings and start with
giving detailed torsion class analysis of all the type II solutions of \cite{chen}. This will help us to state the
heterotic duality in a more precise way.

\subsection{Supersymmetric configurations in geometric transitions}

The issue of supersymmetry for the intermediate configurations
is of course crucial in the geometric transition set-up. We have discussed this in some details in \cite{chen}. Here
we will elaborate the story a bit more, and new details will be presented in section 3.

Our first configuration is in type IIB theory for wrapped D5-branes on a resolved conifold. 
The original construction of \cite{pandoz}, with a conformally Calabi-Yau metric, is
not supersymmetric
The supersymmetric configuration
is given in \cite{chen} where we put a non-K\"ahler metric on the resolved conifold. 
Our starting point in \cite{chen} is the choice of functions $F_i = F_i(r)$, $i = 1, ..., 4$
and $F_0 = F_0(r, \theta_1, \theta_2)$ which are used to write the metric for the 
internal space. Therefore for different choices of $F_0, F_i$ we get different dual gauge theories. 
The complete background in type IIB
then is (see also \cite{chen}):
\bg\label{susybg}
&& F_3 = h~{\rm cosh}~\beta~e^{2\phi} \ast d\left(e^{-2\phi} J\right), ~~~
H_3 = -hF_0^2{\rm sinh}~\beta~e^{2\phi} d\left(e^{-2\phi} J\right), ~~~\phi_{\rm now} = -\phi\nonumber\\
&& F_5 = -{1\over 4} (1 + \ast) dA_0 \wedge dx^0 \wedge dx^1 \wedge  dx^2 \wedge  dx^3, ~~~~ ds^2 = 
F_0 ds^2_{0123} + ds^2_6\nonumber\\
&& ds^2_6 =  F_1~ dr^2 + F_2 (d\psi + {\rm cos}~\theta_1 d\phi_1 + {\rm cos}~\theta_2 d\phi_2)^2
+ \sum_{i = 1}^2 F_{2+i}
(d\theta_i^2 + {\rm sin}^2\theta_i d\phi_i^2)
\nd
where we have defined $\phi, h$ and $A_0$ using $F_0$ and a constant ``boosting'' parameter $\beta$ in the following way:
\bg\label{kdive}
&&h = {F_0 {\rm cosh}^2\beta \over 1 + F_0^2 {\rm sinh}^2\beta}, ~~~~~~~~
e^{-\phi} = {F^{3/2}_0 {\rm cosh}\beta \over \sqrt{1 + F_0^2 {\rm sinh}^2\beta}}\nonumber\\
&&A_0 = (F_0^2 -1 ){\rm tanh}~\beta \left[1 + \left({1-F_0^2\over F_0^2}\right){\rm sech}^2\beta +  
\left({1-F_0^2\over F_0^2}\right)^2 {\rm sech}^4 \beta\right]
\nd
All the coefficients etc are also described in more details in \cite{chen}.
In sec. 3 we will compute all the torsion classes for this
background, and discuss explicitly how supersymmetry is preserved.

The above background is of course the first step in the chain of dualities associated with IR geometric transitions. To
go to the type IIA mirror description using SYZ method \cite{SYZ} we need to make the base very large compared to the
fibre. We achieve the final IIA mirror by making the following steps\footnote{These rules have been derived in  
the local limit in \cite{gtpaper1, gtpaper2}. In \cite{chen} we have shown how in the global picture these rules 
could work.}:

\noindent $\bullet$ Shift of the coordinates ($\psi, \phi_i$) using variables $f_i(\theta_i)$. This shifting of the
coordinates mixes non-trivially all the three isometry directions as described in eq (4.24) of \cite{chen}.

\noindent $\bullet$ Shift the metric along $\psi$ direction by the variable $\epsilon$, as given in the second line
of eq (4.29) of \cite{chen}. This variable doesn't have to be very small in the global limit. The only constraint on 
$\epsilon$ is $\epsilon < 1$ to preserve the signature of the metric. 

\noindent $\bullet$ Make SYZ transformations along the new shifted directions. Thus the three T-dualities are
{\it not} made along the three original isometry directions.

\noindent $\bullet$ In the new metric of IIA make a further rotation along the ($\theta_2, \phi_2$)
directions using a $2\times 2$
matrix given as eq (4.50) of \cite{chen}. The matrix is described using a constant angular variable $\psi_0$.

\noindent $\bullet$ Finally in the transformed metric convert $\psi_0$ to $\psi$ as in eq (4.53) of \cite{chen}.

\noindent The final metric after we perform all the above transformations takes the following form:
\begin{eqnarray}\label{mliftnow}
ds^2_{11}&&= F_0ds_{0123}^2+F_1 dr^2 + {\alpha F_2
\over \Delta_1 \Delta_2} \Big[d\psi -b_{\psi r}dr - b_{\psi\theta_2} d\theta_2\nonumber\\
&&+ \Delta_1 {\rm cos}~\theta_1 \Big(d\phi_1 - b_{\phi_1\theta_1}
d\theta_1-b_{\phi_1 r}dr\Big)+ \Delta_2 {\rm cos}~\theta_2 {\rm cos}~\psi_0
\Big(d\phi_2 - b_{\phi_2\theta_2} d\theta_2-b_{\phi_2
r}dr\Big)\Big]^2\nonumber\\
&& + \alpha j_{\phi_2\phi_2}\Big[d\theta_1^2 + \Big(d\phi_1 -
b_{\phi_1\theta_1} d\theta_1-b_{\phi_1 r}dr\Big)^2\Big] \nonumber\\
&&+ \alpha j_{\phi_1\phi_1}\Big[d\theta_2^2 + \Big(d\phi_2 -
b_{\phi_2\theta_2} d\theta_2-b_{\phi_2
r}dr\Big)^2\Big]\\
&& + 2\alpha j_{\phi_1\phi_2}{\rm cos}~\psi_0\Big[d\theta_1 d\theta_2 - \Big(d\phi_1
- b_{\phi_1\theta_1} d\theta_1-b_{\phi_1 r}dr\Big)\Big(d\phi_2 -
b_{\phi_2\theta_2} d\theta_2-b_{\phi_2
r}dr\Big)\Big]\nonumber\\
&& + 2\alpha j_{\phi_1\phi_2}{\rm sin}~\psi_0\Big[\Big(d\phi_1- b_{\phi_1\theta_1}
d\theta_1-b_{\phi_1 r}dr\Big) d\theta_2
+ \Big(d\phi_2 - b_{\phi_2\theta_2} d\theta_2-b_{\phi_2r}dr\Big)d\theta_1\Big]\nonumber
\end{eqnarray}
The above metric, which we called the {\it symmetric} metric in \cite{chen}, looks very close to the deformed
conifold metric. However due to steps 2, 4 and 5 above, it is not guaranteed that the metric will preserve supersymmetry.
Furthermore one might also question whether the SYZ operation itself could preserve 
supersymmetry\footnote{This is because it is not {\it a~priori} clear whether the fermionic boundary conditions 
are periodic or anti-periodic along the T-duality circles. Sometime when the cycles degenerate we may need to 
put in an additional $(-1)^F$ term to preserve susy. An example of this is given in the third reference of 
\cite{dipole}.}. 
Therefore to verify
this we will evaluate all the torsion classes for this background in sec 3.2. Note that in \cite{chen} we didn't
explicitly derive the fluxes in the mirror. In sec (3.2) we will be able to determine at least the NS three-form
flux that will make the IIA mirror background supersymmetric. This will also help us to fix ($f_1, f_2, \epsilon$).

Once we have the type IIA metric we can lift\footnote{In \cite{chen} we lifted the non-symmetric type IIA metric to
M-theory. This is more generic than the symmetric one.}
this to M-theory using the one-forms ($\sigma_i, \Sigma_i$) as given in
eqs (4.47) and (4.48) of \cite{chen} respectively. The precise flop transformation of the M-theory manifold is described
using a {\it class} of transformations specified by ($a, b$) as in eq (4.59) of
\cite{chen}. The final metric after we reduce the flopped metric to type IIA is:
\begin{eqnarray}\label{IIAmetric}
ds_{10}^2&& =F_0ds_{0123}^2+F_1dr^2+e^{2\phi}\Big[d\psi-b_{\psi
\mu}dx^\mu+\Delta_1 {\rm cos}~\theta_1 \Big(d\phi_1 - b_{\phi_1\theta_1}
d\theta_1-b_{\phi_1 r}dr\Big)\nonumber\\
&&\quad\quad\quad\quad\quad\quad\quad\quad\quad\quad +
\widetilde{\Delta}_2 {\rm cos}~\theta_2 \Big(d\phi_2 - b_{\phi_2\theta_2}
d\theta_2-b_{\phi_2 r}dr\Big)\Big]^2\nonumber\\
&&~+e^{2\phi\over 3}a^2(k^2G_2+kG_3+G_1)\Big[d\theta_1^2+(d\phi_1^2-b_{\phi_1\theta_1}d\theta_1-b_{\phi_1
r}dr)^2\Big]\nonumber\\
&&~+e^{2\phi\over 3}b^2(\mu^2G_2+\mu
G_3+G_1)\Big[d\theta_2^2+(d\phi_2^2-b_{\phi_2\theta_2}d\theta_2-b_{\phi_2
r}dr)^2\Big]
\end{eqnarray}
along with the following one-form charge, but no D6-brane sources:
\bg\label{1ffc}
 A=\Delta_1 {\rm cos}~\theta_1 \Big(d\phi_1 -
b_{\phi_1\theta_1} d\theta_1-b_{\phi_1 r}dr\Big)- \widetilde{\Delta}_2 {\rm
cos}~\theta_2 \Big(d\phi_2 - b_{\phi_2\theta_2} d\theta_2-b_{\phi_2
r}dr\Big)
\nd
A torsion class analysis in sec 3.2 will help us to fix a particular flop transformation i.e fix ($a, b$) so that the
above background remains supersymmetric. Observe again that we haven't determined all the flux components in IIA. As
before, we expect the torsion class analysis to fix at least the NS three-form. The three-form can then be fixed by
EOM or supersymmetry constraints.

In all the above steps we tried to make duality transformations so that we could get {\it geometric} manifolds. However
this is not generic. For a more generic choice of the B-fields in the original type IIB set-up, we could get
non-geometric manifolds both before and after flop in IIA. This non-geometric aspect is also reflected in the
final type IIB mirror configuration.
In fact this tells us that the generic solution spaces we get in type IIB are non-geometric
manifolds. For certain choices of parameters (B-fields, and metric components) we can get geometric manifolds like
Klebanov-Strassler \cite{ks} or Maldacena-Nunez \cite{MN}.
This is almost like the parameter space of \cite{baryonic} but now much bigger,
and allowing {\it both} geometric and non-geometric manifolds that cover various
branches of the dual gauge theories. 

\noindent To make this a little more precise, note that we have analysed the following two scenarios in \cite{chen}:

\noindent $\bullet$ There are various ways to embed wrapped five-branes on a two-cycle in the internal space that 
preserve supersymmetry. For a {\it given} choice of ($F_i, F_0, \epsilon$) we can find the geometry and the 
fluxes that preserve supersymmetry (see the analysis in sec. 4 of \cite{chen}). In the decoupling limit, this is the 
gauge theory side of the story. We called this the scenario {\it before} geometric transition.

\vskip.1in

\noindent $\bullet$ For that particular choice of the background, we followed our duality arguments to give a 
background {\it after} 
geometric transition. We showed that for {\it generic} choices of the fluxes, the dual gravitational 
background become non-geometric. Therefore the fluxes and the geometry in the brane-side of the picture induce 
non-trivial operators in ${\cal N} = 1$ gauge theory that make the dual gravitational background non-geometric. 

\vskip.1in

In this paper we will do an explicit computation to study a {\it geometrical} 
dual for the large $N$ small instantons because 
this case will not be too hard to construct. 
A similar story was also pointed out in \cite{chen}. For example,
if we deliberately restrict ourselves to the special case eq (4.71) of
\cite{chen} i.e make the NS B-fields along ($\phi_1, \phi_2$) and ($\psi, \phi_i$) directions zero,
then the geometric manifold we get has the following metric:
\begin{eqnarray}\label{iibmetfinal}
ds^2 & = & F_0^2ds_{0,1,2,3}^2+g_{rr}dr^2  +g_{\psi\psi}\Big(\widetilde{\cal D}\psi
+ \widehat{\Delta}_1 ~\widetilde{\cal D}\phi_1
+ \widehat{\Delta}_2 ~\widetilde{\cal D}\phi_2\Big)^2\\
&& +g_{\theta_1\theta_1}\Big(d\theta_1^2 + \widetilde{\cal D}\phi_1^2\Big)
+g_{\theta_2\theta_2}\Big(d\theta_2^2 + \widetilde{\cal D}\phi_2^2\Big)
+g_{\theta_1\theta_2}\Big(d\theta_1d\theta_2 + \widehat{\Delta}_3
~\widetilde{\cal D}\phi_1 \widetilde{\cal D}\phi_2\Big)\nonumber
\nd
which looks surprisingly close to the resolved warped-deformed conifold metric. A torsion class analysis can again be
performed for this case (but we will not do so here) that will allow us to put constraints on the parameters from
supersymmetry. This way all the intermediate configurations in the cycle of geometric transition will be supersymmetric.
In our opinion this is probably the first time where explicit supersymmetric configurations for IR geometric transition
in IIB, IIA and M-theory are studied. 
However our analysis also revealed the existence of a much bigger 
picture in the type IIB side where various gauge theory deformations lead to non-geometric duals.
Our aim in this paper is to extend this further to the heterotic and type I cases.

\subsection{Organisation of the paper}

The paper is organised as follows. In section 2 we will give three pieces of 
evidence related to the heterotic gauge/gravity
duality. Some of these have already appeared in \cite{gtpaper2}, but here we will elaborate them in the
global picture. The first evidence, discussed in sec. 2.1,
will come from taking the orientifold limits of the type IIB duality. The issue
of vector bundles, before and after the transition, as well as the Bianchi identity will be discussed therein.
This will be
elaborated further in sec 2.2 where we will briefly study the ADHM sigma model that captures the physics before the
transition. A more direct analysis, using properties of the underlying non-K\"ahler manifolds, will be discussed
in sec 2.3. In section 3 we will study the supersymmetry of these solutions. We will discuss how torsion classes
and supersymmetry put constraints on the warp-factors of the background manifolds. In section 4 we will give a brief
discussion of the interconnections between the torsion classes and the vector bundles both before and after the
transition. Details about the heterotic torsions and the torsion-classes are discussed further in the appendices.
We end with a conclusion and some discussions about future directions.

\section{Three roads to heterotic transitions}

Existence of geometric transition in the heterotic theory was first proposed in \cite{gtpaper2} using various
arguments stemming from U-dualities, orientifold actions and gauge/gravity identifications. However all these
analysis were studied using the so-called {\it local} geometry. Recently in \cite{chen} we have managed to
study the complete global picture for type II theories\footnote{Assuming of course that the UV completions should
follow somewhat similar line exemplified in \cite{fep} albeit now with more non-trivial UV caps. These UV caps should
capture the six-dimensional UV completions of the ${\cal N} = 1$ IR gauge theories.}.
It is therefore time now to extend the local analysis
of \cite{gtpaper2} for the heterotic case to the full global picture. See {\bf figures} {\bf 1} and
{\bf 2} for more details.
In this section we will try to give three evidence related to geometric transition in the heterotic side. Some of these
details have appeared in \cite{gtpaper2, gtpaper1, gtpaper3} for the local case.
However here we will give a somewhat different interpretation for the transition. The configuration before geometric
transition will be identified with the heterotic large $N$ small instantons, where $N$ is the number of small
instantons or heterotic five-branes. The configuration after geometric transition will be identified to the case where
the instantons have all dissolved in the heterotic $SO(32)$ gauge group (in fact the $SO(32)$ group will be broken
by Wilson lines. We will discuss this later). This interpretation is not new, as the small instantons have already been
identified to heterotic five-branes by various authors (see \cite{wittensmall} and citations therein). 
What is new, is probably the whole
{\it interpretation} of heterotic duality as small instanton transitions for some cases.

\noindent Following are the list of steps that could make this duality a bit more precise:

\noindent $\bullet$ Consider IIB on a resolved conifold with $N$ wrapped five-branes.
This is basically the configuration of
\cite{chen}.
We can go to the orientifold limit that keeps the five-branes
but generate seven-branes and orientifold seven-planes. Due to this orientifold
action the gauge theory on five-branes become $Sp(2N)$. This is basically an embedding in the F-theory set-up. We will
discuss this a bit more below.

\vskip.1in

\noindent $\bullet$ We T-dualise twice to go to type I where the five-branes remain five-branes but
the seven-branes become Type I nine-branes. To cancel the nine-brane charges we need orientifold nine-planes. These of
course appear from the orientifold seven-planes.
These five-branes are small instantons on the nine-branes. So the number of these five-branes is very large.

\vskip.1in

\noindent $\bullet$ We have to avoid the Gauss' law constraint as
not all configurations can lead to {\it large} number of five-branes in the Type I
picture. It is crucial that various charge conservation laws are not violated.
For a compact scenario one can easily show that the number of
five-branes are fixed, and in many other cases Gauss' law will not allow too many five-branes. Only in the case with
sufficient number of non-compact directions, the number of five-branes can be made large. Our study will
therefore be based on these
allowed configurations.

\vskip.1in

\noindent $\bullet$ We S-dualise to heterotic theory where they are the Witten's small instantons and the total gauge
symmetry is ${\cal G} \times Sp(2N)$ where ${\cal G}$ is a subgroup of $SO(32)$. The original $SO(32)$ group will be
broken by Wilson lines. These Wilson lines come from the separation of the seven-branes in the type IIB picture.

\vskip.1in

\noindent $\bullet$ In IIB we know that there is a geometric transition that
takes the wrapped five-branes on two-cycle of a
resolved conifold to fluxes on the three-cycles of a deformed conifold i.e the gauge/gravity duality. Embedding
this duality in F-theory will allow us to introduce fundamental matter via seven-branes. Then the geometric
transition will allow us to study the dual geometry in F-theory framework. At the orientifold corner of F-theory
the seven-brane system can be studied using D7-branes and perturbative orientifold planes along with the wrapped
five-branes.
In the dual side 
there would also be an equivalent orientifold corner where we will have fluxes with seven-branes and orientifold
seven-planes but no five-branes.

\vskip.1in

\noindent $\bullet$ So in heterotic theory we expect the dual side to have only torsion
and no heterotic five-branes i.e no small instantons or vector bundles.
Therefore these small instantons have smoothed out and have become geometry! The vector bundle before transition
will come from
the dual of the seven-branes and the separations between these seven-branes will appear as Wilson lines breaking the
$SO(32)$ gauge group to a subgroup. After transition the gauge group is completely broken.
The torsion, on the other hand, will appear from the remnants of the F-theory
three-form RR flux.

\noindent The above arguments show us that there is a possibility to understand gauge/gravity duality in the
heterotic theory as small instanton transitions where after transition the large $N$ small instantons
become torsion.
In the following we will try to
put together these evidence to form a coherent global picture.
\begin{figure}[htb]\label{localvsglobal}
        \begin{center}
\includegraphics[height=6cm]{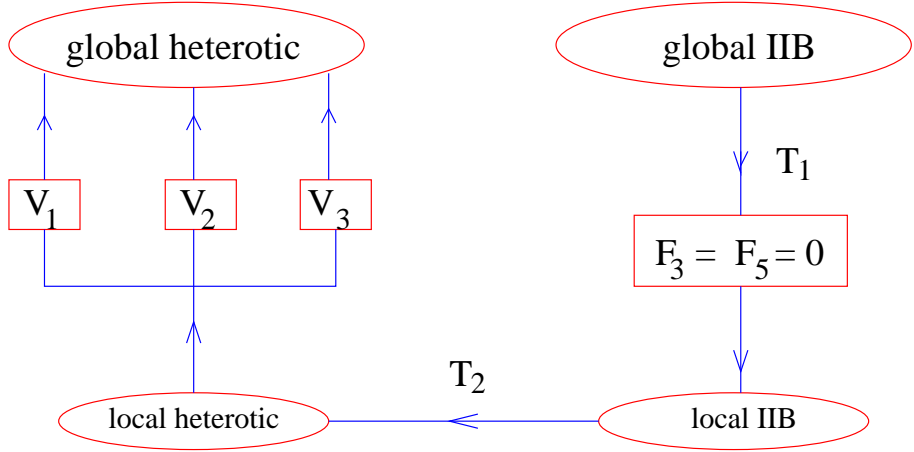}
        \caption{A precise flow diagram to illustrate how a global type IIB background can go to
a global heterotic background using transformations $T_1$ and $T_2$. The transformation
$T_1$ could be an orientifolding operation or something more complicated, as discussed in the text, and similarly
$T_2$ could be T-dualities or something more involved. Thus only local geometric are related by $T_2$ transformations.
Many different global completions with various choices of the vector bundles $V_i$ lead to the same local background
in the heterotic theory.}
        \end{center}
        \end{figure}

\subsection{Evidence from an orientifold action}

For the first step to work we need to go to the orientifold limit.
The simplest orientifold action in the type IIB scenario is given by eq. (4.3) of \cite{chen} i.e 
($x, y$) $\to$ ($-x, -y$) where ($x, y$) are the local coordinates defined in footnote 1. 
This is also the
orientifold action discussed in \cite{gtpaper2}. The proposed local metric before geometric transition in the
heterotic theory is given by\footnote{In terms of local geometry the D5-branes wrap the ($x, \theta_1$) direction
and are spread along the spacetime $x^{0123}$ directions. The seven-branes are {\it points} on the $xy$ torus. This
configuration is supersymetric and survives the orientifold action, and under T and S-dualities lead to the required
heterotic configuration.}:
\bg\label{hetmet1}
ds^2 = && d_1(dy - b_{y\theta_i}d\theta^i)^2 + d_2(dx - b_{x\theta_j}d\theta^j)^2 + d_3 dr^2 \nonumber\\
&& -2d_4 (dx - b_{x\theta_j}d\theta^j)(dy - b_{y\theta_i}d\theta^i) + d_5 dz^2 + d_6\vert d\chi_2\vert^2
\nd
where $d_i$ are the coefficients. In the above metric
one can put non-trivial complex structure on the $\chi_2$ torus also. For the present case we see that the local metric
is given by the following non-zero components:
\bg\label{metcompo}
G  &=& \begin{pmatrix} G_{xx} & G_{xy} &
G_{xz} & G_{x\theta_1} & G_{x\theta_2}\\ {}&{}&{} \\
G_{xy} & G_{yy} & G_{yz} & G_{y\theta_1} & G_{y\theta_2}
\\ {}&{}&{} \\ G_{xz} & G_{yz} & G_{zz} &
G_{z\theta_1} & G_{z\theta_2} \\ {}&{}&{} \\
G_{x\theta_1} & G_{y\theta_1} & G_{z\theta_1} &
G_{\theta_1\theta_1}& G_{\theta_1\theta_2} \\ {}&{}&{} \\
G_{x\theta_2} & G_{y\theta_2} & G_{z\theta_2} &
G_{\theta_1\theta_2} & G_{\theta_2\theta_2} \end{pmatrix}\\
&& {} \nonumber\\
&& {} \nonumber\\
&=& \begin{pmatrix} d_2 & -{d_4}
& 0 & s_1 - d_2~b_{x\theta_1} & s_2 + {d_4}~b_{y\theta_2}
\\ {}&{}&{} \\
 -{d_4} & d_1 & 0 & s_3 + {d_4}~b_{x\theta_1}
 & s_4 - d_1~b_{y\theta_2} \\ {}&{}&{} \\ 0 & 0  & d_5 & 0
& 0 \\ {}&{}&{} \\ s_1 - d_2~b_{x\theta_1}
&  s_3 + {d_4}~b_{x\theta_1} & 0 & d_6 + {\cal A}_1
 & -{d_4}~b_{x\theta_1} b_{y\theta_2} - {d^{-1}_4} s_i s_j \\ {}&{}&{} \\
s_2 + {d_4}~b_{y\theta_2} & s_4 - d_1~b_{y\theta_2} & 0
 & -{d_4}~b_{x\theta_1} b_{y\theta_2} - {d^{-1}_4} s_i s_j & d_6 + {\cal A}_2 \end{pmatrix}\nonumber
\nd
with the various terms $s_i, {\cal A}_i$ are defined in terms of the coefficients $d_i$ in the following way:
\bg\label{cofdef}
&&s_1~=~ {d_4}~b_{y\theta_1}, ~~~~~
s_2~=~ -{d_2}~b_{x\theta_2}, ~~~~~ s_3~=~ -{d_1}~b_{y\theta_1},
~~~~~ s_4~=~ {d_4}~b_{x\theta_2}\\
&& {\cal A}_1~=~
{d_1}~b^2_{y\theta_1} + {d_2}~b^2_{x\theta_1} - 2
d_4~b_{y\theta_1}~b_{x\theta_1}, ~~ {\cal A}_2~=~
{d_1}~b^2_{y\theta_1} + {d_2}~b^2_{x\theta_2} - 2
d_4~b_{y\theta_2}~b_{x\theta_2}\nonumber
\nd
and $s_i s_j~\equiv~ s_1 s_4 - s_2 s_4 - s_1 s_3$. Recall that the $b_{x\theta_i}$ and $b_{y\theta_i}$ are not the
heterotic $B$-fields. The heterotic $B$-fields come from the type IIB $F_3$ field, which will henceforth be called
${\cal H}$. In the presence of a background dilaton $\phi$ we expect \cite{beckerpot, lust1}:
\bg\label{toreq}
d{\cal H} ~= ~d\big[e^{2\phi} \ast d\left(e^{-2\phi} J\right)\big] ~ = ~ {\rm sources}
\nd
with $J$ being the usual fundamental form derived from the above metric and the sources are the heterotic
five-branes or small instantons. Defining:
\bg\label{dxdy}
&&{\cal D}x \equiv dx - b_{x\theta_j}d\theta^j, ~~~~~ {\cal D}y \equiv dy - b_{y\theta_i}d\theta^i\nonumber\\
&& {\cal D}z_1 \equiv {\cal D}x + \tau_1 {\cal D}y, ~~~~~ {\cal D}z_1 \equiv d\chi_2
\nd
we see that the local background for the wrapped heterotic five-branes is given by the following metric:
\bg\label{metmet}
ds^2~=~ d_3 dr^2 + d_5 \big(dz + a~{\rm cot}~\langle\theta_1\rangle~d\widetilde x
 + b~{\rm cot}~\langle\theta_2\rangle~d\widetilde y\big)^2 + d_2 \vert {\cal D}z_1\vert^2 +
d_6 \vert {\cal D}z_2\vert^2
\nd
where we have shifted $dz$ in a suggestive way with ($a, b$) constants,
so that the fibration represents a $U(1)$ fibration over the
two two-tori ${\cal D}z_1$ and ${\cal D}z_2$. We therefore expect the global extension should be:
\bg\label{laddu}
\big({\bf T}^2 \ltimes {\bf T}^2\big) \ltimes {\bf S}^1 ~ \to ~ {\bf S}^2 \times {\bf S}^3
\nd
with $\ltimes$ representing non-trivial fibration
topologically, so that we have heterotic five-branes wrapped on the resolved conifold. Note also that we have used
coordinates ${\widetilde x}, {\widetilde y}$ to denote the $U(1)$ fibration as we expect $d{\widetilde x},
d{\widetilde y}$ to be non-trivially related to ${\cal D}x, {\cal D}y, d\theta_i$. Therefore globally:
\bg
&& d_2 \vert {\cal D}z_1\vert^2 + d_6 \vert {\cal D}z_2\vert^2 ~\to ~ a_1 dS_1^2 + a_2 dS_2^2, ~~~~~
a_1 - a_2 ~=~ {\rm resolution ~parameter}\nonumber\\
&& dz +  a~{\rm cot}~\langle\theta_1\rangle~d\widetilde x + b~{\rm cot}~\langle\theta_2\rangle~d\widetilde y ~\to ~
d\psi +  a~{\rm cos}~ \widetilde\theta_1 ~d\phi_1
 + b~{\rm cos}~\widetilde\theta_2 ~d\phi_2
\nd
$S_i \equiv S_i(\widetilde\theta_i, \phi_i)$ represent squashed spheres with non-trivial complex structures, $a_i$
are functions of the internal coordinates (including $r$). Thus the complete global metric is:
\bg\label{globu}
ds^2_{\rm global} ~=~ a_3 dr^2 + ds^2_{{\bf S}^2 \times {\bf S}^3} ~\equiv~ g_{a, b} ~dR^a dR^b
\nd
where $R^a = (r_2, r_3, r_4, t_2, t_3, t_4)$ are the coordinates and $g_{a, b}$ are a slight variant of
the metric components
considered in the appendix 1 of \cite{chen}.
In the following we give a brief description to implicitly describe
the coordinate change necessary to compare with the metric written in terms of the usual resolved conifold coordinates
($r, \widetilde\theta_i, \phi_i, \psi$).
Noting that the $\bfC^*$ action on the homogeneous coordinates
of the resolved
conifold identifies $(z_1,z_2,z_3,z_4)$ with $(1,z_2/z_1,z_1z_3,z_1z_4)$, we
start with the coordinates $(U,Y,\lambda)$ of
\cite{pandoz}\footnote{With $\theta_i \to \widetilde\theta_i$ therein.},
which we see are related to our coordinates by
\bg\label{pztchange}
U=z_1z_3,\qquad Y=z_1z_4,\qquad \lambda=-\frac{z_2}{z_1}
\nd
due to a sign convention in \cite{pandoz}.  Using eqn (2.5) of \cite{chen} together
with (\ref{pztchange}),
we can express $(U,Y,\lambda)$ in terms of $(z_2,z_3,z_4)$, hence in terms of
the real coordinates $(r_2,r_3,r_4,t_2,t_3,t_4)$.  The desired
change of variables
comes from using (2.13) of \cite{pandoz}, which expresses $(U,Y,\lambda)$
instead in
terms of the desired six real variables ($r,\psi,\phi_i,\widetilde\theta_i$).

An interesting variation of the global scenario is to change the local metric such that the mapping becomes:
\bg\label{laddubaz}
\big({\bf T}^2 \times {\bf T}^2\big) \ltimes {\bf S}^1 ~ \to ~ {\bf S}^2 \times {\bf S}^3
\nd
instead of \eqref{laddu}. One simple way to decouple the two two-tori is to change the background three-form and
the dilaton such that the local metric becomes:
\bg\label{locket}
ds^2  = && {d_4 \over b_{x\theta_1} b_{y\theta_1}}\Big[b_{x\theta_1}^2
(dy - b_{y\theta_i}d\theta^i)^2 + b_{y\theta_1}^2(dx - b_{x\theta_j}d\theta^j)^2\Big] + d_3 dr^2 \nonumber\\
&& -2d_4 (dx - b_{x\theta_j}d\theta^j)(dy - b_{y\theta_i}d\theta^i) + d_5 dz^2 + d_6\vert d\chi_2\vert^2
\nd
with an additional constraint that the matrix:
\bg\label{matnu}
\begin{pmatrix} ~b_{x\theta_1}~ & ~b_{x\theta_2}~ \\ ~ b_{y\theta_1}~ & ~b_{y\theta_2}~\end{pmatrix}
\nd
has a vanishing determinant. This immediately tells us that $z_1$ torus is decoupled from $z_2$ torus, with
\bg\label{tau1}
{\cal D}z_1 ~\to ~ dz_1 ~\equiv ~ dx + \tau_1 dy, ~~~~~
\tau_1 ~=~ -{1-i\sqrt{3}\over 2} \sqrt{d_4 b_{x\theta_1}\over b_{y\theta_1}}
\nd
The global extension of the above local metric is
\bg \label{startmet}
&& ds^2 = h^{1/2} e^\phi d{\widetilde s}^2_{0123} + h^{-1/2} e^\phi ds^2_6, ~~~
{\cal H} = e^{2\phi} \ast d\left(e^{-2\phi} J\right), ~~~ d{\widetilde s}^2_{0123} = F_0 ds^2_{0123}\nonumber\\
&& ds^2_6 = F_1~ dr^2 + F_2 (d\psi + {\rm cos}~\theta_1 d\phi_1 + {\rm cos}~\theta_2 d\phi_2)^2  + \sum_{i = 1}^2 F_{2+i}
(d\theta_i^2 + {\rm sin}^2\theta_i d\phi_i^2)\nonumber\\
\nd
which is of course very close to
the configuration that we studied in \cite{chen} with ($h, \phi, F_0, F_i$) defined as before. 
Note that we have denoted the global three-form
and the dilaton by the earlier notation to avoid clutter. Thus the background \eqref{startmet} with
($g_{\mu\nu}, {\cal H}, \phi$) plus a vector bundle ${\bf V}$ represents the background for the wrapped
heterotic five-branes. We will discuss the vector bundle a little later.

After geometric transition, the local metric takes the following suggestive form (see also \cite{gtpaper1}):
\bg\label{mnview}
ds^2 ~& = ~&{\cal A}_1~\Big(dz + a_1~{\rm
cot}~\langle\theta_1\rangle~dx + b_1~{\rm
cot}~\langle\theta_2\rangle~dy\Big)^2 + {\cal A}_2~\Big[(dy^2 +
d\theta_2^2) + {1\over \vert\tau\vert^2}(dx^2 + d\theta_1^2)\Big] \nonumber\\
&& -2~{\cal A}_2 ~b_{y\theta_1}\Big[{\rm
sin}~\langle\psi\rangle (dy~d\theta_1 +dx~d\theta_2) + {\rm
cos}~\langle\psi\rangle (d\theta_1~d\theta_2 - dx~dy)\Big] +{\cal
A}_5~dr^2\nonumber\\
\nd
where ${\cal A}_i$ are constants locally, but will become non-constant when we extend the metric globally; and
$i\tau$ is the complex structure of $d\chi_2$ torus (the same one that we discussed above before geometric transition).
The B-field that allows for this metric is:
\bg\label{mnb}
B ~ = ~ b_{y\theta_1} \Big(\vert\tau\vert^2 dx \wedge d\theta_2 + dy \wedge d\theta_1\Big)
\nd
and the torsional equation is satisfied with an appropriate dilaton. Note that now we don't expect any five-branes
but ${\cal H}$ will not be closed. More on this soon.

The global extension of the above local metric is typically of the following form:
\bg\label{mnglob}
ds^2_{\rm het} ~ & = &~ ds^2_{0123} + {\cal
A}_1~ \left(d\psi + a_1~{\rm cos}~\theta_1~{\cal D}\phi_1 + b_1~{\rm
cos}~\theta_2~{\cal D}\phi_2\right)^2 + {\cal A}_3~ \left(d\theta_1^2 +
{\rm sin}^2\theta_1~{\cal D}\phi_1^2\right)\nonumber\\
&& + {\cal
A}_2~\left(d\theta_2^2 + {\rm sin}^2\theta_2~{\cal D}\phi_2^2\right) - 2
~{\cal A}_2~b_{y\theta_1}\Big[{\rm cos}~\psi~ (d\theta_1
~d\theta_2 - {\rm sin}~\theta_1 ~ {\rm sin}~\theta_2~{\cal D}\phi_1~
{\cal D}\phi_2)\nonumber\\
&& ~~~~~~~~~~~~~ + {\rm sin}~\psi~ ({\rm
sin}~\theta_1~{\cal D}\phi_1~d\theta_2 + {\rm
sin}~\theta_2~{\cal D}\phi_2~d\theta_1)\Big] + {\cal A}_5~dr^2
\nd
where ${\cal A}_i$ are no longer constants and ${\cal D}\phi_i \equiv d\phi_i + f_{ij} dx^j$ with $dx^j$ being the
internal coordinates and $f_{ij}$ are related to anti-symmetric two-form B-fields. In this paper we will assume
that $f_{ij} \approx 0$; and
for a very special choices of these coefficients:
\bg\label{aichoice}
&&{\cal A}_1 ~=~ {\cal A}_3 ~=~ {{\cal A}_5 \over 4} ~=~ {N \over 4}, ~~~~ {\cal A}_2
~=~ {N(e^{2g}+ a^2)\over 4}\nonumber\\
&& a(r) ~ = ~ -{2r \over {\rm sinh}~2r}, ~~~~~~ e^{2g} ~ = ~4
r~{\rm coth}~2r - {4 r^2 \over {\rm sinh}^2~2r} - 1
\nd
with $N$ being the number of wrapped five-branes, and with the following dilaton \cite{papad}:
\bg\label{dil}
e^{2\phi} ~ = ~ {e^{g + 2\phi_0}\over {\rm sinh}~2r}
\nd
the torsion ${\cal H}_{\rm MN}$ can be easily computed and it takes the following form \cite{gtpaper1}:
\bg\label{torsionnow}
{\cal H}_{\rm MN} ~ & \equiv &~e^{2\phi}~*~d(e^{-2\phi}J) \nonumber\\
  & =&-~{Na' \over 4}~\cos\psi~dr\wedge(d\theta_1\wedge
      d\theta_2-\sin\theta_1\sin\theta_2~d\phi_1\wedge d\phi_2)\nonumber\\
    && -~ {Na' \over 4}~\sin\psi~dr\wedge (\sin\theta_2~d\theta_1\wedge
      d\phi_2-\sin\theta_1~d\theta_2\wedge d\phi_1) \nonumber\\
    && +~{Na \over 4}~\sin\psi~d\theta_1\wedge d\theta_2\wedge
      (d\psi+\cos\theta_1~d\phi_1+\cos\theta_2~d\phi_2) \nonumber\\
    && -~{N\over 4}~(\sin\theta_1\cos\theta_2 - a\cos\psi\cos\theta_1 \sin\theta_2)
      ~d\theta_1\wedge d\phi_1\wedge d\phi_2 \nonumber\\
    && -~{N\over 4}~(\sin\theta_2\cos\theta_1 - a\cos\psi\cos\theta_2 \sin\theta_1)
      ~d\theta_2\wedge d\phi_1\wedge d\phi_2 \nonumber\\
    && -~{N\over 4}~\sin\theta_1~d\theta_1\wedge d\phi_1\wedge d\psi
      +{N\over 4}~\sin\theta_2~d\theta_2\wedge d\phi_2\wedge d\psi \nonumber\\
    && -~ {Na\over 4}\cos\psi~(\sin\theta_2~d\theta_1\wedge d\phi_2\wedge d\psi
      -\sin\theta_1~d\theta_2\wedge d\phi_1\wedge d\psi) \nonumber\\
    && -~ {Na \over 4}~\sin\psi\sin\theta_1\sin\theta_2~d\phi_1\wedge d\phi_2\wedge
      d\psi
\nd
with $J$ being the fundamental form. Interestingly, for this
$d{\cal H}_{\rm MN} = 0$, and we don't expect any five-brane sources.
This is a bit subtle now because in the heterotic theory
we expect:
\bg\label{kadumba}
d{\cal H} ~ = ~ \alpha'\big[{\rm tr}~R_+ \wedge R_+ - {\rm Tr}~F \wedge F\big]
\nd
and since ${\rm tr}~R_+ \wedge R_+$ is independent of $N$, the $N$ dependence of the dissolved small instantons
can come either from the torsion ${\cal H}$ or/and from the bundle ${\rm Tr}~F \wedge F$. (Here $R_+$ is the
curvature tensor with modified connection that will be discussed in sec. 4.) However the gauge group is
completely broken when the small instantons dissolve, and therefore:
\bg\label{kola}
{\rm Tr}~F \wedge F ~ = ~ 0
\nd
so with
$d{\cal H} = 0$ the Bianchi identity will be difficult to satisfy. Thus the only way would be to modify the
torsion \eqref{torsionnow} by a small amount so that both \eqref{kadumba} and \eqref{kola} are satisfied with the
torsion ${\cal H}$ defined as\footnote{We thank Juan Maldacena and Edward Witten for discussions on the above issues.}:
\bg\label{fohs}
{\cal H} ~ = ~ {\cal H}_{\rm MN} ~+ ~ {\cal H}_{\rm small}
\nd
where we presented an explicit form for ${\cal H}$ in {\bf Appendix A}.
Here ${\cal H}_{\rm small}$ is a small $N$-independent shift of the torsion.
For the choice \eqref{aichoice}, the
small $r$ limit is the Maldacena-Nunez background \cite{MN}. The large $r$, i.e the
UV limit of the theory, is given in \cite{baryonic}. Comparing \eqref{torsionnow} with \eqref{mnb} we see that the
coefficient of $d\theta_1 \wedge d\phi_2$ has the following terms:
\bg\label{compar}
&&  -~ {Na' \over 4}~\sin\psi~\sin\theta_2~dr + ~{Na \over 4}~\sin\psi ~\cos\theta_1~d\theta_1
+ ~{Na \over 4}~\cos\psi ~\sin\theta_1~d\psi \nonumber\\
&& -~{N\over 4}~(\sin\theta_2\cos\theta_1 - a\cos\psi\cos\theta_2 \sin\theta_1)~d\phi_2
\nd
and similarly for the coefficient of $d\theta_2 \wedge d\phi_1$. This would have been the natural extension of
\eqref{mnb}, but we see that \eqref{torsionnow} has extra terms that are not there in the local limit.
For example there are no such terms like:
\bg\label{extra}
{N\over 4} \left(dx \wedge d\theta_1 - dy \wedge d\theta_2 - a_0 dx \wedge dy\right)\wedge dz
\nd
with $a_0 \equiv a(r_0)$ in \eqref{mnb}.
Therefore, using all the above arguments, the
full global background is a deformation of the background \eqref{aichoice}, \eqref{dil} and \eqref{torsionnow} that
satisfies the Bianchi identity \eqref{kadumba} with the condition \eqref{kola}.
\begin{figure}[htb]\label{3roads}
        \begin{center}
\includegraphics[height=6cm]{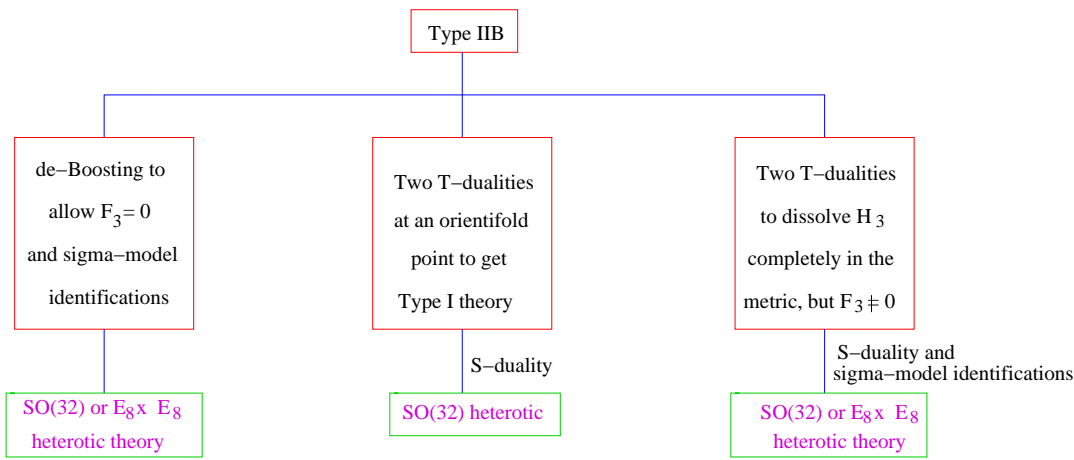}
        \caption{Three possible ways to get local heterotic backgrounds from a given
type IIB local background. These transformations are basically the transformations depicted as $T_1$ and $T_2$ in the
previous figure.
Note that the three paths are not {\it generic}, and in many cases may not exist at all.}
        \end{center}
        \end{figure}


\subsection{Evidence from sigma model identification}

The second evidence comes from the sigma model identification.
For the situation after the transition, one expects a (0,2) worldsheet sigma model for
${\cal N} = 1$ spacetime supersymmetry. The general idea is simple, and can be stated as follows.

To develop (0,2) models in the context of complex
structures we start by considering the following world sheet action for type IIB theory in the presence of $H_{\rm NS}$:
\bg\label{sigma1}
S = {1\over 8\pi \alpha'} \int d^2\sigma \Big[(g_{ij} +
B_{ij}) \partial_+ X^i \partial_- X^j + {1\over 4}S^g_{\rm fermionic}\Big]
\nd
where $S^g_{\rm fermionic}$ contains the standard kinetic term plus the following interaction part:
\bg\label{interact}
S^g_{\rm int}~ = ~ {i\over 8\pi \alpha'}
\int \Bigg(\psi^\rho \omega_+^{ab} \sigma_{ab}^{\rho\sigma} \psi^\sigma +
\psi^{\dot \rho} \omega_+^{ab} \sigma_{ab}^{\dot \rho \dot\sigma} \psi^{\dot\sigma} - {i\over 2} {\cal R}_{ijkl}
\sigma^{ij}_{\dot\rho \dot\sigma}\sigma^{kl}_{\kappa \gamma} \psi^{\dot\rho}
\psi^{\dot\sigma} \psi^{\kappa} \psi^{\gamma} \Bigg)
\nd
where ${\cal R}_{ijkl}$ is the background Riemann tensor.
In this action we have the freedom
to add non--interacting fields. This ruins the
carefully balanced (2,2) supersymmetry of this model. We can
use this to our advantage by adding non--interacting fields {\it
only} in the left--moving sector. This breaks the
left moving supersymmetry, and one might therefore hope
to obtain an action for (0,2) models from \eqref{sigma1}, at least {\it
classically}. On the other hand, a possible (0,2) action is also restricted
because this will be the action for heterotic string. Therefore let us start with
the following {\it naive} steps to find the
classical (0,2) action from a given (2,2) action (see \cite{gtpaper1} for more details):

\vskip.1in

\noindent $\bullet$ Keep the right moving sector unchanged, i.e. $\psi^p$ remain as before.

\noindent $\bullet$ In the left moving sector, replace $\psi^{\dot q}$ by eight
fermions $\Psi^a$, $a = 1, ... 8$. Also add 24 additional non--interacting
fermions $\Psi^b$, $b = 9, ... 32$. In other words:
\bg\label{gobir}
\Psi^A = \begin{pmatrix} ~\psi^{\dot q}~\\ ~\Psi^9~\\ ~...~\\~\Psi^{32}~\end{pmatrix}
\nd
\noindent $\bullet$ Replace $\omega_+$ by gauge fields $A$, i.e. embed the {\it torsional}
spin connection into the gauge connection.

\vskip.1in

The above set of transformations will convert the classical (2,2)
action given in \eqref{sigma1} to a classical (0,2) one. One might,
however, wonder about the Bianchi identity in the heterotic theory in light of the discussions that we had in the
previous subsection.
The type IIB three--form fields are closed, whereas heterotic
three--form fields satisfy the Bianchi identity. One immediate
reconciliation would be that  because of the {\it embedding}
$\omega_+ = A$, the heterotic three--form should be closed. This
may seem like an admissible solution to the problem, but because
of subtleties mentioned earlier\footnote{See also \cite{bbdg, smit, papad2} for
additional subtleties that come from the above
embedding. In fact even in the usual case this embedding  will not allow any compact non--K\"ahler manifolds to appear in
the heterotic theory.} this cannot be the story here. Therefore an
embedding of the
form:
\bg\label{bedd}
A_i^{AB} ~ = ~ \begin{pmatrix} ~\omega_{i+}^{ab} ~ & ~0~ \\ ~0~ & ~{\cal O}(\alpha')~
\end{pmatrix}
\nd
(where for simplicity we have left the off-diagonal part vanishing) cannot quite be the solution for our case as we
require:
\bg\label{bian}
d{\cal H}_{\rm small} = \alpha' \left[d\omega_+ \wedge d\omega_+ + {\cal O}(\omega_+^4)\right]
\nd
Thus one possibility will be to make the gauge field vanishing and replace all connections by the torsional 
connection $\omega_+$. 
Using
this the new action with (0,2) supersymmetry becomes:
\bg\label{daku}
S = {1\over 8\pi \alpha'}&& \int d^2\sigma
\Big[(g_{ij} + B_{ij}) \partial_+ X^i \partial_- X^j + i \psi^p (\Delta_+
\psi)^p + i \Psi^A (\Delta_-\Psi)^A + {\cal
O}(\alpha')\Big]\nonumber
\nd
 where due to the Bianchi identity \eqref{bian} and our choice, there are no $F^a_{ij}$
Yang--Mills field strength.
The fermion indices are $A = 1,..., 32$, which means
there are 32 fermions, and $T^a$ form tensors of
rank 16. The
Laplacians are given as follows:
\bg\label{laplacians}
&& \Delta_-\Psi^A = \partial_-\Psi^A, ~~
\Delta_+\psi^p = \partial_+\psi^p + {1\over
2}(\omega_+)^{ab} \sigma^{pq}_{ab} \psi^q \nonumber\\
&& H_{ijk} = {1\over
2}\left(B_{ij,k} + B_{jk,i} + B_{kj,i}\right)_{({\rm MN})} + {\cal O}(\alpha')
\nd
As expected, this
set of actions determines the (0,1) supersymmetric heterotic sigma
model. This is similar to the (1,1) action for the type
II case. The full (0,2) susy will be determined by additional
actions on the fields (exactly as for the (1,1) case before).

The above discussion is a simple way to see how certain IIB backgrounds can be dragged
directly to the heterotic side by making small modifications to the field contents. The above discussion was
solely for the heterotic background after the transition where the heterotic gauge group is completely broken.
The situation then is completely different for the case before the transition.

To study the heterotic background before geometric transition using our sigma model identification we take the
type IIB background given as equation (4.13) in \cite{chen} and ``de-boost'' the system so that we can have
\bg\label{deboost}
F_3 ~ = ~ F_5 ~ = ~ 0
\nd
The de-boosting procedure follows the {\it reverse} chain of dualities depicted in {\bf figure 1} of \cite{chen}.
Once we have this, then it is obvious that the background before GT is precisely \eqref{startmet}. In the type I
language, which is the S-dual of \eqref{startmet}, the type I D5-branes are the small instantons of the nine-branes
gauge theory. For $N$ D5-branes, or $N$ small instantons in the heterotic theory the gauge symmetry before the
transition can be written succinctly as:
\bg\label{daku}
Sp(2N) ~ \times ~ {\cal G}
\nd
where ${\cal G}$ is a {\it subgroup} of the full $SO(32)$ group in the heterotic theory. The $SO(32)$ is broken
by the Wilson lines. These Wilson lines are related to the distances between the type IIB seven-branes in the full
F-theory picture.

This means that the sigma model before the transition is exactly given by an ADHM sigma model \cite{ADHM}
(much like the one
discussed in \cite{douglasgauge}). One may then understand the geometric transition to be related to an
Affleck-Dine-Seiberg (ADS) \cite{ADS} type superpotential being added to the usual ADHM sigma model superpotential. This
is much like the discussion in Klebanov-Strassler \cite{ks} where the
addition of an equivalent ADS superpotential to the usual
${\cal N} = 1$ quartic superpotential shows how one could go from a conifold to a deformed conifold. More details
of this will appear elsewhere.


\subsection{A more direct analysis}

The third evidence comes from {\it dissolving} the NS three-form completely in the metric in the
IIB picture. We can do it in the
presence of an orientifold action also. First, however, let us try without involving any orientifold action i.e
keep only the wrapped D5-branes. The local geometry is well known to have the following form:
\bg\label{localmet}
ds^2 ~=~ && dr^2 + \Bigg(dz + \sqrt{\gamma' \over \gamma}~r_0 ~{\rm
cot}~\langle\theta_1\rangle~ dx  + \sqrt{\gamma' \over (\gamma+4a^2)~}~r_0~ {\rm
cot}~\langle\theta_2\rangle~dy\Bigg)^2~ + \nonumber\\
&& ~~~~~~~~~~~ + \Bigg[{\gamma\sqrt{h} \over 4}~d\theta_1^2 + dx^2\Bigg] +
\Bigg[{(\gamma + a^2)\sqrt{h} \over 4}~d\theta_2^2 + dy^2\Bigg] + ....
\nd
where all the coefficients are described in \cite{chen, gtpaper1}. There is also a B-field given by
\bg\label{bnslocal}
B_{\rm NS} = b_{x\theta_1} dx \wedge d\theta_1 + b_{y\theta_2} dy \wedge d\theta_2
\nd
plus of course there are $F_3$ and $F_5$ fields whose orientations will be discussed soon. Recall also that all the
coefficients in the above metric are constants. This is going to be useful soon.

Under two T-dualities along ($x, \theta_2$) directions, the B-field \eqref{bnslocal} completely dissolves in the
metric to give us the following non-K\"ahler geometry:
\bg\label{locnkg}
ds^2~=~&&dr^2 + \big[dz + \Delta_1 {\rm cot}\langle\theta_1\rangle (dx - b_{x\theta_1}d\theta_1) +
\Delta_2 {\rm cot}\langle\theta_2\rangle dy\big]^2 \nonumber\\
&& ~+ \big[\alpha_1 d\theta_1^2 + (dx - b_{x\theta_1}d\theta_1)^2\big] + \big[dy^2 + \alpha_2
(d\theta_2 - b_{y\theta_2}dy)^2\big]
\nd
where the coefficients appearing in the metric are defined in terms of the coefficients of \eqref{localmet} appearing
above. The metric \eqref{locnkg} is a non-K\"ahler deformation of \eqref{localmet}. The complex structures of the
base tori change from $\tau_k = i \sqrt{\alpha_k}$ to:
\bg\label{cmpxs}
&&\tau_1 ~=~ -b_{x\theta_1} + i \sqrt{\alpha_1}, ~~~~~~~ \tau_2 ~=~
-{\alpha_2 b_{y\theta_2}\over 1+ \alpha_2 b^2_{y\theta_2}} + i{\sqrt{\alpha_2}\over 1+ \alpha_2
b^2_{y\theta_2}}\\
&&dz_1 ~\equiv~ dx + \tau_1 d\theta_1, ~~~~~~ dz_2~=~ dy + \tau_2 d\theta_2, ~~~~~~ {\cal D}z ~\equiv ~ dz -
\Delta_1 b_{x\theta_1}{\rm cot}\langle\theta_1\rangle ~d\theta_1\nonumber
\nd
so that the metric \eqref{locnkg} takes the following suggestive format:
\bg\label{suggest}
ds^2 ~=~ dr^2 + \big({\cal D}z + \Delta_1 {\rm cot}\langle\theta_1\rangle~dx + \Delta_2 {\rm cot}\langle\theta_2\rangle
~dy\big)^2 + \vert dz_1\vert^2 + {\vert\tau_2\vert^2_0\over \vert\tau_2\vert^2} \vert dz_2\vert^2
\nd
where $\vert\tau_2\vert_0$ is the complex structure of the second torus in the absence of B-fields. Note that the
coefficients in front of the $dz_1$ and $dz_2$ tori are different. This means that the global extension of the
local metric \eqref{locnkg} should have two two-spheres of unequal sizes which, in other words, should be a
resolved conifold. Thus the five-branes wrap two-cycle of a non-K\"ahler resolved conifold, exactly as we have
been considering earlier!

There are still a few loose ends that we need to
tie up before we go to the analysis of the
geometry after the transition. For example: What happens at the orientifold point? How do the seven-branes behave
in the final T-dual set-up? What happened to the RR three and five-forms?

To understand these issues let us analyse the system carefully. As discussed in \cite{chen} there are two possible ways
to perform orientifolding operation here. The first O-action has already been discussed in the previous subsections.
The second O-action is given by eq. (4.6) of \cite{chen} i.e ($x, \theta_1$) $\to$ ($-x, \pi - \theta_1$). 
For this action we can
keep the D5-branes {\it parallel} to the
seven-branes once we are away from the orientifold point.
Such a configuration breaks supersymmetry. However we can form a {\it bound-state} of D5- and
D7-branes that is supersymmetric. This configuration is then different from the one studied before. In the full
global geometry we can assume that the bound state is embedded in a non-trivial way in the non-K\"ahler resolved
conifold space, and the fluxes give rise to a dipole-deformation of the bound state
(see \cite{dipole, gtpaper3} for more details).

This picture can also be understood as an $N$ D5-brane bound states on a single D7-brane with all other
seven-branes moved away in the ($x, \theta_1$) direction. To go to the heterotic side we need to go to the
orientifold point. At the O-point
the world-volume gauge fluxes are all projected out because only dipole deformations are
allowed. However note that in the far IR the resolution cycle is very small and therefore the wrapped D5-branes are
fractional three-branes in the IIB set-up. Thus at the orientifold point we may rotate the three-form fluxes (i.e the
five-brane sources) so that the local RR field is the following (see also \cite{gtpaper2})\footnote{Note that the 
rotation keeps one component of the three-form fluxes along the orbifold direction. As is 
well known, this requirement is enough to 
survive the O-action.}:
\bg\label{ghonti}
B_{RR} ~= &&~{\widetilde b}_{xz} dx \wedge dz + {\widetilde b}_{xy} dx \wedge dy + {\widetilde b}_{x\theta_2}
dx \wedge d\theta_2 +  {\widetilde b}_{y\theta_1} dy \wedge d\theta_1\nonumber\\
 && + {\widetilde b}_{z\theta_1} dz \wedge d\theta_1  + {\widetilde b}_{\theta_1\theta_2} d\theta_1 \wedge d\theta_2
\nd
where ${\widetilde b}_{\alpha\beta} = {\widetilde b}_{\alpha\beta}(r, \theta_1, \theta_2)$, which also means that the
D5-branes at the O-point form a non-trivial surface that could still
be viewed as a bound state with the seven-branes. The $B_{\rm NS}$ in \eqref{bnslocal} will give rise to
dipole deformation at the O-point. The heterotic metric then is \eqref{locnkg} whose global extension should be the
non-K\"ahler resolved conifold mentioned before.

Rest of the analysis follow similar route as outlined earlier. The gravity dual should in general be non-geometric, but
if we concentrate only on the geometric
portion of the moduli space of solutions, the gravity dual is a small deformation
over the Maldacena-Nunez geometry at the far IR. This small deformation cannot be ignored, otherwise Bianchi identity
will not be satisfied.

\begin{figure}[htb]\label{hetlandscape}
        \begin{center}
\includegraphics[height=6cm]{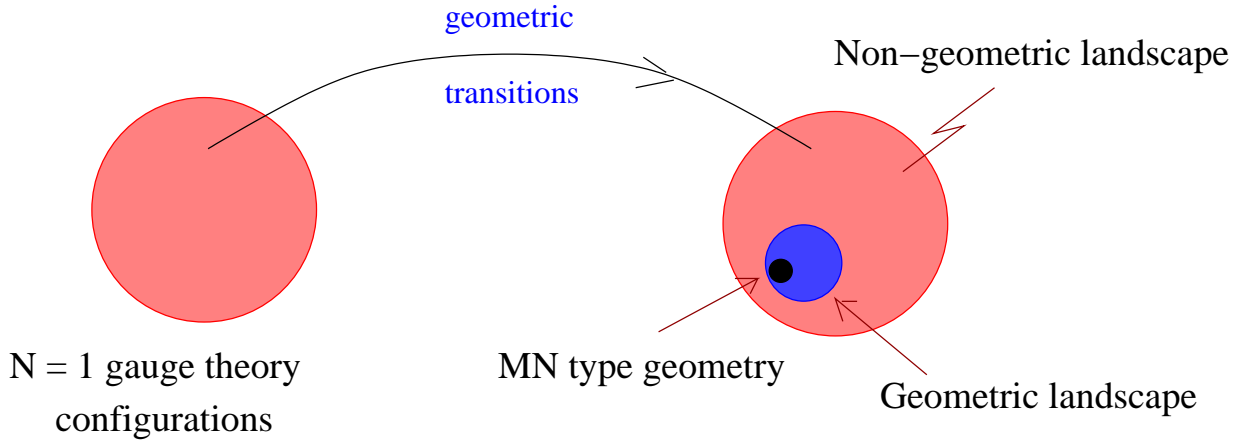}
        \caption{The heterotic transitions corresponding to various gauge theory deformations. We see that
the dual configurations are generically non-geometric. The geometric (but non-Kahler) regions are shown in blue. The
back dot is the warped deformed conifold geometry, or the MN type geometry.}
        \end{center}
        \end{figure}

\section{Analysis of torsion classes}

Now that we have few evidence that suggest that there is a possibility to describe the strongly coupled
theory on the heterotic five-branes using a gravitational description, we should seriously check the supersymmetry
of the underlying gravitational solutions. In fact as pointed out in \cite{chen} and in section 1.1, the issue of
supersymmetry is subtle: there is no immediate guarantee that the backgrounds in IIA, M-theory and in IIB are susy
after performing all the transformations. In this section we will therefore analyse the torsion classes in all the
intermediate theories and ask under what conditions susy could be preserved. Such an analysis will also help us
fix many of the free parameters in the intermediate theories.

A manifold with SU(3)-structure has all the group-theoretical
features of a Calabi-Yau, namely invariant two and three forms, $J$
and $\Omega$ respectively. On a manifold of SU(3) holonomy, not only
$J$ and $\Omega$
 are well defined, but they are also
closed: $dJ = 0 = d\Omega $. If they are not closed, $dJ$ and
$d\Omega$  give a good measure of how far the manifold is from
having SU(3) holonomy \cite{Grana:2004bg}
\begin{eqnarray}\label{structeq}
dJ&=&-\frac{3}{2}\textrm{Im} (W_1\bar{\Omega})+W_4\wedge J+W_3\nonumber\\
d\Omega &=&W_1J^2+W_2\wedge J+ \bar{W_5}\wedge \Omega
\end{eqnarray}
The $W$'s are the $(3\oplus \bar{3}\oplus 1)\otimes(3\oplus
\bar{3})$ components of the intrinsic torsion: $W_1$ is a complex
zero-form in $1\oplus 1$, $W_2$ is a complex primitive two-form, so
it lies in $8\oplus 8$, $W_3$ is a real primitive $(2,1)\oplus(1,2)$
form and it lies in $6\oplus \bar{6}$, $W_4$ is a real one-form in
$3\oplus \bar{3}$, and finally $W_5$ is a complex $(1, 0)$-form, so
its degrees of freedom are again $3\oplus \bar{3}$.

It is sometime convenient to express the torsion classes using another definition.
These have appeared in the literature in the following guise (see also \cite{Lopes Cardoso:2002hd} for
more details):
\begin{eqnarray}
&& d\Omega_{\pm}\wedge J=\bar{W}_{1}^\pm~J\wedge J\wedge
J,\quad\quad d\Omega_{\pm}^{(2,2)}
=\bar{W}_{1}^{\pm}~J\wedge J+\bar{W}_{2}^{\pm}~\wedge J,\nonumber\\
&& dJ^{(2,1)}=[J\wedge \bar{W}_4]^{(2,1)}+\bar{W}_3,\quad\,
\bar{W}_4=\frac{1}{2}J \lrcorner ~dJ, \quad
\bar{W}_5=\frac{1}{2}\Omega_+ \lrcorner ~d\Omega_+.
\end{eqnarray}
where $\bar{W}_1=\bar{W}_1^++\bar{W}_1^-$,
$\bar{W}_2=\bar{W}_2^++\bar{W}_2^-$ and the contraction operator
$\lrcorner$ is defined as
\begin{eqnarray}
&&\lrcorner: \bigwedge^k
T^\ast\otimes\bigwedge^nT^\ast \rightarrow  \bigwedge^{n-k}T^\ast\\
&&(L_{k},M_n) \mapsto
\frac{1}{n!}c_n^kL^{a_1...a_k}M_{a_1...a_n}e^{a_{k+1}...e^{a_n}}
\end{eqnarray}
The two definitions are related as
\begin{eqnarray}
&&W_1=\bar{W}_1^++i\bar{W}_1^-,\quad
W_2=\bar{W}_2^++i\bar{W}_2^-\nonumber\\
&&W_3^{(2,1)}=\bar{W}_3, \quad W_4=\bar{W}_4, \quad
\textrm{Re}(W_5)=-\bar{W}_5
\end{eqnarray}

\subsection{Torsion classes in the heterotic theory}

In the following we will study the torsion classes for the heterotic
string theories before and after geometric transition. Before
geometric transition we have the heterotic string theory as in
\cite{chen}
\begin{eqnarray}
ds^2=k^{-2}e^{2\phi} F_0~ds_{0123}^2+k^2~ds_6^2
\end{eqnarray}
where $k^2(r,\theta_1,\theta_2)=h^{-1/2}e^{\phi}$ and
\begin{eqnarray}
ds_6^2&=&F_1~dr^2+F_2~(d\psi+\cos\theta_1d\phi_1+\cos\theta_2d\phi_2)^2+F_3~(d\theta_1^2+\sin^2\theta_1d\phi_1^2)
\nonumber\\
~~~~~~&&+F_4~(d\theta_2^2+\sin^2\theta_2d\phi_2^2)
\end{eqnarray}
where $F_i$ are functions\footnote{In this paper we will not investigate the case where $F_i$ are more 
generic functions of ($r, \theta_i, \phi_i, \psi$). The analysis with generic 
$F_i$ will definately be techically challenging and will 
give us a bigger moduli space of solutions, but the physics will remain unchanged.} of $r$. 
The metric is of course related to the type IIB metric studied in \cite{chen} and
discussed further in section 1.1.
The NS three-form flux is:
\begin{eqnarray}
H_3&=&e^{2\phi}\ast_6 d(e^{-2\phi}J)\nonumber\\
&=&\frac{F_4\sqrt{F_2}\sin\theta_2}{F_3\sqrt{F_1}\sin\theta_1}\Big[k^2(\sqrt{F_1F_2}\sin\theta_1+2\phi_rF_3\sin\theta_1-F_{3r}\sin\theta_1)\nonumber\\
&&-2kF_3k_r\sin\theta_1\Big]d\theta_2\wedge d\phi_2\wedge
(d\psi+\cos\theta_1d\phi_1+\cos\theta_2d\phi_2)\nonumber\\
&&+\frac{F_3\sqrt{F_2}\sin\theta_1}{F_4\sqrt{F_1}\sin\theta_2}\Big[k^2(\sqrt{F_1F_2}\sin\theta_2+2\phi_rF_4\sin\theta_2-F_{4r}\sin\theta_2)\nonumber\\
&&-2kF_4k_r\sin\theta_2\Big]d\theta_1\wedge
d\phi_1\wedge (d\psi+\cos\theta_1d\phi_1+\cos\theta_2d\phi_2)\nonumber\\
&&+2k\sqrt{F_1F_2}\sin\theta_2(k\phi_{\theta_2}-k_{\theta_2})d\phi_2\wedge
(d\psi+\cos\theta_1d\phi_1+\cos\theta_2d\phi_2)\wedge dr\nonumber\\
&&+2k\sqrt{F_1F_2}\sin\theta_1(k\phi_{\theta_1}-k_{\theta_1})d\phi_1\wedge
(d\psi+\cos\theta_1d\phi_1+\cos\theta_2d\phi_2)\wedge dr\nonumber\\
&&+2kF_4\sin\theta_1\sin\theta_2(k\phi_{\theta_1}-k_{\theta_1})d\phi_1\wedge
d\theta_2\wedge d\phi_2\nonumber\\
&&+2kF_3\sin\theta_1\sin\theta_2(k\phi_{\theta_2}-k_{\theta_2})d\theta_1\wedge
d\phi_1\wedge d\phi_2
\end{eqnarray}
which in turn will soon be related to $W_3$. Here we have defined $k_\alpha \equiv \partial_\alpha k$
and $F_{n\alpha} \equiv \partial_\alpha F_n$, with $\alpha$ being any of the internal coordinates.
Now to see the precise connection of $H_3$ with $W_3$ we first need to write the vielbeins
for the internal space. They are given by:
\begin{eqnarray}
e^1&=&k\sqrt{F_3}(\cos\psi_1
d\theta_1+\sin\psi_1\sin\theta_1d\phi_1),\quad
e^2=k\sqrt{F_3}(-\sin\psi_1
d\theta_1+\cos\psi_1\sin\theta_1d\phi_1),\nonumber\\
e^3&=&k\sqrt{F_4}(\cos\psi_2
d\theta_2+\sin\psi_2\sin\theta_2d\phi_2),\quad
e^4=k\sqrt{F_4}(-\sin\psi_2,
d\theta_2+\cos\psi_2\sin\theta_2d\phi_2)\nonumber\\
e^5&=&k\sqrt{F_2}~(d\psi+\cos\theta_1d\phi_1+\cos\theta_2d\phi_2),\quad
e^6=k\sqrt{F_1}~dr
\end{eqnarray}
The fundamental two-form $J$ and holomorphic three-form $\Omega$ are
defined in terms of these vielbeins  as
\begin{eqnarray}
J&=&e^1\wedge e^2+e^3\wedge e^4+ e^5\wedge e^6\\
\Omega &=&(e^1+i~e^2)\wedge (e^3+i~e^4)\wedge (e^5+i~e^6)
\end{eqnarray}
which will immediately give us the following values for $W_1$ and $W_2$:
\bg\label{w12}
W_1~ = ~ W_2 ~ = ~0
\nd
implying that the internal manifold is a complex manifold. The three-form NS flux would be directly related to $W_3$
if we demand $d(\ast_6 H_3)=0$. In that case $W_3=\ast_6 H_3$. Thus for our case $W_3$ is:
\bg\label{w3}
W_3=&&\frac{k^2}{2}(F_4F_{3r}-F_3F_{4r}+F_3\sqrt{F_1F_2}-F_4\sqrt{F_1F_2})\nonumber\\
&&\quad\quad\quad\quad\left(\frac{\sin\theta_1}{F_4}dr\wedge
d\theta_1\wedge d\phi_1
-\frac{\sin\theta_2}{F_3}dr\wedge d\theta_2 \wedge d\phi_2\right)
\nd
The above three torsion classes told us how to put three-form flux on a complex manifold. However supersymmetry is still
not guaranteed. To demand supersymmetry we first require to compute $W_4$ and ${\rm Re}~W_5$. For our case they are
given by:
\bg\label{w45}
&&W_4=\left(\frac{F_{3r}-\sqrt{F_1F_2}}{2F_3\sqrt{F_1}}+\frac{F_{4r}-\sqrt{F_1F_2}}{2F_4\sqrt{F_1}}+\frac{2k_r}{k}\right)
dr
+\frac{2k_{\theta_1}}{k}d\theta_1+\frac{2k_{\theta_2}}{k}d\theta_2,\nonumber\\
&&\textrm{Re}~W_5 =\left(\frac{3k_r}{2k}+\frac{F_{2r}}{4F_2}
+\frac{F_{3r}}{4F_3}+\frac{F_{4r}}{4F_4}-\frac{1}{2}\sqrt{\frac{F_1}{F_2}}\right)dr
+\frac{3k_{\theta_1}}{k}d\theta_1+\frac{3k_{\theta_2}}{k}d\theta_2
\nd
The SUSY condition requires $2W_4=\textrm{Re}~ W_5$ so $k$ is a
function of only $r$. The susy requirement from the torsion classes gives rise to the following constraint equation
for the warp factors $F_i$:
\bg\label{ghontida}
&& \left({1\over \sqrt{F_1}} - {1\over 4}\right)\partial_r {\rm log}~F_3  +
\left({1\over \sqrt{F_1}} - {1\over 4}\right)\partial_r {\rm log}~F_4 - {1\over 4}\partial_r {\rm log}~F_2
+{5\over 2} \partial_r {\rm log}~k\nonumber\\
&& = \left({1\over F_3} + {1\over F_4}\right) \sqrt{F_2} -{1\over 2}\sqrt{F_1\over F_2}
\nd
This constraint equation would also be the one that we will need to impose on the
 type IIB side. From \cite{bbdg} we expect $e^{\phi} = h^{-1/2} F_0^{-1}$ to also come out of the torsional 
constraint \eqref{ghontida}. However observe that \eqref{ghontida} is independent of $F_0$, so $F_0$ is a 
{\it free} parameter for the background. Therefore without loss of generality one may choose 
\bg\label{dokir}
e^{-\phi} ~=~ h^{1/2} F_0 ~=~ {F^{3/2}_0 {\rm cosh}\beta \over \sqrt{1 + F_0^2 {\rm sinh}^2\beta}}
\nd
as in \eqref{kdive},
which will keep the spacetime part of the metric independent of any warp factor. This would then be consistent 
with the solutions of \cite{bbdg}.

After the geometric transition, as we discussed in details earlier, we expect the generic solution to look like
\eqref{mnglob}. The
torsion classes for this case is a special case of \eqref{mliftnow}
with all the fibration vanishing, i.e., $b_{ij}=0$ where
$i,j= r, \psi, \theta_i, \phi_i$ (see also {\bf Appendix B} for more details).

\subsection{Torsion classes in type IIA theory}

We want to make sure that during our duality chain discussed in \cite{chen} the solutions we
take are all supersymmetric, as the final heterotic metric after the transition will come from
dualising the type IIB case after geometric transition.
The starting point of IIB solution is
obviously supersymmetric (provided the warp factors satisfy the constraint equation \eqref{ghontida}).
However, after the coordinate
transformation and shift in the metric it is not obvious that the
IIA solutions before and after flop are still supersymmetric. Therefore in this
section we will
calculate the torsion classes for the IIA solutions and
impose constraints to make them supersymmetric.

Before the flop the IIA metric takes the similar form as the
deformed conifold as discussed in \cite{chen}:
\begin{eqnarray}\label{mliftnow}
ds^2_{10}=&&F_0ds_{0123}^2+F_1 dr^2 + {\alpha F_2
\over \Delta_1 \Delta_2} \Big[d\psi -b_{\psi r}dr - b_{\psi\theta_2} d\theta_2\nonumber\\
&&+ \Delta_1 {\rm cos}~\theta_1 \Big(d\phi_1 - b_{\phi_1\theta_1}
d\theta_1-b_{\phi_1 r}dr\Big)+ \Delta_2 {\rm cos}~\theta_2 {\rm
cos}~\psi_0 \Big(d\phi_2 - b_{\phi_2\theta_2} d\theta_2-b_{\phi_2
r}dr\Big)\Big]^2\nonumber\\
&& + \alpha j_{\phi_2\phi_2}\Big[d\theta_1^2 + \Big(d\phi_1 -
b_{\phi_1\theta_1} d\theta_1-b_{\phi_1 r}dr\Big)^2\Big] \nonumber\\
&&+ \alpha j_{\phi_1\phi_1}\Big[d\theta_2^2 + \Big(d\phi_2 -
b_{\phi_2\theta_2} d\theta_2-b_{\phi_2
r}dr\Big)^2\Big]\nonumber\\
&& + 2\alpha j_{\phi_1\phi_2}{\rm cos}~\psi\Big[d\theta_1 d\theta_2
- \Big(d\phi_1 - b_{\phi_1\theta_1} d\theta_1-b_{\phi_1
r}dr\Big)\Big(d\phi_2 - b_{\phi_2\theta_2} d\theta_2-b_{\phi_2
r}dr\Big)\Big]\nonumber\\
&& + 2\alpha j_{\phi_1\phi_2}{\rm sin}~\psi\Big[\Big(d\phi_1-
b_{\phi_1\theta_1} d\theta_1-b_{\phi_1 r}dr\Big) d\theta_2
+ \Big(d\phi_2 - b_{\phi_2\theta_2} d\theta_2-b_{\phi_2r}dr\Big)d\theta_1\Big]\nonumber\\
\end{eqnarray}
As before, to compute the torsion classes we need the veilbeins. For our case, they are:
\begin{eqnarray}
&&e^1=\sqrt{F_1}dr, \quad e^3=K~d\theta_1, \quad
e^4=-K~(d\phi_1-b_{\phi_1\theta_1}d\theta_1-b_{\phi_1
r}dr),\nonumber\\
&&e^2=G(d\psi-b_{\psi
r}dr+\Delta_1\cos\theta_1(d\phi_1-b_{\phi_1\theta_1}d\theta_1-b_{\phi_1
r}dr)+\Delta_2\cos\theta_2(d\phi_2-b_{\phi_2\theta_2}d\theta_2-b_{\phi_2
r}dr)),\nonumber\\
&&e^5=L~\Big[\sin\psi(d\phi_2-b_{\phi_2\theta_2}d\theta_2-b_{\phi_2
r}dr)+\cos\psi d\theta_2-a~d\theta_1\Big],\nonumber\\
&&e^6=L~\Big[\cos\psi(d\phi_2-b_{\phi_2\theta_2}d\theta_2-b_{\phi_2
r}dr)-\sin\psi
d\theta_2-a~(d\phi_1-b_{\phi_1\theta_1}d\theta_1-b_{\phi_1
r}dr)\Big]\nonumber\\
\end{eqnarray}
where $G, L, K$ and $a$ are defined as:
\begin{eqnarray}
G&=&\sqrt{\frac{\alpha F_2}{\Delta_1\Delta_2}}, \quad L=\sqrt{\alpha
j_{\phi_1\phi_1}}, \quad K=\sqrt{\alpha
\Big(j_{\phi_2\phi_2}-\frac{j_{\phi_1\phi_2}^2}{j_{\phi_1\phi_1}}\Big)},
\quad a=\frac{j_{\phi_1\phi_2}}{j_{\phi_1\phi_1}}~~~~~~
\end{eqnarray}
From the above vielbeins we can write the complex vielbeins as:
\begin{eqnarray}
E_1=e^1+i~e^2, \quad E_2=e^3+i~(Ae^4+Be^6),\quad
E_3=e^5+i~(Be^4-Ae^6)
\end{eqnarray}
with $A$ and $B$ as functions of the radial direction $r$ which in turn are
determined by the $SU(3)$ structure of the underlying manifold satisfying $A^2+B^2=1$.

With all these preparations, we are now ready to write the torsion classes. They are given by:
\begin{eqnarray}
&&W_1=-\frac{1}{6L^2K^2\sqrt{F_1}GA}\Big(-i L^2GK^2
b_{\phi_2\theta_2,r}AB -2i  L^2AG b_{\psi r}
  K^2B +2 L^2A\sqrt{F1}K^2B \nonumber\\
&&~~~+2iG  L^3K  \cos\psi^2 b_{\phi_2\theta_2 ,r}aB^2-i  L^3GK
b_{\phi_2\theta_2,r}aB^2 +i L^3G b_{\phi_1\theta_1,r}Ka\nonumber\\
&&~~~ -i L^3G b_{\phi_1\theta_1,r}KaB^2-i L^2G
b_{\phi_1\theta_1,r}K^2AB+G^2B \sqrt{F1}A L^2 \Delta_1\sin
 \theta_1  a^2\nonumber\\
 &&~~~ +G^2B  \sqrt{F_1}B \Delta_2\sin\theta_2L^2 +G^2B \sqrt{F_1}A \Delta_2 \sin \theta_2
  K^2+2G  L^3K\sin\psi b_{\phi_2\theta_2,r}\cos  \psi  A a \nonumber\\
  &&~~~ +2
  L^2GBK^2-2iG  L^3K  \cos  \psi
  ^2 b_{\phi_2\theta_2,r}a +i  L^3GK b_{\phi_2\theta_2,r}a
L^2K^2\sqrt{F_1}GB\Big)\nonumber\\
&& W_4=w_{4r}dr+w_{4\theta_1}d\theta_1+w_{4\theta_2}d\theta_2+w_{4\phi_1}d\phi_1+w_{4\phi_2}d\phi_2\nonumber\\
&&{\rm Re}~W_5 =w_{5e^1}e^1+
w_{5e^2}e^2+w_{5e^3}e^3+w_{5e^4}e^4+w_{5e^5}e^5+w_{5e^6}e^6
\end{eqnarray}
where $w_i$ are given in {\bf Appendix B}. Once we know $W_1$, $W_4$
and $W_5$, it is easy to calculate $W_2$ and $W_3$ from
\ref{structeq}. We will not give the explicit expressions here. Note
also that since $W_1, W_2$ are not zero, the type IIA manifold is
not complex. This is of course consistent with our earlier works
\cite{gtpaper1, gtpaper2, gtpaper3, chen}. The supersymmetry
condition imposes the following constraints: \bg\label{susconstr}
&&2w_{4\theta_1} = Kw_{5e^3} + K b_{\phi_1\theta_1} w_{5e^4} - a L w_{5e^5}+ a L b_{\phi_1\theta_1} w_{5e^6}\nonumber\\
&&2w_{4r} = \sqrt{F_1}~w_{5e^1}+K b_{\phi_1 r} w_{5e^4} - L {\rm
sin}~\psi b_{\phi_2 r} w_{5e^5} - L(a b_{\phi_1 r} - {\rm cos}~\psi
b_{\phi_2 r})w_{5e^6}\nonumber\\
&&2 w_{4\theta_2} = L(-{\rm sin}~\psi b_{\phi_2 \theta_2} + {\rm
cos}~\psi) w_{5e^5} -
L ({\rm cos}~\psi b_{\phi_2 \theta_2} + {\rm sin}~\psi) w_{5e^6}\nonumber\\
&&2w_{4\phi_2} = L( {\rm sin}~\psi ~w_{5e^5} + {\rm cos}~\psi
~w_{5e^6}), ~~~~ 2w_{4\phi_1} = -a L ~w_{5e^6} - K~w_{5e^4}
 \nd
with $w_{5e^2} = 0$.
These conditions are in addition to the condition \eqref{ghontida},
and therefore would constrain the warp factors further.

After the flop the IIA metric takes the similar form as the resolved
conifold \cite{chen}:
\begin{eqnarray}\label{IIAmetric}
ds_{10}^2&& =F_0ds_{0123}^2+F_1dr^2+e^{2\phi}\Big[d\psi-b_{\psi
\mu}dx^\mu+\Delta_1 {\rm cos}~\theta_1 \Big(d\phi_1 -
b_{\phi_1\theta_1}
d\theta_1-b_{\phi_1 r}dr\Big)\nonumber\\
&&\quad\quad\quad\quad\quad\quad\quad\quad\quad\quad +
\widetilde{\Delta}_2 {\rm cos}~\theta_2 \Big(d\phi_2 -
b_{\phi_2\theta_2}
d\theta_2-b_{\phi_2 r}dr\Big)\Big]^2\nonumber\\
&&~+e^{2\phi\over
3}a^2(k^2G_2+kG_3+G_1)\Big[d\theta_1^2+(d\phi_1^2-b_{\phi_1\theta_1}d\theta_1-b_{\phi_1
r}dr)^2\Big]\nonumber\\
&&~+e^{2\phi\over 3}b^2(\mu^2G_2+\mu
G_3+G_1)\Big[d\theta_2^2+(d\phi_2^2-b_{\phi_2\theta_2}d\theta_2-b_{\phi_2
r}dr)^2\Big]
\end{eqnarray}
the definitions of coefficients are the same as in
\cite{chen}. We define
\begin{eqnarray}
\mathcal{F}_1^2&=&e^{2\phi\over 3}a^2(k^2G_2+kG_3+G_1)\nonumber\\
\mathcal{F}_2^2&=&e^{2\phi\over 3}b^2(\mu^2G_2+\mu G_3+G_1)
\end{eqnarray}
To determine the supersymmetry condition now, we follow the same procedure, namely, compute the torsion classes.
In the following we give the general torsion classes assuming that
the fields and metric can depend on angular coordinate
$\theta_1$ and $\theta_2$ also. But first we need the vielbeins for the metric \eqref{IIAmetric}. They are:
\bg\label{jkalo}
e^1&=&\mathcal{F}_1d\theta_1, ~~ e^3=\mathcal{F}_2d\theta_2, ~~ e^6=\sqrt{F_1}dr \nonumber\\
e^2&=&\mathcal{F}_1(d\phi_1-b_{\phi_1\theta_1}d\theta_1-b_{\phi_1
r}dr), ~~ e^4=\mathcal{F}_2(d\phi_2-b_{\phi_2\theta_2}d\theta_2-b_{\phi_2
r}dr)\nonumber\\
e^5&=&e^{\phi}[d\psi-b_{\psi
r}dr+\Delta_1\cos\theta_1(d\phi_1-b_{\phi_1\theta_1}d\theta_1-b_{\phi_1
r}dr)+\Delta_2\cos\theta_2(d\phi_2-b_{\phi_2\theta_2}d\theta_2-b_{\phi_2
r}dr)]\nonumber\\
\nd
Using these vielbeins it is a straightforward (but nevertheless tedious) exercise to
determine the torsion classes. This time we find the following values for the torsion classes:
\bg\label{torsa} && W_1=\frac{1}{6\mathcal {F}_1\mathcal
{F}_2\sqrt{F}_1}(\mathcal {F}_1^2b_{\phi_1r,\theta_2}+\mathcal
{F}_2^2b_{\phi_2r,\theta_1}-e^{2\phi}\Delta_2\cos\theta_2\sqrt{F}_1b_{\phi_2\theta_2,\theta_1}\nonumber\\
&&~~~+e^{2\phi}\Delta_1\cos\theta_1\sqrt{F}_1b_{\phi_1\theta_1,\theta_2}+i~e^{2\phi}\sqrt{F}_1(\Delta_{2,\theta_1}\cos\theta_2-\Delta_{1,\theta_2}\cos\theta_1))\nonumber\\
&&W_4=\frac{\mathcal {F}_{2,\theta_1}+\mathcal
{F}_{2}\phi_{\theta_1}}{\mathcal {F}_2}d\theta_1+\frac{\mathcal
{F}_{1,\theta_2}+\mathcal {F}_{1}\phi_{\theta_2}}{\mathcal
{F}_1}d\theta_2-\frac{1}{2\mathcal {F}_1\mathcal {F}_2}(\mathcal
{F}_1^2e^{2\phi}\Delta_{2}\sin\theta_2\sqrt{F}_1\nonumber\\
&&~~~-\mathcal{F}_2^2e^{2\phi}\Delta_{1}\sin\theta_1\sqrt{F}_1-\mathcal
{F}_1^2e^{2\phi}\Delta_{2,\theta_2}\cos\theta_2\sqrt{F}_1-\mathcal
{F}_2^2e^{2\phi}\Delta_{1,\theta_1}\cos\theta_1\sqrt{F}_1\nonumber\\
&&~~~-2\mathcal {F}_1^2\mathcal {F}_2\mathcal {F}_{2,r}-2\mathcal
{F}_2^2\mathcal {F}_1\mathcal {F}_{1,r})dr\nonumber\\
&&{\rm Re}~W_5 =\frac{e_1}{\sqrt{F_1}\mathcal {F}_1^2\mathcal
{F}_2} \Big(e^{\phi}\mathcal
{F}_1\mathcal{F}_2\Delta_{1,r}\cos\theta_1-2\mathcal{F}_1\mathcal{F}_{2,\theta_1}\sqrt{F_1}
-2\mathcal{F}_{1,\theta_1}\mathcal{F}_2\sqrt{F}_1-\mathcal{F}_1\mathcal{F}_2\phi_{\theta_1}
\Big)\nonumber\\
&&~~~+\frac{-e^2}{\mathcal{F}_1^2\mathcal{F}_2\sqrt {{\it
F_1}}}\Big({\it \Delta_2}\,\cos  {\it \theta_2}
e^{\phi}\mathcal{F}_1\mathcal{F}_2 {\it b_{\phi_2 r}}_{{{\it
\theta_1}}}-{\it \Delta_1}\,\cos {\it \theta_1}
e^{\phi}\mathcal{F}_1{\it b_{\phi_1
\theta_1}}_{{r}}\mathcal{F}_2\nonumber\\
&&~~~+{\it \Delta_1}\,\cos {\it \theta_1} e^{\phi}\mathcal{F}_1{\it
b_{\phi_1 r}}_{{{\it
\theta_1}}}\mathcal{F}_2+e^{\phi}\mathcal{F}_1\mathcal{F}_2{\it
b_{\psi r}}_{{{ \it \theta_1}}}-\sqrt {{\it
F_1}}\mathcal{F}_1\mathcal{F}_2{\it b_{\phi_2 \theta_2}}_{{{\it
\theta_1}}}\Big)\nonumber\\
&&~~~+\frac {e^3}{\sqrt {{\it F_1}}\mathcal{F}_2^{2
}\mathcal{F}_1}\Big(-2\,\mathcal{F}_{{{1,
\theta_2}}}\mathcal{F}_2\sqrt {{\it
F_1}}-2\,\mathcal{F}_1\mathcal{F}_{{{2, \theta_2} }}\sqrt {{\it
F_1}}-\mathcal{F}_1\mathcal{F}_2\phi_{{{\it \theta_2}}}\sqrt {{\it
F_1}}+e^{\phi}\mathcal{F}_1\mathcal{F}_2{\it \Delta_2}_{{r}}\cos
{\it \theta_2} \Big)\nonumber\\
&&~~~+\frac{-e^4}{\sqrt { {\it
F_1}}\mathcal{F}_2^2\mathcal{F}_1}\Big(-{\it \Delta_2}\,\cos {\it
\theta_2} e^{\phi}\mathcal{F}_1\mathcal{F}_2{ \it b_{\phi_2
\theta_2}}_{{r}}+{\it \Delta_2}\,\cos  {\it \theta_2}
e^{\phi}\mathcal{F}_1\mathcal{F}_2{\it b_{\phi_2 r}}_{{{\it
\theta_2}}}\nonumber\\
&&~~~+{\it \Delta_1}\,\cos {\it \theta_1} e^{\phi}\mathcal{F}_1{\it
b_{\phi_1 r}}_{{{\it
\theta_2}}}\mathcal{F}_2+e^{\phi}\mathcal{F}_1\mathcal{F}_2{\it
b_{\psi r}}_{{{ \it \theta_2}}}-\sqrt {{\it F_1}}\mathcal{F}_1{\it
b_{\phi_1 \theta_1}}_{{{\it
\theta_2}}}\mathcal{F}_2\Big)\nonumber\\
&&~~~+\frac{e^5}{\sqrt {{\it
F_1}}\mathcal{F}_1\mathcal{F}_2}\Big(-k{\it b_{\phi_1
\theta_1}}_{{r}}\mathcal{F}_2+k{\it b_{\phi_1 r}}_{{{\it
\theta_1}}}\mathcal{F}_2-\mathcal{F}_1\mathcal{F}_2{\it b_{\phi_2
\theta_2}}_{{r}}+\mathcal{F}_1\mathcal{F}_2{\it b_{\phi_2
r,\theta_2}}\Big)\nonumber\\
&&~~~+\frac{-e^6}{\sqrt {{\it
F_1}}\mathcal{F}_1\mathcal{F}_2}\Big(2\,\mathcal{F}_{{1,r}}\mathcal{F}_2+2\,\mathcal{F}_1\mathcal{F}_2\phi_r+2\,\mathcal{F}_1\mathcal{F}_{2,r}\Big)
\nd
where $W_2$ and $W_3$ can be easily determined from the above information. Note that the type IIA manifold after
geometric transition is again a non-complex non-K\"ahler manifold as we would have expected. The supersymmetry
condition:
$$2W_4 ~ = ~ {\rm Re}~W_5$$
will put further constrains on the parameters of the background. Combining the other two set of constraints:
\eqref{ghontida} and \eqref{susconstr} we can fix most of the parameters of our background. The remaining parameters,
which are not fixed by our constraint equations, will give rise to a class of backgrounds corresponding to various
gauge theory deformations, as shown in {\bf figure 3}.

Combining all the ideas together, we see that a careful analysis of the torsion classes for various intermediate
configurations allow us to present explicit supersymmetric solutions for the geometric transitions. This would then
not only justify the supersymmetric cases studied in \cite{chen}, but also the new heterotic configurations that
we present in this paper. Therefore combining the above set of arguments, stemming from torsion classes and
explicit backgrounds analysis, we believe, should  strongly justify the new
heterotic duality that we conjecture in this paper.

\section{Vector bundles through conifold transition}

To study the vector bundles we will start from the anomaly condition for the heterotic theory with torsion.
As emphasised in \cite{bbdg} a more useful way to express the anomaly condition is to {\it complexify} the
heterotic three-form $H$ to $G$, and write the anomaly condition as:
\bg\label{hetcond}
G ~=~ dB + \alpha'\left[\Omega_3({\omega_+}) - \Omega_3(A)\right]
\nd
where ${\omega_+}$ is the modified spin-connection, now described using the one-form $\widetilde{G}$, defined
in the following way:
\bg\label{modspin}
{\omega_+} = \omega - {1\over 2}\widetilde{G}, ~~~~{\rm with}~~~
\widetilde{G}  \equiv G e^{-2} \equiv {G}_{ijk}e^{aj}
e^{bk}
\nd
The complexified three-form is very useful in many analysis, for example writing the superpotential \cite{beckerpot}
or the action, and can be expressed in terms of the real three-form 
${\cal H}$ of the heterotic theory complexified with the
geometrical data \cite{bbdg}.
One may easily show that the Chern-Simons term related to the
torsional-spin connection $\omega_+$ is given by 
\bg\label{torpo}
\Omega_3(\omega_+) = \Omega_3(\omega) + {1\over
4} \Omega_3(\widetilde{G}) -{1\over 2} (\omega \wedge {R}_{\widetilde{G}} + {\widetilde{G}}\wedge {R}_\omega)
\nd
where we define $\Omega_3(\widetilde{G})$ in somewhat
similar way as $\Omega_3(A)$
or $\Omega_3(\omega)$:
\bg\label{rhdefo}
\Omega_3(\widetilde{G})=\widetilde{G} \wedge d{\widetilde{G}}
- {1\over 3} \widetilde{G}\wedge \widetilde{G} \wedge
\widetilde{G}
\nd
The quantity ${R}_{\widetilde{G}}$ is the curvature
polynomial due to the torsion and is defined as
\bg\label{ruha}
{R}_{\widetilde{G}}= d\widetilde{G} - {1\over 3}
\widetilde{G} \wedge \widetilde{G}
\nd
whereas ${R}_\omega$ differs from the usual curvature
polynomial by $-{1\over 3}\omega \wedge \omega$. In fact, we can
write in a more compact form as
\bg\label{honu}
\Omega_3(\omega_+) = {\Big (}\omega - {1\over 2}
\widetilde{G}{\Big )} {\Big (}{R}_\omega - {1\over 2} {R}_{\widetilde{G}}{\Big )}
\nd
with the curvature polynomials defined
above\footnote{In this form it is instructive to compare with the other
choice of torsional-spin connection ${\omega_-}$
$$\Omega_3(\omega_-) ={\Big (}\omega +{1\over 2}
\tilde{\cal H}{\Big )} {\Big (}{\cal R}_\omega + {1\over 2}{\cal
R}_{\tilde{\cal H}} + {1\over 3} \tilde{\cal H} \wedge \tilde{\cal
H}{\Big ),}$$
which differs from in relative signs and an additional
term.}.

{}From the above analysis it is easy to infer, what the background
torsion is. If we concentrate only to the lowest order in
$\alpha'$ and linear order in ${G}$, the three-form
background is given by
\bg\label{thfobgis}
{G} = dB {\Big (} 1 - {\alpha'\over 2} {
R}_\omega e^{-2}  {\Big )} + \alpha'\Omega_3(\omega) + {\cal O}(\alpha'^2)
\nd
where we have imposed $\Omega_3(A) = 0$ because the gauge group is completely broken.
To all orders in ${G}$
and $\alpha'$ the equation, that we need to solve is
\bg\label{mexi}
{G} +{\alpha'\over 2}{\Bigg [} \omega \wedge {
R}_{\widetilde{G}} + {\widetilde{G}}\wedge {\cal R}_\omega
-{1\over 2} {\widetilde{G}} \wedge {R}_{\widetilde{
G}}{\Bigg ]}
 = dB + \alpha'\Omega_3(\omega)\equiv f
\nd
Thus $f$ will
have a term linear in $\alpha'$. Using this,
the solution for $G$ is written in terms of powers of $\alpha'$ in the following way:
\bg\label{jimkaal}
G ~= ~ \sum \alpha'^n H_n ~ +~ {i\over \sqrt{\alpha'}}\sum \alpha'^n h_n
\nd
where $n$ goes from zero onwards. As discussed in \cite{bbdg}, the various terms in $G$ can be presented as:
\bg\label{gterms}
&&h_0 - {1\over 12} \widetilde{h_0}\wedge \widetilde{h_0}\wedge \widetilde{h_0} = 0\nonumber\\
&& H_0 = -{f\over 2} + {1\over 4} {\rm Tr}~\left(\widetilde{h_0}\wedge d\widetilde{h_0}\right) + {1\over 6} {\rm Tr} ~
\left(\omega_0\wedge \widetilde{h_0}\wedge \widetilde{h_0}\right)\nonumber\\
&& h_1 - {1\over 4}{\rm Tr}~\left(\widetilde{h_0}\wedge \widetilde{h_0}\wedge
\widetilde{h_1}\right) = -{1\over 2} {\rm Tr}~\left(\omega_0\wedge d\widetilde{h_0}\right) + {1\over 3} {\rm Tr} ~\left(\omega_0\wedge \widetilde{H_0}\wedge
\widetilde{h_0}\right)\\
&&~~~~ -{1\over 2} {\rm Tr}~\left(\widetilde{h_0}\wedge R_{\omega_0}\right) +
{1\over 4}{\rm Tr}~\left(\widetilde{H_0}\wedge d\widetilde{h_0} + \widetilde{h_0}\wedge d\widetilde{H_0}\right) - {1\over 4} {\rm Tr}~\left(\widetilde{H_0}\wedge \widetilde{H_0}\wedge \widetilde{h_0}\right)\nonumber
\nd
where the tilde terms are one-forms constructed out of three-forms using vielbeins as in \eqref{modspin} and the subscript
$0$ denotes zeroth order\footnote{We write $\omega$ and $J$ as
$\omega = \sum \alpha'^n \omega_n, ~~ J = \sum \alpha'^n J_n$ to compare terms order by order in $\alpha'$. This is
discussed in more details in \cite{bbdg}.}
 in $\alpha'$.
Solving the
above set of equations give us the following:
\bg\label{lbazo}
G = -{1\over 2} d(B+iJ_0) + \alpha'\Omega_3(\omega_0) + ~{\rm corrections}
\nd
where, as one would have expected, the complexified K\"ahler form appears naturally from our analysis. The corrections are
both to $J_0$ as well as well as to higher orders in $\alpha'$. The $-{1\over 2}$ coefficient can be absorbed by redefining
$G$. Once we do that, we could rewrite the real part of $G$, i.e ${\cal H}$, in the following
way:
\bg\label{realG}
{\cal H} = f + {\alpha'\over 2}\left(\omega_0\wedge \widetilde{f} \wedge \widetilde{f}
+ \widetilde{f}\wedge R_{\omega_0} +
{1\over 2} \widetilde{f}\wedge d\widetilde{f} - {1\over 6} \widetilde{f}\wedge
\widetilde{f}\wedge \widetilde{f}\right)
\nd
where $f = dB + \alpha'\Omega_3(\omega)$ as defined in \eqref{mexi} above. Since we know ${\cal H}$ from 
{\bf Appendix A}, we can determine $f$ or ${\widetilde f}$ to lowest order in $\alpha'$ by solving the cubic 
equation \eqref{realG}. 

Therefore the story after the transition is simple: the torsion and the metric are the only information needed to
specify the dual geometry. On the other hand before the transition the situation is more involved. There
is a non-trivial vector bundle:
\bg\label{lcal}
Sp(2N) ~\times ~ {\cal G}
\nd
where, as mentioned earlier, the gauge group ${\cal G}$ comes from the type IIB seven-branes. Various distributions
of the F-theory seven-branes {\it a-la} \cite{ftheory} will give various ${\cal G}$. If $\widetilde{\cal H}$ denotes the
torsion before the transition, we expect $d\widetilde{\cal H}$ to have contributions from ${\rm Tr}~F_{\cal G} \times
F_{\cal G}$ as well as from ${\rm Tr}~ F_{Sp(2N)} \times F_{Sp(2N)}$. As before, the torsion $\widetilde{\cal H}$
will have two parts: one proportional to $N$ and the other independent of $N$. The part independent of $N$ could be
balanced by the torsional curvature and the ${\cal G}$-bundle. It will be interesting to work out the full picture as
both sides, before and after the transition, require careful analysis of the Bianchi identity.
A more detailed analysis of how to pull the
bundle \eqref{lcal} through the conifold point will be discussed elsewhere.

\section{Conclusion and discussions}

In this paper we gave some evidence for the gravity dual of large $N$ heterotic small instantons. We pointed out that
geometric transition in the heterotic set-up is related to small instanton transition under which the
small instantons smoothen out. This way the $Sp(2N)$ gauge symmetry before the transition is completely broken and
therefore in the dual side we no longer have branes or vector bundles, but only torsion. For certain
cases the gravity duals are {\it deformations} of the corresponding type II cases because of the underlying Bianchi
identity.

\noindent We left many questions unanswered. For example:

\vskip.1in

\noindent $\bullet$ Can this way of thinking be extended to the type IIB case also? Recall that the IIB D3-branes
are small instantons on the seven-branes in the full F-theory set-up. Therefore before the transition we can move the
seven-branes along the Coulomb branch so that the susy remains unbroken at low scales. Then presumably the
large $N$ limit of D3-branes could be studied via this mechanism. It would be miraculous to recover AdS target 
space from ADHM sigma model. 

\vskip.1in

\noindent $\bullet$ In the heterotic side the vector bundle is completely broken. So to satisfy the Bianchi identity
we cannot allow a {\it closed} three-form. However in IIB there might be a situation where all the D3-brane small
instantons smoothen out on a {\it subset} of the allowed seven-branes. The gauge fields on these seven-branes become all
massive, but we can still have non zero ${\rm Tr} ~F \wedge F$ from the other seven-branes.
Therefore we might be able to allow for a
closed three-form and still satisfy the Bianchi identity.

\vskip.1in

\noindent $\bullet$ One of the issue that we skimmed over is the ADHM sigma model and possible contributions to
the world-sheet superpotential. The precise question is as follows.
Could there be an Affleck-Dine-Seiberg (ADS) like contribution to
the ADHM sigma model that can tell us how the target space changes from a non-K\"ahler resolved conifold to
a non-K\"ahler deformed conifold (or even to a non-geometric one)?

\vskip.1in

\noindent $\bullet$ In the type IIB case the effect of 
world-volume quartic potential plus the ADS contribution can also be seen
from the Gukov-Vafa-Witten type {\it bulk} superpotential \cite{GVW}. Now that we know the heterotic superpotential
\cite{beckerpot, lust1}, we should be able to see the connection between this superpotential and the total
ADHM superpotential.

\vskip.1in

\noindent $\bullet$ Is it possible to understand the full cascading dualities from this perspective? This may be
more tricky because in type I we don't have D3-brane degrees of freedom. But maybe they all can be understood directly
from the F-theory viewpoint where the D3-branes are small instantons on the seven-branes, and the D5-branes (that are
not parallel to the seven-branes) are in fact T-dual to type I small 
instantons\footnote{In fact there is already a hint that such deformation of the instantons that we see in the 
heterotic side should have an equivalent story in the T-dual of the IIB geometric transition. This construction 
has appeared in the second reference of \cite{kyungho} some time back, and here we will elaborate the story 
very briefly.  
The IIA brane construction after the last cascade will be $M$ D4-branes in the interval {\it between} the 
two orthogonal  NS5-branes, and no D4-branes on the other side. Once we shrink the ${\bf P^1}$ to zero size, the 
two NS5-branes in the T-dual picture come together. To see the subsequent behavior,
we lift this to M-theory. There the SUSY condition is preserved only when the shrunk D4-branes (or M5-branes now) 
deform to form a {\it diamond} structure between the two M5-branes (see \cite{ohta} for more details about this 
construction). Therefore the final configuration is like two intersecting M5-branes 
with the intersection ``point" blown up and 
the M5-branes between the two orthogonal M5-branes virtually dissolved.
The T-dual type IIB configuration will give us a deformed conifold and no D3-branes. Note that the deformation of the 
shrunken M5-branes exactly create the extra metric components required to convert the resolved conifold geometry to a
deformed one in the T-dual IIB side. 
This is almost like the small instanton story: the small instantons deform and become geometry. The only difference is
that in IIA/M-theory the curved M5-branes become $M$ planar M5-branes and 
consequently lose the $U(M)$ gauge symmetry to end up with $M$ $U(1)$'s.
In heterotic theory the $k$ instantons blow up to lose the $Sp(2k)$ gauge symmetry, 
but now due to background ${\cal G}$-subgroup of $SO(32)$ all the gauge symmetries are completely broken.}.

\vskip.1in

\noindent $\bullet$ We discussed how the MN solution \cite{MN} should be deformed slightly to satisfy the Bianchi
identity. However we did not compute the actual deviations of the components of the metric or the three-form
in this paper. Although this is technically challenging as
the components of ${\rm tr}~ R_+ \wedge R_+$ from the metric \eqref{mnglob} are rather 
unwieldy (see also {\bf Appendix A} for the form for ${\cal H}$), 
it will
nevertheless be an interesting exercise to get the background precisely. This will then 
provide another confirmation of the heterotic duality.

\vskip.1in

\noindent $\bullet$ Last but not the least, we haven't said anything about the $E_8 \times E_8$ case. As
discussed in the introduction, here the story may follow similar line of thought as in \cite{ganorhanany}. We will
discuss about this case and hopefully some of the above mentioned points in the sequel.

\vskip.1in

\noindent {\bf Note:} As we were writing this draft two interesting papers appeared in the archive that studied 
some aspect of the story in a slightly different guise \cite{martspark}. The second paper in \cite{martspark} 
studied some aspect of heterotic/CFT ($0, 2$) sigma model. It would be interesting to relate them to our results.
Some other papers with some indirect relations to our work can be found in \cite{halmagyi}. 

\vskip.1in

\centerline{\bf Acknowledgement}

\noindent Its is our pleasure to thank
Andrew Frey, Ori Ganor, Marc Grisaru, Sheldon Katz,
Juan Maldacena and Edward Witten for many helpful discussions
and correspondences. We would also like to thank Sheldon Katz for initial participation on this project. The
works of FC,  KD and PF are supported in parts by NSERC grants. The work of RT is supported by the PPARC grant.


\newpage

\appendix

\section{The torsion for the dual gravitational background}\label{Hmdfd}

The modified ${\cal H}$ for the heterotic background \eqref{mnglob} is rather involved because of the non-trivial
fibration structure in the definition of ${\cal D}\phi_i$. However if we consider the simpler case where
${\cal D}\phi_i \approx d\phi_i$ the three-form, or the torsion, can be easily found. Under this simplification
${\cal H}$ is given by:
\begin{eqnarray}
{\cal H}&=&\frac {1}{\sqrt {{\it A_5}}{K}^{2}L\sqrt
{1-{B}^{2}}}(-BK{L}^{2}a_{{r}}+{B}^{3}K{L}^{2}a_{{r}}-{L}^{2}a
\left( 1-{B}
^{2} \right) BK_{{r}}\nonumber\\
&&-2\,{L }^{2}{a}^{3}\sqrt {1-{B}^{2}}\sin \left( \psi \right)
L_{{r}}{B}^{2}-{ L}^{2}{a}^{2} \left( 1-{B}^{2} \right) \sin \left(
\psi \right) B_{{r}
}K\nonumber\\
&&- \left( \cos \left( \psi \right)  \right) ^{2}{L}^{3}a\sqrt
{1-{B}^ {2}}a_{{r}}+{L}^{3}{a}^{2}\sqrt {1-{B}^{2}}\sin \left( \psi
\right) a_
{{r}}\nonumber\\
&&-aB{K}^{2}\sin \left( \psi \right) LB_{{r}}\sqrt {1-{B}^{2}}-{L}^
{2}{a}^{2} \left( 1-{B}^{2} \right) \sin \left( \psi \right) BK_{{r}}\nonumber\\
&&+
 \left( \cos \left( \psi \right)  \right) ^{2}La \left( 1-{B}^{2}
 \right) L_{{r}}BK+ \left( \cos \left( \psi \right)  \right) ^{2}{L}^{
3}a\sqrt {1-{B}^{2}}a_{{r}}{B}^{2}\nonumber\\
&&-2\, \left( \cos \left( \psi
 \right)  \right) ^{2}{L}^{2}{a}^{2}\sqrt {1-{B}^{2}}L_{{r}}+2\,a{B}^{
2}{K}^{2}\sin \left( \psi \right) \phi_{{r}}L\sqrt {1-{B}^{2}}\nonumber\\
&&+2\,{L}^{3} {a}^{2}\sqrt
{1-{B}^{2}}\phi_{{r}}{B}^{2}-2\,{L}^{3}{a}^{3}\sqrt {1-{B}^{ 2}}\sin
\left( \psi \right) \phi_{{r}}+2\, \left( \cos \left( \psi
 \right)  \right) ^{2}{L}^{3}{a}^{2}\sqrt {1-{B}^{2}}\phi_{{r}}\nonumber\\
&&-L{a}^{2}
 \left( 1-{B}^{2} \right) \sin \left( \psi \right) L_{{r}}BK+ \left(
\cos \left( \psi \right)  \right) ^{2}{L}^{3}{a}^{2}\sqrt
{1-{B}^{2}}B
B_{{r}}\nonumber\\
&&-L{a}^{2} \left( 1-{B}^{2} \right) \sqrt {{\it A_5}}G{\it \Delta_2
}\,\sin \left( {\it \theta_2} \right) -2\, \left( \cos \left( \psi
 \right)  \right) ^{2}{L}^{2}a \left( 1-{B}^{2} \right) \phi_{{r}}BK\nonumber\\
&&-a{B} ^{2}{K}^{2}\sin \left( \psi \right) L_{{r}}\sqrt
{1-{B}^{2}}- \left( 1 -{B}^{2} \right) L\sqrt {{\it A_5}}G{\it
\Delta_1}\,\sin \left( {\it
\theta_1} \right) \nonumber\\
&&+ \left( \cos \left( \psi \right)  \right) ^{2}{L}^{2} a \left(
1-{B}^{2} \right) B_{{r}}K-La \left( 1-{B}^{2} \right) L_{{r}
}BK\nonumber\\
&&+ \left( \cos \left( \psi \right)  \right) ^{2}{L}^{2}a \left(
1-{B }^{2} \right) BK_{{r}}-a{B}^{3}K\sin \left( \psi \right)
{L}^{2}a_{{r}
}-{a}^{2}{B}^{2}K\sin \left( \psi \right) {L}^{2}B_{{r}}\nonumber\\
&&+2\,{L}^{2}{a} ^{2} \left( 1-{B}^{2} \right) \sin \left( \psi
\right) \phi_{{r}}BK+aBK \sin \left( \psi \right)
{L}^{2}a_{{r}}-\sqrt {1-{B}^{2}}{L}^{3}aa_{{r
}}{B}^{2}\nonumber\\
&& -{L}^{3}{a}^{3}\sqrt {1-{B}^{2}}\sin \left( \psi \right)
BB_{
{r}}-{L}^{2}a \left( 1-{B}^{2} \right) B_{{r}}K\nonumber\\
&&-2\,{L}^{3}{a}^{2}\sqrt {1-{B}^{2}}\phi_{{r}}+2\,\sqrt
{1-{B}^{2}}LKK _{{r}}+2\,{L}^{2}{a}^{2}\sqrt
{1-{B}^{2}}L_{{r}}\nonumber\\
&&-2\,\sqrt {1-{B}^{2}}L
\phi_{{r}}{K}^{2}-{B}^{2}{K}^{2}L_{{r}}\sqrt {1-{B}^{2}}\nonumber\\
&&+\sqrt {1-{B}^{2} }{L}^{3}aa_{{r}}-2\,\sqrt
{1-{B}^{2}}LBB_{{r}}{K}^{2}-2\, \left( \cos
 \left( \psi \right)  \right) ^{2}{L}^{3}{a}^{2}\sqrt {1-{B}^{2}}\phi_{{r
}}{B}^{2}\nonumber\\
&&-aBK\sqrt {{\it A_5}}G{\it \Delta_2}\,\sin \left( {\it \theta_2}
 \right) \sqrt {1-{B}^{2}}-{L}^{3}{a}^{2}\sqrt {1-{B}^{2}}\sin \left(
\psi \right) a_{{r}}{B}^{2}\nonumber\\
&&+2\,{L}^{2}{a}^{3}\sqrt {1-{B}^{2}}\sin
 \left( \psi \right) L_{{r}}+2\,{L}^{2}a \left( 1-{B}^{2} \right) \phi_{{
r}}BK+2\, \left( \cos \left( \psi \right)  \right)
^{2}{L}^{2}{a}^{2}
\sqrt {1-{B}^{2}}L_{{r}}{B}^{2}\nonumber\\
&&+2\,{a}^{2}{B}^{3}K\sin \left( \psi
 \right) \phi_{{r}}{L}^{2}-2\,{a}^{2}BK\sin \left( \psi \right) \phi_{{r}}{L
}^{2}-2\,{a}^{2}{B}^{3}K\sin \left( \psi \right) L_{{r}}L\nonumber\\
&&+2\,{a}^{2}BK \sin \left( \psi \right)
L_{{r}}L-{L}^{3}{a}^{2}\sqrt {1-{B}^{2}}BB_{{ r}}+4\,\sqrt
{1-{B}^{2}}L\phi_{{r}}{K}^{2}{B}^{2}\nonumber\\
&&+2\,{L}^{3}{a}^{3}\sqrt {1-{B}^{2}}\sin \left( \psi \right)
\phi_{{r}}{B}^{2}-a{B}^{2}K\sin
 \left( \psi \right) LK_{{r}}\sqrt {1-{B}^{2}}\nonumber\\
&&-3\,\sqrt {1-{B}^{2}}LKK
_{{r}}{B}^{2}-2\,{L}^{2}{a}^{2}\sqrt {1-{B}^{2}}L_{{r}}{B}^{2})~E_1\wedge E_3 \wedge E_4\nonumber\\
&&+ \frac {1}{\sqrt {{\it A_5}}{K}^{2}L\sqrt {1-{B}^{2}}}(-2\,
\left( \cos \left( \psi \right)  \right) ^{2}{B}^{3}{L}^{
3}{a}^{2}\phi_{{r}}+2\, \left( \cos \left( \psi \right)  \right)
^{2}B{L} ^{3}{a}^{2}\phi_{{r}}\nonumber\\
&&+ \left( \cos \left( \psi \right)  \right) ^{2}{B}^{2}{L}^{
3}{a}^{2}B_{{r}}+2\, \left( \cos \left( \psi \right)  \right)
^{2}{B}^ {3}{L}^{2}{a}^{2}L_{{r}}-2\, \left( \cos \left( \psi
\right)  \right)
^{2}B{L}^{2}{a}^{2}L_{{r}}\nonumber\\
&&+a \left( 1-{B}^{2} \right) K\sin \left( \psi \right) LBK_{{r}}+a
\left( 1-{B}^{2} \right) {K}^{2}\sin \left(
\psi \right) LB_{{r}}-{B}^{2}LB_{{r}}{K}^{2}\nonumber\\
&&-a\sqrt {1-{B}^{2}}K{L}^{2
}BB_{{r}}-2\,BL\phi_{{r}}{K}^{2}+2\,BLKK_{{r}}+2\,{B}^{3}L\phi_{{r}}{K}^{2}-
2\,{B}^{3}LKK_{{r}}
\nonumber\\
&&+a \left( 1-{B}^{2} \right) K\sqrt {{\it A_5}}G{\it
\Delta_2}\,\sin
 \left( {\it \theta_2} \right) -BL\sqrt {{\it A_5}}G{\it \Delta_1}\,\sin
 \left( {\it \theta_1} \right) \sqrt {1-{B}^{2}}+2\,{B}^{3}{L}^{3}{a}^{2
}\phi_{{r}}\nonumber\\
&&-2\, \left( 1-{B}^{2} \right) {K}^{2}\phi_{{r}}LB-2\,B{L}^{3}{a}^
{2}\phi_{{r}}+ \left( 1-{B}^{2} \right)
{K}^{2}LB_{{r}}+2\,B{L}^{2}{a}^{2 }L_{{r}}\nonumber\\
&&+2\,a\sqrt {1-{B}^{2}}K\phi_{{r}}{L}^{2}{B}^{2}+ \left( \cos
 \left( \psi \right)  \right) ^{2}{B}^{3}{L}^{3}aa_{{r}}+\sqrt {1-{B}^
{2}}K{L}^{2}a_{{r}}\nonumber\\
&&+ \left( \cos \left( \psi \right)  \right) ^{2}{B}^
{2}LaL_{{r}}\sqrt {1-{B}^{2}}K-{B}^{2}{L}^{2}aK_{{r}}\sqrt
{1-{B}^{2}}\nonumber\\
&& -2\,{B}^{3}{L}^{2}{a}^{2}L_{{r}}-a\sqrt {1-{B}^{2}}K\sin \left(
\psi
 \right) {L}^{2}a_{{r}}-2\,{a}^{2}\sqrt {1-{B}^{2}}K\sin \left( \psi
 \right) L_{{r}}L\nonumber\\
&&+ \left( 1-{B}^{2} \right) {K}^{2}L_{{r}}B-{B}^{3}{L}
^{3}{a}^{2}\sin \left( \psi \right) a_{{r}}-BL{a}^{2}\sqrt {{\it
A_5}}G
{\it \Delta_2}\,\sin \left( {\it \theta_2} \right) \sqrt {1-{B}^{2}}\nonumber\\
&&+a \sqrt {1-{B}^{2}}K\sin \left( \psi \right)
{L}^{2}a_{{r}}{B}^{2}-2\,a
 \left( 1-{B}^{2} \right) {K}^{2}\sin \left( \psi \right) \phi_{{r}}LB+
 \left( \cos \left( \psi \right)  \right) ^{2}B{L}^{2}aB_{{r}}K\sqrt {
1-{B}^{2}}\nonumber\\
&&+{a}^{2}\sqrt {1-{B}^{2}}K\sin \left( \psi \right) L_{{r}}L{
B}^{2}-\sqrt
{1-{B}^{2}}K{L}^{2}a_{{r}}{B}^{2}+2\,{B}^{3}{L}^{3}{a}^{3
}\sin \left( \psi \right) \phi_{{r}}\nonumber\\
&&+a \left( 1-{B}^{2} \right) {K}^{2} \sin \left( \psi \right)
L_{{r}}B-{B}^{2}{L}^{3}{a}^{3}\sin \left(
\psi \right) B_{{r}}+ \left( 1-{B}^{2} \right) KLBK_{{r}}\nonumber\\
&&+2\,{a}^{2} \sqrt {1-{B}^{2}}K\sin \left( \psi \right)
\phi_{{r}}{L}^{2}+B{L}^{3}aa_{
{r}}-{B}^{3}{L}^{3}aa_{{r}}-a\sqrt {1-{B}^{2}}KLL_{{r}}{B}^{2}\nonumber\\
&&-2\,{B}^ {3}{L}^{2}{a}^{3}\sin \left( \psi \right)
L_{{r}}+2\,B{L}^{2}{a}^{3} \sin \left( \psi \right)
L_{{r}}+B{L}^{3}{a}^{2}\sin \left( \psi
 \right) a_{{r}}-{B}^{2}{L}^{3}{a}^{2}B_{{r}}\nonumber\\
&&-2\, \left( \cos \left( \psi \right)  \right)
^{2}{B}^{2}{L}^{2}a\phi_{{r}}\sqrt {1-{B}^{2}}K-{B}
^{2}{L}^{2}{a}^{2}\sin \left( \psi \right) K_{{r}}\sqrt {1-{B}^{2}}\nonumber\\
&&+
 \left( \cos \left( \psi \right)  \right) ^{2}{B}^{2}{L}^{2}aK_{{r}}
\sqrt {1-{B}^{2}}- \left( \cos \left( \psi \right) \right) ^{2}B{L}^
{3}aa_{{r}}-2\,B{L}^{3}{a}^{3}\sin \left( \psi \right) \phi_{{r}}
)E_1\wedge E_3 \wedge E_6\nonumber\\
&&+ \frac {1}{ \sqrt {{\it A_5}}{L}^{2}K\sqrt
{1-{B}^{2}}}(-2\,{B}^{3}K\phi_{{r}}{L}^{2}+{B}^{2}K{L}^{2}B_{{r}}+2\,BK\sin
 \left( \psi \right) \phi_{{r}}{L}^{2}a\nonumber\\
&&-2\,{B}^{3}K\sin \left( \psi
 \right) \phi_{{r}}{L}^{2}a-2\,{L}^{2}a \left( 1-{B}^{2} \right) \sin
 \left( \psi \right) \phi_{{r}}BK+2\,{B}^{3}KLL_{{r}}\nonumber\\
&&-2\,BK\sin \left( \psi \right) L_{{r}}La+La \left( 1-{B}^{2}
\right) \sin \left( \psi
 \right) L_{{r}}BK+2\,{B}^{3}K\sin \left( \psi \right) L_{{r}}La\nonumber\\
&&-BK \sin \left( \psi \right) {L}^{2}a_{{r}}+{B}^{2}{K}^{2}\sin
\left( \psi
 \right) L_{{r}}\sqrt {1-{B}^{2}}-2\,{L}^{3}{a}^{2}\sqrt {1-{B}^{2}}
\sin \left( \psi \right) \phi_{{r}}{B}^{2}\nonumber\\
&&- \left( \cos \left( \psi
 \right)  \right) ^{2}\sqrt {1-{B}^{2}}{L}^{3}a_{{r}}{B}^{2}+2\,{L}^{2 }{a}^{2}\sqrt {1-{B}^{2}}\sin \left( \psi \right)
L_{{r}}{B}^{2}+2\,
 \left( \cos \left( \psi \right)  \right) ^{2}\sqrt {1-{B}^{2}}{L}^{2}
L_{{r}}a\nonumber\\
&&- \left( \cos \left( \psi \right)  \right) ^{2} \left( 1-{B}^{ 2}
\right) {L}^{2}B_{{r}}K+{B}^{2}K\sin \left( \psi \right) {L}^{2}aB_
{{r}}+{B}^{2}K\sin \left( \psi \right) LK_{{r}}\sqrt {1-{B}^{2}}\nonumber\\
&&+2\,
 \left( \cos \left( \psi \right)  \right) ^{2} \left( 1-{B}^{2}
 \right) {L}^{2}\phi_{{r}}BK- \left( \cos \left( \psi \right)  \right) ^{
2} \left( 1-{B}^{2} \right) LL_{{r}}BK\nonumber\\
&&+{L}^{3}{a}^{2}\sqrt {1-{B}^{2}} \sin \left( \psi \right)
BB_{{r}}+{L}^{2}a \left( 1-{B}^{2} \right) \sin \left( \psi \right)
B_{{r}}K- \left( \cos \left( \psi \right)
 \right) ^{2}\sqrt {1-{B}^{2}}{L}^{3}aBB_{{r}}\nonumber\\
&&+2\,{L}^{2}a\sqrt {1-{B} ^{2}}L_{{r}}{B}^{2}+2\, \left( \cos
\left( \psi \right)  \right) ^{2} \sqrt
{1-{B}^{2}}{L}^{3}\phi_{{r}}a{B}^{2}\nonumber\\
&&+{L}^{3}a\sqrt {1-{B}^{2}}\sin
 \left( \psi \right) a_{{r}}{B}^{2}-{L}^{3}a\sqrt {1-{B}^{2}}\sin
 \left( \psi \right) a_{{r}}+La \left( 1-{B}^{2} \right) \sqrt {{\it A_5}}G{\it
\Delta_2}\,\sin \left( {\it \theta_2} \right)\nonumber\\
&& -2\, \left( \cos
 \left( \psi \right)  \right) ^{2}\sqrt {1-{B}^{2}}{L}^{2}L_{{r}}a{B}^
{2}+ \left( \cos \left( \psi \right)  \right) ^{2}\sqrt
{1-{B}^{2}}{L}
^{3}a_{{r}}\nonumber\\
&&+2\,{L}^{3}a\sqrt {1-{B}^{2}}\phi_{{r}}-2\,{L}^{2}a\sqrt {1-{B
}^{2}}L_{{r}}+{L}^{3}a\sqrt {1-{B}^{2}}BB_{{r}}+{B}^{3}K\sin \left(
\psi \right)
{L}^{2}a_{{r}}\nonumber\\
&&+2\,BK\phi_{{r}}{L}^{2}-2\,BKLL_{{r}}+B{K}^{2} \sin \left( \psi
\right) LB_{{r}}\sqrt {1-{B}^{2}}+{L}^{2}a \left( 1-{ B}^{2} \right)
\sin \left( \psi \right) BK_{{r}}\nonumber\\
&&-2\,{L}^{3}a\sqrt {1-{ B}^{2}}\phi_{{r}}{B}^{2}- \left( \cos
\left( \psi \right)  \right) ^{2}
 \left( 1-{B}^{2} \right) {L}^{2}BK_{{r}}-2\,{L}^{2}{a}^{2}\sqrt {1-{B }^{2}}\sin \left( \psi \right)
L_{{r}}\nonumber\\
&&+2\,{L}^{3}{a}^{2}\sqrt {1-{B}^{ 2}}\sin \left( \psi \right)
\phi_{{r}}-2\,{B}^{2}{K}^{2}\sin \left( \psi
 \right) \phi_{{r}}L\sqrt {1-{B}^{2}}\nonumber\\
&&+BK\sqrt {{\it A_5}}G{\it \Delta_2}\, \sin \left( {\it \theta_2}
\right) \sqrt {1-{B}^{2}}-2\, \left( \cos
 \left( \psi \right)  \right) ^{2}\sqrt {1-{B}^{2}}{L}^{3}\phi_{{r}}a)E_1\wedge E_4 \wedge E_5\nonumber\\
&& -\frac {\sin \left( \psi \right) \cos \left( \psi \right)}{\sqrt
{{\it A_5}}KL\sqrt {1- {B}^{2}}}  \Big( -2
\,\phi_{{r}}{L}^{2}a+2\,\phi_{{r}}{L}^{2}a{B}^{2}+2\,\phi_{{r}}L\sqrt
{1-{B}^{2 }}BK+2\,L_{{r}}La\nonumber\\
&&-2\,L_{{r}}La{B}^{2}-L_{{r}}\sqrt {1-{B}^{2}}BK-{L}^{
2}aBB_{{r}}+{L}^{2}a_{{r}}-{L}^{2}a_{{r}}{B}^{2}-LB_{{r}}K\sqrt
{1-{B}
^{2}}\nonumber\\
&&-LBK_{{r}}\sqrt {1-{B}^{2}} \Big)E_1\wedge E_4 \wedge E_6 \nonumber\\
&& -\frac {1}{\sqrt {{\it A_5}}{L}^{2}K\sqrt {1-{B} ^{2}}}\Big(-
\left( 1-{B}^{2} \right) {K}^{2}\sin \left( \psi \right) L_
{{r}}B-2\,B{L}^{2}aL_{{r}}-2\,\sqrt
{1-{B}^{2}}KLL_{{r}}{B}^{2}\nonumber\\
&&-{B}^{2 }La\sin \left( \psi \right) L_{{r}}\sqrt {1-{B}^{2}}K-
\left( \cos
 \left( \psi \right)  \right) ^{2}{B}^{3}{L}^{3}a_{{r}}-\sqrt {1-{B}^{ 2}}K\sin \left( \psi \right)
{L}^{2}a_{{r}}{B}^{2}\nonumber\\
&&+BLa\sqrt {{\it A_5}} G{\it \Delta_2}\,\sin \left( {\it \theta_2}
\right) \sqrt {1-{B}^{2}}+2\, \sqrt {1-{B}^{2}}KLL_{{r}}-2\,\sqrt
{1-{B}^{2}}K\phi_{{r}}{L}^{2}\nonumber\\
&&+{B}^{2}{L}^{3}aB_{{r}}+2\,{B}^{3}{L}^{2}aL_
{{r}}+{B}^{2}{L}^{2}a\sin \left( \psi \right) K_{{r}}\sqrt
{1-{B}^{2}}
-2\,{B}^{3}{L}^{3}a\phi_{{r}}\nonumber\\
&&- \left( 1-{B}^{2} \right) {K}^{2}\sin
 \left( \psi \right) LB_{{r}}+ \left( \cos \left( \psi \right)
 \right) ^{2}B{L}^{3}a_{{r}}- \left( \cos \left( \psi \right)
 \right) ^{2}{B}^{2}{L}^{2}K_{{r}}\sqrt {1-{B}^{2}}\nonumber\\
&&+\sqrt {1-{B}^{2}}K \sin \left( \psi \right)
{L}^{2}a_{{r}}+2\,\sqrt {1-{B}^{2}}K\sin
 \left( \psi \right) L_{{r}}La+2\, \left( \cos \left( \psi \right)
 \right) ^{2}{B}^{3}{L}^{3}\phi_{{r}}a\nonumber\\
&&+{B}^{2}{L}^{3}{a}^{2}\sin \left( \psi \right) B_{{r}}+2\, \left(
\cos \left( \psi \right)  \right) ^{2}
B{L}^{2}L_{{r}}a+2\,{B}^{3}{L}^{2}{a}^{2}\sin \left( \psi \right)
L_{{
r}}\nonumber\\
&&- \left( \cos \left( \psi \right)  \right) ^{2}B{L}^{2}B_{{r}}K
\sqrt {1-{B}^{2}}+2\, \left( \cos \left( \psi \right)  \right)
^{2}{B}
^{2}{L}^{2}\phi_{{r}}\sqrt {1-{B}^{2}}K\nonumber\\
&&+2\,B{L}^{3}{a}^{2}\sin \left( \psi \right) \phi_{{r}}-2\, \left(
\cos \left( \psi \right)  \right) ^{2} B{L}^{3}\phi_{{r}}a- \left(
\cos \left( \psi \right)  \right) ^{2}{B}^{2}
LL_{{r}}\sqrt {1-{B}^{2}}K\nonumber\\
&&-2\,\sqrt {1-{B}^{2}}K\sin \left( \psi
 \right) \phi_{{r}}{L}^{2}a+{B}^{3}{L}^{3}a\sin \left( \psi \right) a_{{r
}}- \left( 1-{B}^{2} \right) K\sin \left( \psi \right) LBK_{{r}}\nonumber\\
&&-2\,
 \left( \cos \left( \psi \right)  \right) ^{2}{B}^{3}{L}^{2}L_{{r}}a-2
\,{B}^{3}{L}^{3}{a}^{2}\sin \left( \psi \right) \phi_{{r}}-\sqrt {
1-{B}^{2}}K{L}^{2}BB_{{r}}\nonumber\\
&&+2\, \left( 1-{ B}^{2} \right) {K}^{2}\sin \left( \psi \right)
\phi_{{r}}LB- \left( 1-{B} ^{2} \right) K\sqrt {{\it A_5}}G{\it
\Delta_2}\,\sin \left( {\it \theta_2}
 \right) \nonumber\\
&&+2\,B{L}^{3}a\phi_{{r}}-2\,B{L}^{2}{a}^{2}\sin \left( \psi
 \right) L_{{r}}+2\,\sqrt {1-{B}^{2}}K\phi_{{r}}{L}^{2}{B}^{2}\nonumber\\
&&- \left( \cos \left( \psi \right)  \right)
^{2}{B}^{2}{L}^{3}aB_{{r}}-B{L}^{3}a
\sin \left( \psi \right) a_{{r}}\Big)E_1 \wedge E_5 \wedge E_6,\nonumber\\
&&
-{\frac {\sqrt {1-{B}^{2}}La+BK}{GK}}E_2 \wedge E_3 \wedge E_5,\nonumber\\
&&
-{\frac {-\sqrt {1-{B}^{2}}La+BK}{GK}}E_2 \wedge E_4 \wedge E_6,\nonumber\\
&&+ \frac {1}{L{K}^{2}} \left( \sqrt {1-{B}^{2}}La+BK \right) \Big(
\cos \left( \psi \right) {\it \Delta_2}\,\cos \left( {\it \theta_2}
\right) La\sqrt {1 -{B}^{2}}\nonumber\\
&&+\cos \left( \psi \right) {\it \Delta_2}\,\cos \left( {\it
\theta_2} \right) BK+{\it \Delta_1}\,\cos \left( {\it \theta_1}
\right)
\sqrt {1-{B}^{2}}L \Big)E_3\wedge E_4 \wedge E_5 \nonumber\\
&&+ {\frac {a{\it \Delta_2}\,\cos \left( {\it \theta_2} \right)
\left( -
\sqrt {1-{B}^{2}}La+BK \right) \sin \left( \psi \right) }{{K}^{2}}}E_3\wedge E_4\wedge E_6\nonumber\\
&&+ \frac {1}{L{K}^{2}} \left( \sqrt {1-{B}^{2}}La+BK \right) \Big(
-\cos \left( \psi \right) {\it \Delta_2}\,\cos \left( {\it \theta_2}
\right) BLa\nonumber\\
&&+\cos
 \left( \psi \right) {\it \Delta_2}\,\cos \left( {\it \theta_2} \right)
\sqrt {1-{B}^{2}}K-B{\it \Delta_1}\,\cos \left( {\it \theta_1}
\right) L
 \Big) E_3\wedge E_5\wedge E_6\nonumber\\
&& -{\frac { \left( -\sqrt {1-{B}^{2}}La+BK \right) {\it
\Delta_2}\,\cos
 \left( {\it \theta_2} \right) \sin \left( \psi \right) }{LK}}E_4\wedge E_5\wedge E_6
\end{eqnarray}
where the function $B$ is a function of radial direction $r$ and is
determined by the $SU(3)$ structure of the manifold.
\begin{eqnarray}\label{mathmetric}
G=\sqrt{A_1},\quad L=\sqrt{A_2},\quad
K=\sqrt{A_3-A_2b^2_{y\theta_1}},\quad
a=b_{y\theta_1},\quad\Delta_1=a_1,\quad \Delta_2=b_1\nonumber
\end{eqnarray}
and $E_i$ are defined in the following way:
\begin{eqnarray} &&E_1=\sqrt{A_5}dr, \quad E_3=K~d\theta_1,
\quad
E_4=-K~d\phi_1,\quad E_2=G(d\psi+\Delta_1\cos\theta_1d\phi_1+\Delta_2\cos\theta_2d\phi_2),\nonumber\\
&&E_5=L~\Big(\sin\psi d\phi_2+\cos\psi
d\theta_2-a~d\theta_1\Big),\quad E_6=L~\Big(\cos\psi
d\phi_2-\sin\psi
d\theta_2-a~d\phi_1\Big)\nonumber\\
\end{eqnarray}

\newpage

\section{General torsion classes for type IIA}\label{torsionIIA}

The type IIA torsion classes before geometric transition are given by the following expressions:
\begin{eqnarray}
&&W_1=\frac{-1}{6L^2K^2\sqrt{F_1}GA}\Big(-i L^2GK^2
b_{\phi_2\theta_2,r}AB -2i  L^2AG b_{\psi r}
  K^2B +2 L^2A\sqrt{F1}K^2B \nonumber\\
&&~~~+2iG  L^3K  \cos\psi^2 b_{\phi_2\theta_2 ,r}aB^2-i  L^3GK
b_{\phi_2\theta_2,r}aB^2 +i L^3G b_{\phi_1\theta_1,r}Ka\nonumber\\
&&~~~ -i L^3G b_{\phi_1\theta_1,r}KaB^2-i L^2G
b_{\phi_1\theta_1,r}K^2AB+G^2B \sqrt{F1}A L^2 \Delta_1\sin
 \theta_1  a^2\nonumber\\
 &&~~~ +G^2B  \sqrt{F_1}B \Delta_2\sin\theta_2L^2 +G^2B \sqrt{F_1}A \Delta_2 \sin \theta_2
  K^2+2G  L^3K\sin\psi b_{\phi_2\theta_2,r}\cos  \psi  A a \nonumber\\
  &&~~~ +2
  L^2GBK^2-2iG  L^3K  \cos  \psi
  ^2 b_{\phi_2\theta_2,r}a +i  L^3GK b_{\phi_2\theta_2,r}a
L^2K^2\sqrt{F_1}GB\Big)\nonumber\\
&& W_4=w_{4r}dr+w_{4\theta_1}d\theta_1+w_{4\theta_2}d\theta_2+w_{4\phi_1}d\phi_1+w_{4\phi_2}d\phi_2\nonumber\\
&&{\rm Re}~W_5 = w_{5e^1}e^1 + w_{5e^2}e^2+w_{5e^3}e^3+w_{5e^4}e^4+w_{5e^5}e^5+w_{5e^6}e^6
\end{eqnarray}
where $w_i$'s appearing above are defined in the following way:
\bg\label{hinsu}
w_{4r}&=&-\frac{1}{2L^{2}K^{2}}\,\Big[-\sin  \psi  { \Delta_2}
  \cos  { \theta_2}    L
  ^{4}{ b_{\phi_1 r}}    a
  ^{3}  B    ^{2}-\cos  {
 \theta_1}  { \Delta_1}    L
    ^{4}\sin  \psi  { b_{\phi_2 r}}
  a    B    ^{2}\nonumber\\
  &&-
\sin  \psi  { \Delta_2}  \cos  {
 \theta_2}    L    ^{2}  K
    ^{2}{ b_{\phi_1 r}}  a
    B    ^{2}-\sin  \psi
  { \Delta_2}  \cos  { \theta_2}
    L    ^{3}K  {
 b_{\phi_1 r}}  \sqrt {1-  B
  ^{2}}B    a    ^
{2}\nonumber\\
&&+\cos  { \theta_1}  { \Delta_1}
  L    ^{3}\sin  \psi  {
b_{\phi_2 r}}  K  \sqrt {1-  B
    ^{2}}B  -\sin  \psi  {
 \Delta_2}  \cos  { \theta_2}  L
    K    ^{3}{ b_{\phi_1 r}}
  \sqrt {1-  B    ^{2}}B
 \nonumber\\
 && +\sin  \psi  { \Delta_2}
  \cos  { \theta_2}    L
  ^{2}  K    ^{2}{ b_{\phi_1 r}}
  a  -2\,L    L_r      K
  ^{2}+\cos  { \theta_1}  { \Delta_1}
    L    ^{4}\sin  \psi
  { b_{\phi_2 r}}  a  \nonumber\\
  &&+\sin
\psi  { \Delta_2}  \cos  { \theta_2}
    L    ^{4}{ b_{\phi_1 r}}
    a    ^{3}+\sqrt {{ F_1}
  }G  { \Delta_2}
\sin  { \theta_2}  \sqrt {1-  B
  ^{2}}  L    ^{2}  a
    ^{2}\nonumber\\
    &&+  L    ^{
2}\sqrt {{ F_1}  }G  { \Delta_1}
  \sin  { \theta_1}  \sqrt {1-  B
    ^{2}}+\sqrt {{ F_1}  }G
  { \Delta_2}  \sin  {
\theta_2}    K    ^{2}\sqrt {1-
  B    ^{2}}-2\,  L
    ^{2}K K_r
  \Big]  \nonumber\\
  w_{4\theta_1}&=&\frac{1}{2 K^{2}\sqrt
{1-  B^{2}}L  }\Big[{ \Delta_2} \cos  \theta_2  \cos  \psi    B
    ^{3}K    L
    ^{2}  a    ^{2}-{
\Delta_2}  \cos  { \theta_2}  \sin \psi    B    ^{3}K
  { b_{\phi_1 \theta_1}}    L
  ^{2}  a    ^{2}\nonumber\\
  &&-{ \Delta_2}
  \cos  { \theta_2}  \sin  \psi
    B    ^{3}  K
    ^{3}{ b_{\phi_1 \theta_1}}  +{ \Delta_2}
  \cos  { \theta_2}  \cos  \psi
    B    ^{3}  K
    ^{3}+2\,  L    ^{2}a
    B    ^{3}K
  { \Delta_1}  \cos  { \theta_1}\nonumber\\
    &&
  +{ \Delta_2}  \cos  { \theta_2}
  L  \sin  \psi  \sqrt {1-
B    ^{2}}a  { b_{\phi_1 \theta_1}}
    K    ^{2}  B
    ^{2}x +{ \Delta_2}  \cos  { \theta_2}
  L  \cos  \psi  \sqrt {1-
B    ^{2}}a    K
    ^{2}  B    ^{2}\nonumber\\
    && -{
\Delta_2}  \cos  { \theta_2}  \cos \psi  B  K    L
    ^{2}  a    ^{2}+{ \Delta_2}  \cos
  { \theta_2}    L    ^{3}
\sin  \psi  \sqrt {1-  B ^{2}}  a    ^{3}{ b_{\phi_1 \theta_1}}
    B    ^{2}\nonumber\\
    &&+  L
    ^{3}\sqrt {1-  B    ^{2
}}  a    ^{2}{ \Delta_1}
  \cos  { \theta_1}    B
  ^{2}+{ \Delta_2}  \cos  { \theta_2}
  L  \cos  \psi  \sqrt {1-
B    ^{2}}a    K
    ^{2}  B    ^{2} \nonumber\\
&&-{ \Delta_2}  \cos  { \theta_2}  \cos \psi  B  K    L
    ^{2}  a    ^{2}+{
\Delta_2}  \cos  { \theta_2}  \sin \psi  B  K  { b_{\phi_1
\theta_1}} r    L    ^{2}  a
    ^{2}+{ \Delta_2}  \cos  { \theta_2}  \sin  \psi  B    K
    ^{3}{ b_{\phi_1 \theta_1}} \nonumber\\
&& -{ \Delta_2}  \cos  { \theta_2}  \cos \psi  B    K    ^{ 3}-2\, L
^{2}a  B
  K  { \Delta_1}
\cos  { \theta_1} -  L ^{3}\sqrt {1-  B    ^{2}}  a
    ^{2}{ \Delta_1}  \cos
  { \theta_1} \nonumber\\
&& -{ \Delta_2}  \cos
  { \theta_2}    L    ^{3}
\cos  \psi  \sqrt {1-  B ^{2}}  a    ^{3}-{ \Delta_2}
  \cos  { \theta_2}    L
  ^{3}\sin  \psi  \sqrt {1-  B
    ^{2}}  a    ^{3}{
b_{\phi_1 \theta_1}}  \nonumber\\
&&-{ \Delta_2}  \cos  {
 \theta_2}  L  \cos  \psi  \sqrt
{1-  B    ^{2}}a
  K    ^{2}-{ \Delta_2}
  \cos  { \theta_2}  L  \sin
  \psi  \sqrt {1-  B    ^{2}
}a  { b_{\phi_1 \theta_1}}    K
    ^{2}\Big]\nonumber\\
w_{4\theta2}&=&\frac{1}{2\sqrt {1-  B    ^{2}}
  K    ^{2}}\Big[-  L    ^{2}  a r    ^{2}\sqrt {1-
B ^{ 2}}{ \Delta_2}  \cos  { \theta_2}  +{
 \Delta_2}  \cos  { \theta_2}
B    ^{2}  K    ^{ 2}\sqrt {1-  B    ^{2}} \nonumber\\
&&+L
  \cos  \psi    B
^{3}K  { \Delta_1}  \cos  { \theta_1} -L  { \Delta_1} \cos
  { \theta_1}  \sin  \psi  { b_{\phi_2\theta_2}}
    B    ^{3}K
  -L  \cos  \psi  B
  K  { \Delta_1}  \cos
{ \theta_1}  \nonumber\\
&& -\cos  \psi  \sqrt {1-  B
    ^{2}}  L    ^{
2}a  { \Delta_1}  \cos  { \theta_1} +  L    ^{2}{ \Delta_1}
  \cos  { \theta_1}  \sin  \psi
  { b_{\phi_2\theta_2}}  \sqrt {1-  B
    ^{2}}a  \nonumber\\
&& +L  { \Delta_1}  \cos  { \theta_1}  \sin \psi  {
b_{\phi_2\theta_2}}  B K-  L    ^{2}{ \Delta_1}
  \cos  { \theta_1}  \sin  \psi
  { b_{\phi_2\theta_2}}  a    B
    ^{2}\sqrt {1-  B
  ^{2}}\nonumber\\
  &&+\cos  \psi    L
  ^{2}a  { \Delta_1}  \cos
  { \theta_1}    B    ^{2}
\sqrt {1-  B    ^{2}}+{ \Delta_2}
  \cos  { \theta_2}    L
    ^{2}  a    ^{2}
B    ^{2}\sqrt {1-  B
  ^{2}}\Big]\nonumber\\
  w_{4\phi_1}&=&\frac{1}{2K    ^{2}\sqrt {1-
  B    ^{2}}L}\Big[(
  B    ^{3}  K
  ^{3}+  B    ^{3}K
    L    ^{2}  a
    ^{2}-L  \sqrt {1-  B
    ^{2}}a    K
  ^{2}  B    ^{2}-  L
    ^{3}\sqrt {1-  B
  ^{2}}  a    ^{3}  B
    ^{2}\nonumber\\
    &&-B    K
    ^{3}-B  K    L
    ^{2}  a    ^{2
}+L  \sqrt {1-  B    ^{2} }a    K    ^{2}+  L
    ^{3}\sqrt {1-  B
  ^{2}}  a    ^{3} ) \cos
  { \theta_2}  \sin  \psi  { \Delta_2}
  \Big]\nonumber\\
  w_{r\phi_2}&=&\frac{1}{2}\Big[\sqrt {1-  B    ^{2}} (\sqrt {1-  B    ^{2}}L  a
  +B  K )   {
 \Delta_1}  L  \cos  {
\theta_1}  \sin  \psi    K
  ^{2}\Big]\\
w_{5e^1}&=& \frac {1}{\sqrt {{  F_1}}G{K}^{2}{L}^{2}}\Big[
\Omega_{r\psi\theta_2\phi_2}\,B{K}^{2}-{
\Omega_{r\psi\theta_1\phi_2}}\,L\sin  \psi  {
b_{\phi_2\theta_2}}\,\sqrt {1-{B}^{2}}K\nonumber\\
&&-{ \Omega_{r\psi\theta_1\theta_2}}\,\sin  \psi  L\sqrt {
1-{B}^{2}}K-{ \Omega_{\psi\theta_2\phi_1\phi_2}}\,K\cos  \psi  \sqrt
{1-{B}^{2}}{ b_{\phi_2r}}\,L\nonumber\\
&&-{ \Omega_{r\psi\theta_1\phi_1}}\,B{L}^{2}+{
\Omega_{r\psi\phi_1\phi_2}}\,{L}^{2}{ b_{\phi_1\theta_1}}\,\sin
  \psi  { b_{\phi_2\theta_2}}\,Ba\nonumber\\
  &&+{ \Omega_{r\psi\theta_1\phi_2}}\,L\cos  \psi
  \sqrt {1-{B}^{2}}K+{ \Omega_{r\psi\phi_1\phi_2}}\,L{ b_{\phi_1\theta_1}}\,\cos  \psi
  \sqrt {1-{B}^{2}}K\nonumber\\
  &&-{ \Omega_{r\psi\theta_2\phi_1}}\,{L}^{2}\cos  \psi
Ba-{ \Omega_{r\psi\phi_1\phi_2}}\,L\sqrt {1-{B}^{2}}K{
b_{\phi_2\theta_2}}\,\cos  \psi\nonumber\\
&&  -{ \Omega_{r\psi\phi_1\phi_2}}\,L\sqrt {1-{B}^{2}}K\sin  \psi -{
\Omega_{\psi\theta_1\theta_2\phi_2}}\, \sin  \psi  {
b_{\phi_2r}}\,{L}^{2}Ba\nonumber\\
&&+{ \Omega_{\psi\theta_2\phi_1\phi_2}}\,{ b_{\phi_2r}}\, \sin  \psi
{ b_{\phi_1\theta_1}}\,B{L}^{2}a-{
\Omega_{\psi\theta_2\phi_1\phi_2}}\,L{ b_{\phi_2r}} \,\sin  \psi  {
b_{\phi_1\theta_1}}\,\sqrt {1-{B}^{2}}K\nonumber\\
&&+{ \Omega_{r\psi\theta_2\phi_1}}\,L \sin  \psi  {
b_{\phi_1\theta_1}}\,\sqrt {1-{B}^{2}}K+2\,{
\Omega_{r\psi\theta_2\phi_2}}\, aL\sqrt {1-{B}^{2}}K\nonumber\\
&&+{ \Omega_{r\psi\phi_1\phi_2}}\,{L}^{2}Ba{
b_{\phi_2\theta_2}}\,\cos \psi
  -{ \Omega_{\psi\theta_1\phi_1\phi_2}}\,L{ b_{\phi_1r}}\,\sin  \psi  { b_{\phi_2\theta_2}}
\,\sqrt {1-{B}^{2}}K\nonumber\\
&&-{ \Omega_{r\psi\phi_1\phi_2}}\,L{ b_{\phi_1\theta_1}}\,\sin  \psi
{ b_{\phi_2\theta_2}}\,\sqrt {1-{B}^{2}}K-{
\Omega_{r\psi\phi_1\phi_2}}\,{L}^{2}{ b_{\phi_1\theta_1}}\,\cos
  \psi  Ba\nonumber\\
  &&+{ \Omega_{\psi\theta_1\phi_1\phi_2}}\,{L}^{2}{ b_{\phi_1r}}\,\sin  \psi
  { b_{\phi_2\theta_2}}\,Ba+{ \Omega_{\psi\theta_1\theta_2\phi_2}}\,\sin  \psi  { b_{\phi_2r}
}\,L\sqrt {1-{B}^{2}}K\nonumber\\
&&+{ \Omega_{\psi\theta_1\phi_1\phi_2}}\,L{ b_{\phi_1r}}\,\cos  \psi
  \sqrt {1-{B}^{2}}K-{ \Omega_{r\psi\theta_2\phi_1}}\,{L}^{2}\sin  \psi
{ b_{\phi_1\theta_1}}\,Ba+{ \Omega_{r\psi\theta_2\phi_1}}\,L\cos
\psi  \sqrt {1-{B}^{2}}K \nonumber\\
&&+2\,{ \Omega_{\psi\theta_2\phi_1\phi_2}}\,L{ b_{\phi_1r}}\,a\sqrt
{1-{B}^{2}}K+{ \Omega_{\psi\theta_1\phi_1\phi_2}}\,{L}^{2}B{
b_{\phi_2r}}\nonumber\\
&&-{ \Omega_{r\psi\theta_2\phi_2}}\,{a}^{2}{L}^{2}B+{
\Omega_{r\psi\theta_1\phi_2}}\,{L}^{2}\sin  \psi
  { b_{\phi_2\theta_2}}\,Ba
  -{ \Omega_{\psi\theta_1\theta_2\phi_1}}\,\sin  \psi  { b_{\phi_1r}
}\,{L}^{2}Ba\nonumber\\&&+{ \Omega_{\psi\theta_2\phi_1\phi_2}}\,{
b_{\phi_1r}}\,B{K}^{2}+{ \Omega_{r\psi\theta_1\theta_2}}\,\sin \psi
 {L}^{2}Ba-{ \Omega_{\psi\theta_1\phi_1\phi_2}}\,{L}^{2}{
b_{\phi_1r}}\,\cos  \psi
  Ba\nonumber\\
  &&+{ \Omega_{r\psi\phi_1\phi_2}}\,{L}^{2}Ba\sin  \psi  -{ \Omega_{r\psi\theta_1\phi_2}}\,{L}
^{2}\cos  \psi  Ba+{ \Omega_{\psi\theta_1\theta_2\phi_1}}\,\sin \psi
 {  b_{\phi_1r}}\,L\sqrt {1-{B}^{2}}K\nonumber\\
&&-{ \Omega_{\psi\theta_2\phi_1\phi_2}}\,{
b_{\phi_1r}}\,{a}^{2}B{L}^{2}+ {
\Omega_{\psi\theta_2\phi_1\phi_2}}\,{ b_{\phi_2r}}\,\cos  \psi
B{L}^{2}a\Big]
\nonumber\\
w_{5e^2}&=&\frac {1}{\sqrt {{ F_1}    }G    K
    L    }\Big[ \Omega_{r\psi\theta_1\theta_2}\,\cos  \psi  +{ \Omega_{r\psi\theta_1\phi_2}}\,{ b_{\phi_2\theta_2}}
    \cos  \psi  +{ \Omega_{r\psi\theta_1\phi_2}}\,\sin  \psi
  -{ \Omega_{r\psi\theta_2\phi_1}}\,\cos  \psi  { b_{\phi_1\theta_1}}
  \nonumber\\
  &&+{ \Omega_{r\psi\theta_2\phi_1}}\,\sin  \psi  +{ \Omega_{r\psi\phi_1\phi_2}}\,{ b_{\phi_1\theta_1}}
    { b_{\phi_2\theta_2}}    \cos  \psi
  +{ \Omega_{r\psi\phi_1\phi_2}}\,{ b_{\phi_1\theta_1}}    \sin  \psi
  +{ \Omega_{r\psi\phi_1\phi_2}}\,\cos  \psi  \nonumber\\&&-{ \Omega_{r\psi\phi_1\phi_2}}\,\sin
\psi  { b_{\phi_2\theta_2}}    -{
\Omega_{\psi\theta_1\theta_2\phi_1}}\,\cos  \psi
  { b_{\phi_1 r}}    -{ \Omega_{\psi\theta_1\theta_2\phi_2}}\,\cos  \psi
  { b_{\phi_2 r}}    +{ \Omega_{\psi\theta_1\phi_1\phi_2}}\,{ b_{\phi_1 r}}
  { b_{\phi_2\theta_2}}    \cos  \psi  \nonumber\\&&+{
\Omega_{\psi\theta_1\phi_1\phi_2}}\,{ b_{\phi_1 r}}    \sin  \psi +{
\Omega_{\psi\theta_2\phi_1\phi_2}} \,{ b_{\phi_2 r}}    \cos \psi  {
b_{\phi_1\theta_1}}
    -{ \Omega_{\psi\theta_2\phi_1\phi_2}}\,{ b_{\phi_2 r}}    \sin
\psi \Big]\\
w_{5e^3}&=&\frac {1}{  K      ^{2}\sqrt {{ F_1}
    }  L      ^{2}G
  }\Big[-{ \Omega_{r\psi\theta_1\theta_2}}\,\sin  \psi  { \Delta_1}
  \cos  { \theta_1}  L    G
  -{ \Omega_{r\psi\theta_1\phi_1}}\,{ \Delta_2}    \cos  {
\theta_2}  \cos  \psi  L    G
  \nonumber\\&&+{ \Omega_{r\psi\theta_2\phi_2}}\,a    { \Delta_1}
\cos  { \theta_1}  L    G
  -{ \Omega_{\theta_1\theta_2\phi_1\phi_2}}\,\sin  \psi  { b_{\phi_2 r}}
  L    G    -{ \Omega_{r\theta_1\phi_1\phi_2}}\,L
  G    \sin  \psi  { b_{\phi_2\theta_2}}
    \nonumber\\&&+{ \Omega_{r\psi\theta_1\phi_2}}\,{ \Delta_1}    \cos
  { \theta_1}  L    G
\cos  \psi  -{ \Omega_{r\psi\theta_1\phi_2}}\,{ \Delta_1}    \cos
  { \theta_1}  L    G
\sin  \psi  { b_{\phi_2\theta_2}}    \nonumber\\&&+{
\Omega_{r\psi\theta_2\phi_1}}\,L
    G    \sin  \psi  {
\Delta_1}    \cos  { \theta_1}  { b_{\phi_1\theta_1}}
    +{ \Omega_{\psi\theta_1\theta_2\phi_1}}\,L    \sin  \psi
  G    { b_{\psi r}}    \nonumber\\&&-{ \Omega_{r\psi\theta_2\phi_1}}\,L
    G    a    { \Delta_2}
    \cos  { \theta_2}  +{ \Omega_{r\psi\phi_1\phi_2}}\,L
    G    { b_{\phi_1\theta_1}}    {
 \Delta_1}    \cos  { \theta_1}  \cos
  \psi  \nonumber\\&&-{ \Omega_{r\psi\phi_1\phi_2}}\,L    G
{ b_{\phi_1\theta_1}}    { \Delta_1}    \cos
  { \theta_1}  \sin  \psi  { b_{\phi_2\theta_2}}
    +{ \Omega_{r\psi\phi_1\phi_2}}\,L    G    {
 \Delta_2}    \cos  { \theta_2}  {
b_{\phi_2\theta_2}}    a    \nonumber\\&&+{
\Omega_{\psi\theta_1\theta_2\phi_1}}\,L
  \sin  \psi  G    { \Delta_1}
    \cos  { \theta_1}  { b_{\phi_1 r}}
  -{ \Omega_{\psi\theta_1\theta_2\phi_1}}\,L    \cos  \psi
\sqrt {-  B      ^{2}}\sqrt {{ F_1}
    }\nonumber\\&&-{ \Omega_{\psi\theta_1\theta_2\phi_2}}\,\sqrt {{ F_1}    }L
    a    \sqrt {1-  B
    ^{2}}+{ \Omega_{\psi\theta_1\theta_2\phi_2}}\,\sin  \psi  {
\Delta_1}    \cos  { \theta_1}  { b_{\phi_2 r}}
    L    G    \nonumber\\&&+{ \Omega_{\psi\theta_1\phi_1\phi_2}}\,L
    G    { \Delta_2}
\cos  { \theta_2}  \cos  \psi  { b_{\phi_2 r}}
    +{ \Omega_{\psi\theta_1\phi_1\phi_2}}\,L    G    {
 b_{\psi r}}    \cos  \psi  +{ \Omega_{\psi\theta_1\phi_1\phi_2}}\,L
    G    { b_{\phi_1 r}}    {
\Delta_1}    \cos  { \theta_1}  \cos \psi  \nonumber\\&&-{
\Omega_{\psi\theta_1\phi_1\phi_2}}\,L    G    { b_{\phi_1 r}}    {
\Delta_1}    \cos  { \theta_1}  \sin  \psi  { b_{\phi_2\theta_2}} +{
\Omega_{\psi\theta_1\phi_1\phi_2}}\,L    \sqrt {1- B
  ^{2}}\sqrt {{ F_1}    }\sin  \psi
  \nonumber\\&&-{ \Omega_{\psi\theta_1\phi_1\phi_2}}\,L    G    { b_{\psi r}}
    \sin  \psi  { b_{\phi_2\theta_2}}
  +{ \Omega_{\psi\theta_1\phi_1\phi_2}}\,L    \sqrt {1-  B
    ^{2}}\sqrt {{ F_1}    }{ b_{\phi_2\theta_2}}
    \cos  \psi  \nonumber\\&& -{ \Omega_{\psi\theta_2\phi_1\phi_2}}\,L
  G    { b_{\phi_2 r}}    \sin
\psi  { \Delta_1}    \cos  { \theta_1}
  { b_{\phi_1\theta_1}}    +{ \Omega_{\psi\theta_2\phi_1\phi_2}}\,L    G
    { b_{\phi_2 r}}    a    {
\Delta_2}    \cos  { \theta_2}  \nonumber\\&&+{
\Omega_{\psi\theta_2\phi_1\phi_2}}\, { b_{\phi_1\theta_1}}    \sqrt
{{ F_1}    }L
    a    \sqrt {1-  B
    ^{2}}+{ \Omega_{\psi\theta_2\phi_1\phi_2}}\,L    G
  a    { \Delta_1}    \cos
{ \theta_1}  { b_{\phi_1 r}}    +{
\Omega_{\psi\theta_2\phi_1\phi_2}}\,L
  G    a    { b_{\psi r}}
  \nonumber\\&&+{ \Omega_{\psi\theta_2\phi_1\phi_2}}\,{ b_{\phi_1\theta_1}}    \sqrt {{ F_1}
    }B    K    +{ \Omega_{r\theta_1\theta_2\phi_1}}\,
\sin  \psi  L    G    -{ \Omega_{\psi\theta_1\theta_2\phi_2}}\,\sqrt
{{ F_1}    }B    K
  \nonumber\\&&+{ \Omega_{r\theta_2\phi_1\phi_2}}\,a    L    G
  +{ \Omega_{r\theta_1\phi_1\phi_2}}\,L    G    \cos
\psi  \Big]\\
 w_{5e^4}&=&\frac{1}{  K
      ^{2}\sqrt {{ F_1}    }
  L      ^{2}G    } \Big[ { \Omega_{r\psi\theta_1\theta_2}}\,G    \cos  \psi  {
\Delta_1}    \cos  { \theta_1}  L
  \sqrt {1-  B      ^{2}}+{ \Omega_{r\psi\theta_1\phi_2}}\,
G    { \Delta_2}    \cos  { \theta_2}  { b_{\phi_2\theta_2}}    L a
    \sqrt {1-  B      ^{2}}\nonumber\\&&+{
 \Omega_{r\psi\theta_1\theta_2}}\,G    { \Delta_2}    \cos
{ \theta_2}  L    a    \sqrt {1-
  B      ^{2}}+{ \Omega_{r\psi\theta_1\theta_2}}\,G
  { \Delta_2}    \cos  { \theta_2}
  B    K    +{ \Omega_{r\theta_1\phi_1\phi_2}}\,\sqrt {1-
  B      ^{2}}L    G
  \sin  \psi  \nonumber\\&&-{ \Omega_{r\psi\theta_1\phi_1}}\,\sin  \psi
{ \Delta_2}    \cos  { \theta_2}  \sqrt { 1-  B      ^{2}}L    G
    -{ \Omega_{r\theta_1\theta_2\phi_1}}\,\cos  \psi  \sqrt {1-
  B      ^{2}}L    G
  \nonumber\\&&-{ \Omega_{r\theta_1\theta_2\phi_2}}\,G    L    a
  \sqrt {1-  B      ^{2}}+{ \Omega_{r\psi\theta_1\phi_2}}\,
G    { \Delta_2}    \cos  { \theta_2}  { b_{\phi_2\theta_2}}    B K
    \nonumber\\&&+{ \Omega_{\psi\theta_2\phi_1\phi_2}}\,{ b_{\phi_1\theta_1}}    G
  { b_{\phi_2 r}}    \cos  \psi  {
\Delta_1}    \cos  { \theta_1}  L
  \sqrt {1-  B      ^{2}}-{ \Omega_{r\psi\theta_2\phi_1}}\,
{ b_{\phi_1\theta_1}}    G    { \Delta_2}
  \cos  { \theta_2}  L    a
  \sqrt {1-  B      ^{2}}\nonumber\\&&-{ \Omega_{r\psi\theta_2\phi_1}}
\,{ b_{\phi_1\theta_1}}    G    { \Delta_2}
  \cos  { \theta_2}  B    K
    +{ \Omega_{\theta_1\theta_2\phi_1\phi_2}}\,G    \cos  \psi
  \sqrt {1-  B      ^{2}}{ b_{\phi_2 r}}
    L    \nonumber\\&&-{ \Omega_{\theta_1\theta_2\phi_1\phi_2}}\,G    {
 b_{\phi_1 r}}    L    a
\sqrt {1-  B      ^{2}}+{ \Omega_{r\theta_1\phi_1\phi_2}}\,\sqrt {1-
  B      ^{2}}L    G
  { b_{\phi_2\theta_2}}    \cos  \psi  \nonumber\\&&+{
\Omega_{\psi\theta_2\phi_1\phi_2}}\,{ b_{\phi_1\theta_1}}    G    {
b_{\psi r}}
    L    a    \sqrt {1-
  B      ^{2}}+{ \Omega_{\psi\theta_2\phi_1\phi_2}}\,{ b_{\phi_1\theta_1}}
    G    { b_{\psi r}}    B
    K    \nonumber\\&&-{ \Omega_{\theta_1\theta_2\phi_1\phi_2}}\,G    {
 b_{\phi_1 r}}    B    K    +{
 \Omega_{r\psi\theta_1\phi_2}}\,G    { \Delta_1}    \cos
{ \theta_1}  L    \sqrt {1-  B
    ^{2}}{ b_{\phi_2\theta_2}}    \cos  \psi
  \nonumber\\&&+{ \Omega_{r\psi\theta_1\phi_2}}\,G    { \Delta_1}
\cos  { \theta_1}  L    \sqrt {1-  B
      ^{2}}\sin  \psi  -{ \Omega_{r\psi\theta_2\phi_1}}\,{
 b_{\phi_1\theta_1}}    G    \cos  \psi
  { \Delta_1}    \cos  { \theta_1}
  L    \sqrt {1-  B
  ^{2}}\nonumber\\&&+{ \Omega_{r\psi\phi_1\phi_2}}\,{ b_{\phi_1\theta_1}}    G
  { \Delta_2}    \cos  { \theta_2}
  { b_{\phi_2\theta_2}}    L    a
  \sqrt {1-  B      ^{2}}+{ \Omega_{r\psi\phi_1\phi_2}}\,
{ b_{\phi_1\theta_1}}    G    { \Delta_1}
  \cos  { \theta_1}  L    \sqrt {1-
  B      ^{2}}\sin  \psi  \nonumber\\&&+{
 \Omega_{r\psi\phi_1\phi_2}}\,{ b_{\phi_1\theta_1}}    G    { \Delta_2}
    \cos  { \theta_2}  { b_{\phi_2\theta_2}}
  B    K    +{ \Omega_{r\psi\phi_1\phi_2}}\,{ b_{\phi_1\theta_1}}
    G    { \Delta_1}
\cos  { \theta_1}  L    \sqrt {1-  B
      ^{2}}{ b_{\phi_2\theta_2}}    \cos
  \psi  \nonumber\\&&+{ \Omega_{r\theta_2\phi_1\phi_2}}\,{ b_{\phi_1\theta_1}}    G
    L    a    \sqrt {1-
  B      ^{2}}+{ \Omega_{r\theta_2\phi_1\phi_2}}\,{ b_{\phi_1\theta_1}}
    G    B    K
  -{ \Omega_{\psi\theta_1\theta_2\phi_1}}\,G    { b_{\phi_1 r}}    {
 \Delta_2}    \cos  { \theta_2}  B
    K    \nonumber\\&&-{ \Omega_{\psi\theta_1\theta_2\phi_1}}\,G    {
 b_{\psi r}}    \cos  \psi  \sqrt {1-
B      ^{2}}L    -{ \Omega_{\psi\theta_1\theta_2\phi_1}}\,G
    { b_{\phi_1 r}}    \cos  \psi
  { \Delta_1}    \cos  { \theta_1}
  L    \sqrt {1-  B
  ^{2}}\nonumber\\&&-{ \Omega_{\psi\theta_1\theta_2\phi_1}}\,G    { b_{\phi_1 r}}
  { \Delta_2}    \cos  { \theta_2}
  L    a    \sqrt {1-  B
      ^{2}}-{ \Omega_{\psi\theta_1\theta_2\phi_2}}\,G    {
b_{\psi r}}    L    a    \sqrt {1 -  B      ^{2}}\nonumber\\&&-{
\Omega_{\psi\theta_1\theta_2\phi_2}}\,G
  { b_{\phi_2 r}}    { \Delta_2}
\cos  { \theta_2}  L    a
  \sqrt {1-  B      ^{2}}+{ \Omega_{\psi\theta_1\phi_1\phi_2}}
\,G    { b_{\psi r}}    \sqrt {1-  B
      ^{2}}L    { b_{\phi_2\theta_2}}
  \cos  \psi  \nonumber\\&&+{ \Omega_{\psi\theta_1\phi_1\phi_2}}\,G    {
 b_{\psi r}}    \sqrt {1-  B
  ^{2}}L    \sin  \psi  +{ \Omega_{\psi\theta_1\phi_1\phi_2}}\,
\sqrt {{ F_1}    }L    \sin  \psi
  { b_{\phi_2\theta_2}}    \nonumber\\&&+{ \Omega_{\psi\theta_1\phi_1\phi_2}}\,G    {
 \Delta_2}    \cos  { \theta_2}  {
b_{\phi_1 r}}    { b_{\phi_2\theta_2}}    B
  K    \nonumber\\&&+{ \Omega_{\psi\theta_1\phi_1\phi_2}}\,G    { b_{\phi_1 r}}
    { \Delta_1}    \cos  {
\theta_1}  L    \sqrt {1-  B
    ^{2}}{ b_{\phi_2\theta_2}}    \cos  \psi
  +{ \Omega_{\psi\theta_1\phi_1\phi_2}}\,G    { \Delta_2}
\cos  { \theta_2}  { b_{\phi_1 r}}    { b_{\phi_2\theta_2}}    L a
\sqrt {1 -  B      ^{2}}\nonumber\\&&+{
\Omega_{\psi\theta_1\phi_1\phi_2}}\,G
  { b_{\phi_1 r}}    { \Delta_1}
\cos  { \theta_1}  L    \sqrt {1-  B
      ^{2}}\sin  \psi  -{ \Omega_{\psi\theta_1\theta_2\phi_2}}\,G
    { b_{\psi r}}    B    K
    \nonumber\\&&+{ \Omega_{\psi\theta_1\phi_1\phi_2}}\,G    \sin  \psi
  { \Delta_2}    \cos  { \theta_2}
  \sqrt {1-  B      ^{2}}{ b_{\phi_2 r}}
    L    -{ \Omega_{\psi\theta_1\theta_2\phi_2}}\,G    {
 b_{\phi_2 r}}    { \Delta_2}    \cos  {
 \theta_2}  B    K    \nonumber\\&&-{ \Omega_{\psi\theta_1\theta_2\phi_2}}\,
G    { b_{\phi_2 r}}    \cos  \psi
  { \Delta_1}    \cos  { \theta_1}
  L    \sqrt {1-  B
  ^{2}}\nonumber\\&&+{ \Omega_{\psi\theta_2\phi_1\phi_2}}\,{ b_{\phi_1\theta_1}}    G
  { b_{\phi_2 r}}    { \Delta_2}
\cos  { \theta_2}  L    a
  \sqrt {1-  B      ^{2}}+{ \Omega_{\psi\theta_2\phi_1\phi_2}}
\,{ b_{\phi_1\theta_1}}    G    { b_{\phi_2 r}}
  { \Delta_2}    \cos  { \theta_2}
  B    K    \nonumber\\&&-{ \Omega_{\psi\theta_1\theta_2\phi_1}}\,\sin
  \psi  \sqrt {{ F_1}    }L
  -{ \Omega_{\psi\theta_2\phi_1\phi_2}}\,a    \sqrt {{ F_1}
  }L    \nonumber\\&&-{ \Omega_{\psi\theta_1\phi_1\phi_2}}\,\sqrt {{ F_1}
  }L    \cos  \psi  -{ \Omega_{r\theta_1\theta_2\phi_2}}\,G
    B    K    \Big]\\
w_{5e^5}&=&\frac {1}{  K      ^{2}\sqrt {{
 F_1}    }  L      ^{2}G
    }\Big[-{ \Omega_{r\psi\theta_2\phi_1}}\,{ \Delta_2}    \cos  {
\theta_2}  K    G    +{ \Omega_{r\psi\theta_2\phi_2}}\,{ \Delta_1}
\cos  { \theta_1}  K
  G    +{ \Omega_{r\psi\phi_1\phi_2}}\,{ \Delta_2}
\cos  { \theta_2}  { b_{\phi_2\theta_2}}    K
    G    \nonumber\\&&+{ \Omega_{r\theta_2\phi_1\phi_2}}\,K    G
    -{ \Omega_{\psi\theta_1\theta_2\phi_1}}\,\cos  \psi  B
  \sqrt {{ F_1}    }L    -{
\Omega_{\psi\theta_1\theta_2\phi_2}}\,\sqrt {{ F_1}    }B    L
  a    \nonumber\\&&+{ \Omega_{\psi\theta_1\theta_2\phi_2}}\,\sqrt {{ F_1}
  }\sqrt {1-  B      ^{2}}K
  +{ \Omega_{\psi\theta_1\phi_1\phi_2}}\,B    \sqrt {{ F_1}
  }L    { b_{\phi_2\theta_2}}    \cos
\psi  +{ \Omega_{\psi\theta_1\phi_1\phi_2}}\,B    \sqrt {{ F_1}
  }L    \sin  \psi  \nonumber\\&&+{ \Omega_{\psi\theta_2\phi_1\phi_2}}\,K
    G    { \Delta_2}
\cos  { \theta_2}  { b_{\phi_2 r}}    +{
\Omega_{\psi\theta_2\phi_1\phi_2} }\,K    G    { b_{\psi r}} +{
\Omega_{\psi\theta_2\phi_1\phi_2}}\,K    G    { \Delta_1}
  \cos  { \theta_1}  { b_{\phi_1 r}}
  \nonumber\\&&+{ \Omega_{\psi\theta_2\phi_1\phi_2}}\,{ b_{\phi_1\theta_1}}    \sqrt {{ F_1}
    }B    L    a
  -{ \Omega_{\psi\theta_2\phi_1\phi_2}}\,{ b_{\phi_1\theta_1}}    \sqrt {{ F_1}
    }\sqrt {1-  B      ^{2}}K
    \Big]\\
w_{5e^6}&=&\frac{1}{
  K      ^{2}\sqrt {{ F_1}
  }  L      ^{2}G    }\Big[  { \Omega_{r\theta_1\theta_2\phi_2}}\,G    \sqrt {1-  B
    ^{2}}K    -{ \Omega_{\psi\theta_2\phi_1\phi_2}}\,{ b_{\phi_1\theta_1}}
    G    { b_{\phi_2 r}}    {
\Delta_2}    \cos  { \theta_2}  \sqrt {1-
  B      ^{2}}K    \nonumber\\&&+{ \Omega_{\psi\theta_2\phi_1\phi_2}}
\,{ b_{\phi_1\theta_1}}    G    { b_{\psi r}}
  B    L    a    -{
 \Omega_{\psi\theta_2\phi_1\phi_2}}\,{ b_{\phi_1\theta_1}}    G    { b_{\psi r}}
    \sqrt {1-  B      ^{2}}K
    +{ \Omega_{\psi\theta_2\phi_1\phi_2}}\,{ b_{\phi_1\theta_1}}    G
  { b_{\phi_2 r}}    { \Delta_2}
\cos  { \theta_2}  B    L
  a    \nonumber\\&&+{ \Omega_{\theta_1\theta_2\phi_1\phi_2}}\,G    \cos
  \psi  B    { b_{\phi_2 r}}    L
    -{ \Omega_{\theta_1\theta_2\phi_1\phi_2}}\,G    { b_{\phi_1 r}}
  B    L    a    +{
 \Omega_{\theta_1\theta_2\phi_1\phi_2}}\,G    { b_{\phi_1 r}}    \sqrt {1-
  B      ^{2}}K    \nonumber\\&&-{ \Omega_{r\psi\theta_1\phi_1}}
\,\sin  \psi  { \Delta_2}    \cos  {
 \theta_2}  B    L    G
  -{ \Omega_{r\theta_1\theta_2\phi_1}}\,\cos  \psi  B    L
    G    +{ \Omega_{r\psi\theta_1\phi_2}}\,G    {
 \Delta_2}    \cos  { \theta_2}  {
b_{\phi_2\theta_2}}    B    L    a
    \nonumber\\&&-{ \Omega_{r\psi\theta_1\theta_2}}\,G    { \Delta_2}
  \cos  { \theta_2}  \sqrt {1-  B
    ^{2}}K    +{ \Omega_{r\psi\theta_1\theta_2}}\,G
  \cos  \psi  B    { \Delta_1}
    \cos  { \theta_1}  L
+{ \Omega_{r\psi\theta_1\theta_2}}\,G    { \Delta_2}    \cos
  { \theta_2}  B    L    a
    \nonumber\\&&+{ \Omega_{r\psi\theta_1\phi_2}}\,G    B    {
 \Delta_1}    \cos  { \theta_1}  L
    \sin  \psi  +{ \Omega_{r\psi\theta_1\phi_2}}\,G
  B    { \Delta_1}    \cos
{ \theta_1}  L    { b_{\phi_2\theta_2}} \cos  \psi  \nonumber\\&&-{
\Omega_{r\psi\theta_1\phi_2}}\,G    { \Delta_2}
    \cos  { \theta_2}  { b_{\phi_2\theta_2}}
  \sqrt {1-  B      ^{2}}K
  -{ \Omega_{r\psi\theta_2\phi_1}}\,{ b_{\phi_1\theta_1}}    G
\cos  \psi  B    { \Delta_1}
  \cos  { \theta_1}  L    \nonumber\\&&-{ \Omega_{r\psi\theta_2\phi_1}}
\,{ b_{\phi_1\theta_1}}    G    { \Delta_2}
  \cos  { \theta_2}  B    L
    a    +{ \Omega_{r\psi\theta_2\phi_1}}\,{ b_{\phi_1\theta_1}}
  G    { \Delta_2}    \cos
{ \theta_2}  \sqrt {1-  B      ^{2} }K    \nonumber\\&&+{
\Omega_{r\psi\phi_1\phi_2}}\,{ b_{\phi_1\theta_1}}    G
  B    { \Delta_1}    \cos
{ \theta_1}  L    { b_{\phi_2\theta_2}} \cos  \psi  +{
\Omega_{r\psi\phi_1\phi_2}}\,{ b_{\phi_1\theta_1}}    G
    B    { \Delta_1}
\cos  { \theta_1}  L    \sin  \psi
  \nonumber\\&&+{ \Omega_{r\psi\phi_1\phi_2}}\,{ b_{\phi_1\theta_1}}    G    {
 \Delta_2}    \cos  { \theta_2}  {
b_{\phi_2\theta_2}}    B    L    a
    -{ \Omega_{r\psi\phi_1\phi_2}}\,{ b_{\phi_1\theta_1}}    G
  { \Delta_2}    \cos  { \theta_2}
  { b_{\phi_2\theta_2}}    \sqrt {1-  B
    ^{2}}K    \nonumber\\&&-{ \Omega_{r\theta_1\theta_2\phi_2}}\,G
  B    L    a    +{
 \Omega_{r\theta_1\phi_1\phi_2}}\,B    L    G    {
 b_{\phi_2\theta_2}}    \cos  \psi  +{ \Omega_{r\theta_1\phi_1\phi_2}}\,B
    L    G    \sin
\psi  \nonumber\\&&+{ \Omega_{r\theta_2\phi_1\phi_2}}\,{
b_{\phi_1\theta_1}}    G
  B    L    a    -{
 \Omega_{r\theta_2\phi_1\phi_2}}\,{ b_{\phi_1\theta_1}}    G    \sqrt {1-
  B      ^{2}}K    \nonumber\\&&-{ \Omega_{\psi\theta_1\theta_2\phi_1}}
\,G    { b_{\phi_1 r}}    \cos  \psi
  B    { \Delta_1}    \cos
{ \theta_1}  L    -{ \Omega_{\psi\theta_1\theta_2\phi_1}}\,G
  { b_{\phi_1 r}}    { \Delta_2}
\cos  { \theta_2}  B    L
  a    \nonumber\\&&+{ \Omega_{\psi\theta_1\theta_2\phi_1}}\,G    { b_{\phi_1 r}}
    { \Delta_2}    \cos  {
\theta_2}  \sqrt {1-  B      ^{2}}K
    -{ \Omega_{\psi\theta_1\theta_2\phi_1}}\,G    { b_{\psi r}}
  \cos  \psi  B    L
  \nonumber\\&&-{ \Omega_{\psi\theta_1\theta_2\phi_2}}\,G    { b_{\phi_2 r}}
\cos  \psi  B    { \Delta_1}
  \cos  { \theta_1}  L    +{ \Omega_{\psi\theta_1\theta_2\phi_2}
}\,G    { b_{\phi_2 r}}    { \Delta_2}
  \cos  { \theta_2}  \sqrt {1-  B
    ^{2}}K    \nonumber\\&&-{ \Omega_{\psi\theta_1\theta_2\phi_2}}\,G
  { b_{\psi r}}    B    L
  a    +{ \Omega_{\psi\theta_1\theta_2\phi_2}}\,G    { b_{\psi r}}
    \sqrt {1-  B      ^{2}}K
    \nonumber\\&&-{ \Omega_{\psi\theta_1\theta_2\phi_2}}\,G    { b_{\phi_2 r}}
  { \Delta_2}    \cos  { \theta_2}
  B    L    a    +{
 \Omega_{\psi\theta_1\phi_1\phi_2}}\,G    \sin  \psi  { \Delta_2}
    \cos  { \theta_2}  B
{ b_{\phi_2 r}}    L    \nonumber\\&&+{
\Omega_{\psi\theta_1\phi_1\phi_2}}\,G
  { b_{\phi_1 r}}    B    { \Delta_1}
    \cos  { \theta_1}  L
{ b_{\phi_2\theta_2}}    \cos  \psi  +{
\Omega_{\psi\theta_1\phi_1\phi_2}}\,G
    { b_{\phi_1 r}}    B    {
\Delta_1}    \cos  { \theta_1}  L
  \sin  \psi \nonumber\\&& +{ \Omega_{\psi\theta_1\phi_1\phi_2}}\,G    {
b_{\psi r}}    B    L    { b_{\phi_2\theta_2}}    \cos  \psi  +{
\Omega_{\psi\theta_1\phi_1\phi_2}}\,G
  { b_{\psi r}}    B    L
  \sin  \psi  \nonumber\\&&+{ \Omega_{\psi\theta_1\phi_1\phi_2}}\,G    {
\Delta_2}    \cos  { \theta_2}  { b_{\phi_1 r}}
    { b_{\phi_2\theta_2}}    B    L
    a    -{ \Omega_{\psi\theta_1\phi_1\phi_2}}\,G    {
 \Delta_2}    \cos  { \theta_2}  {
b_{\phi_1 r}}    { b_{\phi_2\theta_2}}    \sqrt {1-  B
      ^{2}}K    \nonumber\\&&+{ \Omega_{\psi\theta_2\phi_1\phi_2}}\,{
b_{\phi_1\theta_1}}    G    { b_{\phi_2 r}}
  \cos  \psi  B    { \Delta_1}
    \cos  { \theta_1}  L
-{ \Omega_{\psi\theta_2\phi_1\phi_2}}\,\sqrt {{ F_1}    }K    \Big]
\nd
\noindent where $\Omega_{ijkl}$ are now given by the following components:
\bg\label{jkaal}
\Omega_{r\psi\theta_1\theta_2}&=&\frac {1}{\sqrt {1-B^{2}}}\Big[G  L
K_r \sin  \psi
  B    ^{2}+  G_r
    L  { b_{\phi_1\theta_1}}
  K  \sin  \psi  { b_{\phi_2\theta_2}}
  -  G_r
L  { b_{\phi_1\theta_1}}  K \sin  \psi  { b_{\phi_2\theta_2}}    B
    ^{2}\nonumber\\&&+G  L
{ b_{\phi_1\theta_1}}    K_r
    \sin  \psi  { b_{\phi_2\theta_2}}
  -G  L    { b_{\phi_1\theta_1,r}}    K  \cos
\psi  -  G_r    L
  K  \cos  \psi  {
b_{\phi_2\theta_2}}  \nonumber\\&&-G  L  { b_{\phi_1\theta_1}}
K_r
  \sin  \psi  { b_{\phi_2\theta_2}}
  B    ^{2}+G   L_r    { b_{\phi_1\theta_1}}
  K  \sin  \psi  { b_{\phi_2\theta_2}}
  +G    L_r
   K  \sin  \psi
  B    ^{2}\nonumber\\&&-L  G
  { \Delta_2}  \cos  { \theta_2}
  { b_{\phi_2 r}}  { b_{\phi_1\theta_1}}  K
  \sin  \psi    B
  ^{2}+L  G  { b_{\psi r}}
  K  \sin  \psi  {
b_{\phi_2\theta_2}}  +L  G  { b_{\psi r}}  K  \cos  \psi
  B    ^{2}\nonumber\\&&-L  G
  { \Delta_2}  \cos  { \theta_2}
  { b_{\phi_2 r}}  K  \cos
\psi  +L  G  { \Delta_2}
  \cos  { \theta_2}  { b_{\phi_2 r}}
  K  \cos  \psi    B
   ^{2}-L  G  {
b_{\psi r}}  K  \sin  \psi  {
 b_{\phi_2\theta_2}}    B    ^{2}\nonumber\\&&-L
  G  { \Delta_1}
\cos  { \theta_1}  { b_{\phi_1 r}}  K
 \cos  \psi  +L  G
  { \Delta_1}  \cos  { \theta_1}
  { b_{\phi_1 r}}  K  \cos
\psi    B    ^{2}-G
  L    { b_{\phi_1\theta_1,r}}
   K  \sin  \psi  {
b_{\phi_2\theta_2}}    B    ^{2}\nonumber\\&&+G
  L  { b_{\phi_1\theta_1}}  K
  \sin  \psi  { b_{\phi_2\theta_2,r}}
  +  G_r
L  K  \sin  \psi
  B    ^{2}-G   L_r    { b_{\phi_1\theta_1}}
  K  \sin  \psi  { b_{\phi_2\theta_2}}
    B    ^{2}\nonumber\\&&+G
  L    { b_{\phi_1\theta_1,r}}
   K  \sin  \psi  {
b_{\phi_2\theta_2}}  -  G_r
  L  { b_{\phi_1\theta_1}}  K
  \cos  \psi  -G  L
  K  \cos  \psi  {
 b_{\phi_2\theta_2},r}  \nonumber\\&&+G  L  K
  \cos  \psi    {
 b_{\phi_2\theta_2},r}      B
  ^{2}+G  L  { b_{\phi_1\theta_1}}
    K_r
\cos  \psi    B    ^{2}-G
  L    K_r
    \cos  \psi  { b_{\phi_2\theta_2}}
  \nonumber\\&&+G  L    K_r    \cos  \psi  { b_{\phi_2\theta_2}}
    B    ^{2}-G
    L_r    K
  \cos  \psi  { b_{\phi_2\theta_2}}  +G
    L_r    K
  \cos  \psi  { b_{\phi_2\theta_2}}
    B    ^{2}\nonumber\\&&-G  L
  { b_{\phi_1\theta_1}}
K_r    \cos  \psi  +G
    L    { b_{\phi_1\theta_1,r}
}  K  \cos  \psi
  B    ^{2}+  G_r
    L  K  \cos
  \psi  { b_{\phi_2\theta_2}}    B
    ^{2}\nonumber\\&&+G  L    { b_{\phi_1\theta_1,r}}    K
  \cos  \psi    B
^{2}-G  L  { b_{\phi_1\theta_1}}
  K  \sin  \psi    {\frac {d
}{dr}}{ b_{\phi_2\theta_2}}      B
    ^{2}+  G_r
  L  { b_{\phi_1\theta_1}}  K
  \cos  \psi    B
^{2}\nonumber\\&&-G    L_r
  { b_{\phi_1\theta_1}}  K  \cos
\psi  +L  \sqrt {1-  B
  ^{2}}\sqrt {{ F1}  }K
\sin  \psi  -L  \sqrt {1-  B
    ^{2}}\sqrt {{ F1}  }{
 b_{\phi_1\theta_1}}  K  \sin  \psi
  { b_{\phi_2\theta_2}}  \nonumber\\&&-  L
  ^{2}\sqrt {1-  B    ^{2}}B
  G  { \Delta_1}
\cos  { \theta_1}  { b_{\phi_1 r}}  \sin
  \psi  { b_{\phi_2\theta_2}}  a  +
L  \sqrt {1-  B    ^{2}} \sqrt {{ F1}  }{ b_{\phi_1\theta_1}}  K
  \cos  \psi  +  \nonumber\\&&G_r
      L    ^{2}
\sqrt {1-  B    ^{2}}{ b_{\phi_1\theta_1}}
  B  a  \sin  \psi
  { b_{\phi_2\theta_2}}  +  G_r
      L    ^{2}
\sqrt {1-  B    ^{2}}B \cos  \psi  { b_{\phi_2\theta_2}}  a
  \nonumber\\&&-  G_r      L
    ^{2}\sqrt {1-  B
  ^{2}}{ b_{\phi_1\theta_1}}  B  a
  \cos  \psi  -G  L
  { b_{\phi_1\theta_1}}  K  \sin
  \psi  { b_{\phi_2\theta_2}}  B  B_r \nonumber\\&& -  L
^{2}\sqrt {1-  B    ^{2}}G
  { b_{\psi r}}  B  a
  \sin  \psi  { b_{\phi_2\theta_2}}  +
  L    ^{2}\sqrt {1-  B
    ^{2}}G  { b_{\psi r}}
  { b_{\phi_1\theta_1}}  B  a
  \sin  \psi  \nonumber\\&&+  L
  ^{2}\sqrt {1-  B    ^{2}}G
  { b_{\psi r}}  { b_{\phi_1\theta_1}}
  B  a  \cos  \psi
  { b_{\phi_2\theta_2}}  +  L
  ^{2}\sqrt {1-  B    ^{2}}B
  G  { \Delta_1}
\cos  { \theta_1}  { b_{\phi_1 r}}  \cos
  \psi  a \nonumber\\&& +  L
  ^{2}\sqrt {1-  B    ^{2}}G
  { \Delta_2}  \cos  {
\theta_2}  { b_{\phi_2 r}}  { b_{\phi_1\theta_1}}
  B  a  \sin  \psi
  +  L    ^{2}\sqrt {1-  B
    ^{2}}G  { \Delta_2}
  \cos  { \theta_2}  { b_{\phi_2 r}}
  B  a  \cos  \psi
  \nonumber\\&&+  L    ^{2}\sqrt {1-  B
    ^{2}}G  { b_{\psi r}}
 B  a  \cos  \psi
  +L  \sqrt {1-  B
  ^{2}}\sqrt {{ F1}  }K
\cos  \psi  { b_{\phi_2\theta_2}} \nonumber\\&& -G
  L    K_r
  \sin  \psi  +G  L
  K  \cos  \psi  { b_{\phi_2\theta_2}}
  B  B_r
  \nonumber\\&&+G  L  K
\sin  \psi  B  B_r
  +2\,G    L_r
    \sqrt {1-  B    ^{2}}{
 b_{\phi_1\theta_1}}  B  L  a
  \sin  \psi  { b_{\phi_2\theta_2}}
 \nonumber\\&& +G  L  { b_{\phi_1\theta_1}}
  K  \cos  \psi  B
  B_r  +2\,G
  L_r    \sqrt {1-  B
    ^{2}}B  L
a  \sin  \psi  \nonumber\\&&+2\,G
  L_r    \sqrt {1-  B
    ^{2}}B  \cos  \psi
  { b_{\phi_2\theta_2}}  L  a
  -2\,G    L_r
    \sqrt {1-  B    ^{2}}{
 b_{\phi_1\theta_1}}  B  L  a
  \cos  \psi  \nonumber\\&&+G
  L    ^{2}{ b_{\phi_1\theta_1}}
  B_r    a
\sin  \psi  { b_{\phi_2\theta_2}}  \sqrt {1-
  B    ^{2}}-G    L_r    K  \sin
  \psi  \nonumber\\&&+G    L
  ^{2}{ b_{\phi_1\theta_1}}  B    a_r    \sin  \psi  {
 b_{\phi_2\theta_2}}  \sqrt {1-  B
  ^{2}}-G    L
^{2}  { b_{\phi_1\theta_1,r}}    B
  a  \cos  \psi  \sqrt {1
-  B    ^{2}}\nonumber\\&&+G L    ^{2}B    a_r    \sin  \psi  \sqrt
{1-
  B    ^{2}}+G    L
    ^{2}  { b_{\phi_1\theta_1,r}}
    B  a  \sin
  \psi  { b_{\phi_2\theta_2}}  \sqrt {1-  B
    ^{2}}\nonumber\\&&+L  G
  { \Delta_1}  \cos  { \theta_1}
  { b_{\phi_1 r}}  K  \sin
\psi  { b_{\phi_2\theta_2}}  -L  G
  { \Delta_1}  \cos  {
\theta_1}  { b_{\phi_1 r}}  K  \sin
  \psi  { b_{\phi_2\theta_2}}    B
    ^{2}\nonumber\\&&+L  G  {
b_{\psi r}}  { b_{\phi_1\theta_1}}  K
  \sin  \psi  -L  G
  { b_{\psi r}}  { b_{\phi_1\theta_1}}  K
  \sin  \psi    B
  ^{2}\nonumber\\&&+G    L    ^
{2}  B_r    a
  \sin  \psi  \sqrt {1-  B
  ^{2}}+L  G  { b_{\psi r}}
  { b_{\phi_1\theta_1}}  K  \cos
  \psi  { b_{\phi_2\theta_2}}  \nonumber\\&&-L
G  { b_{\psi r}}  { b_{\phi_1\theta_1}}
  K  \cos  \psi  { b_{\phi_2\theta_2}}
    B    ^{2}+L
  G  { \Delta_2}  \cos
{ \theta_2}  { b_{\phi_2 r}}  { b_{\phi_1\theta_1}}
  K  \sin  \psi  \nonumber\\&&+G
    L    ^{2}  B_r    \cos  \psi  { b_{\phi_2\theta_2}}
  a  \sqrt {1-  B
    ^{2}}+G    L
  ^{2}B  \cos  \psi    { b_{\phi_2\theta_2,r}}    a
  \sqrt {1-  B    ^{2}}\nonumber\\&&-G
    L    ^{2}{ b_{\phi_1\theta_1}}
    B_r    a
  \cos  \psi  \sqrt {1-  B
  ^{2}}-G    L
^{2}{ b_{\phi_1\theta_1}}  B    a_r    \cos  \psi  \sqrt {1-
  B    ^{2}}\nonumber\\&&+G    L
    ^{2}B  \cos  \psi
  { b_{\phi_2\theta_2}}    a_r
   \sqrt {1-  B    ^{2}}+
G    L    ^{2}{ b_{\phi_1\theta_1}}
  B  a  \sin
\psi    { b_{\phi_2\theta_2,r}}
  \sqrt {1-  B    ^{2}}\nonumber\\&&-G
  L  \cos  \psi  { b_{\psi r}}
  K  +  G_r
     L    ^{2}\sqrt {1-
  B    ^{2}}B  a
  \sin  \psi  -  G_rLK\sin\psi\Big]\nonumber\\
\Omega_{r\psi\theta_1\phi_1}&=&  G_r    B
    L    ^{2}  a
    ^{2}+G    B_r      L    ^{
2}  a    ^{2}+2\,G  B
  L    a
  ^{2}L_r  +2\,G
B    L    ^{2}a
  a_r  \nonumber\\&&+  G_r
    B    K
    ^{2}+G    B_r
      K    ^{2}+2
\,G  B  K K_r  -G  { \Delta_1}
  \cos  { \theta_1}    L
  ^{2}{ b_{\phi_2 r}}  B  a
  \sin  \psi  \nonumber\\&&+G  { \Delta_1}
  \cos  { \theta_1}  L
{ b_{\phi_2 r}}  \sqrt {1-  B
  ^{2}}K  \sin  \psi\nonumber\\
\Omega_{r\psi\theta_1\phi_2}&=&\frac {1}{\sqrt {1-  B ^{2}}}\Big[-
G_r
  L  K  \cos  \psi
    B    ^{2}-G  L
    K_r
\cos  \psi    B    ^{2}+G
    L_r    K
  \cos  \psi  \nonumber\\&&-G
  L_r    K
\cos  \psi    B    ^{2}-
  G_r    L
{ b_{\phi_1\theta_1}}  K  \sin  \psi
  +  G_r    L
 { b_{\phi_1\theta_1}}  K  \sin
\psi    B    ^{2}+G
  L    { b_{\phi_1\theta_1,r}}
   K  \sin  \psi
  B    ^{2}\nonumber\\&&-L  G
  { \Delta_1}  \cos  { \theta_1}
  { b_{\phi_1 r}}  K  \sin
\psi  -G  L  \sin  \psi
    { b_{\phi_1\theta_1,r}}
K  -L  G  { b_{\psi r}}
  K  \sin  \psi  +L
  G  { b_{\psi r}}  K
  \sin  \psi    B
  ^{2}\nonumber\\&&+L  G  { b_{\psi r}}
  { b_{\phi_1\theta_1}}  K  \cos
  \psi    B    ^{2}+L
  G  { \Delta_1}
\cos  { \theta_1}  { b_{\phi_1 r}}  K
 \sin  \psi    B
  ^{2}+  G_r    L
  K  \cos  \psi  +G
  L    K_r
    \cos  \psi  \nonumber\\&&-  G_r
      L    ^{2}
\sqrt {1-  B    ^{2}}B \cos  \psi  a  -  G_r
      L    ^{2}
\sqrt {1-  B    ^{2}}{ b_{\phi_1\theta_1}}
  B  a  \sin  \psi
  \nonumber\\&&-G    L    ^{2}
  B_r    \cos  \psi
  a  \sqrt {1-  B
  ^{2}}-G    L
^{2}B  \cos  \psi    a_r    \sqrt {1-  B
  ^{2}}\nonumber\\&&-G    L
^{2}{ b_{\phi_1\theta_1}}    B_r
    a  \sin  \psi  \sqrt {1
-  B    ^{2}}-G L    ^{2}{ b_{\phi_1\theta_1}}  B
   a_r    \sin
  \psi  \sqrt {1-  B    ^{2}
}\nonumber\\&&-2\,G    L_r
  \sqrt {1-  B    ^{2}}B
  \cos  \psi  L  a
  -2\,G    L_r
    \sqrt {1-  B    ^{2}}{
 b_{\phi_1\theta_1}}  B  L  a
  \sin  \psi  \nonumber\\&&-G
  L    ^{2}  {
b_{\phi_1\theta_1,r}}    B  a
  \sin  \psi  \sqrt {1-  B
  ^{2}}-\sqrt {{ F1}  }K
\sqrt {1-  B    ^{2}}L \cos  \psi  \nonumber\\&&+G  L  {
b_{\phi_1\theta_1}}  K \sin  \psi  B
  B_r  -G
  L  K  \cos  \psi
  B  B_r  +
  L    ^{2}\sqrt {1-  B
    ^{2}}G  { b_{\psi r}}
  B  a  \sin  \psi
  \nonumber\\&&+  L    ^{2}\sqrt {1-  B
    ^{2}}B  G
{ \Delta_1}  \cos  { \theta_1}  { b_{\phi_1 r}}  \sin  \psi  a  +L
  \sqrt {1-  B    ^{2}}
\sqrt {{ F1}  }{ b_{\phi_1\theta_1}}  K
  \sin  \psi  \nonumber\\&&-  L
    ^{2}\sqrt {1-  B    ^{2
}}G  { b_{\psi r}}  { b_{\phi_1\theta_1}}
  B  a  \cos  \psi
  -G    L_r
    \sin  \psi  { b_{\phi_1\theta_1}}
  K  +G  L  {
 b_{\phi_1\theta_1}}    K_r
  \sin  \psi    B
^{2}\nonumber\\&&+G    L_r
  { b_{\phi_1\theta_1}}  K  \sin
\psi    B    ^{2}-G
  L  { b_{\phi_1\theta_1}}    K_r    \sin  \psi  -G
  L  \cos  \psi  {
b_{\psi r}}  K  { b_{\phi_1\theta_1}}
  \Big]\nonumber\\
\Omega_{r\psi\theta_2\phi_1}&=&\frac {1}{ \sqrt {1-  B^{2}}} \Big[
G_r
  L  K  \cos  \psi
    B    ^{2}+G  L
    K_r
\cos  \psi    B    ^{2}-G
    L_r    K
  \cos  \psi  +G
  L_r    K
\cos  \psi    B    ^{2}\nonumber\\&&+L
  G  { b_{\psi r}}  K
  \sin  \psi  -L  G
  { b_{\psi r}}  K  \sin
  \psi    B    ^{2}-
G_r    L  K
  \cos  \psi  -G  L
    K_r
\cos  \psi  \nonumber\\&&-  G_r
    L    ^{2}\sqrt {1-  B
    ^{2}}B  \cos  \psi
  a  -G    L
    ^{2}  B_r
  \cos  \psi  a  \sqrt {1-
B    ^{2}}\nonumber\\&&-G    L
   ^{2}B  \cos  \psi
  a_r    \sqrt {1-  B
    ^{2}}-2\,G    {\frac
{d}{dr}}L    \sqrt {1-  B
    ^{2}}B  \cos  \psi  L
  a  \nonumber\\&&+\sqrt {{ F1}
  }K  \sqrt {1-  B
  ^{2}}L  \cos  \psi  +G
  L  K  \cos  \psi
  B  B_r  -G
  L  K  \sin
\psi  { b_{\phi_2\theta_2}}  B  {\frac {d }{dr}}B \nonumber\\&&+2\,G
L_r \sqrt {1-  B
  ^{2}}B  L  a
  \sin  \psi  { b_{\phi_2\theta_2}}  +
  G_r      L
    ^{2}\sqrt {1-  B    ^{2
}}B  a  \sin  \psi  { b_{\phi_2\theta_2}}  \nonumber\\&&-  G_r
  L  K  \sin  \psi
  { b_{\phi_2\theta_2}}    B
  ^{2}+G    L_r
    K  \sin  \psi  {
b_{\phi_2\theta_2}}  -G    L_r
    K  \sin  \psi
  { b_{\phi_2\theta_2}}    B
  ^{2}+G  L    K_r    \sin  \psi  { b_{\phi_2\theta_2}}
  \nonumber\\&&-G  L    K_r    \sin  \psi  {
 b_{\phi_2\theta_2}}    B    ^{2}+G
  L  K  \sin
\psi  { b_{\phi_2\theta_2,r}}  -G
  L  K  \sin  \psi
    { b_{\phi_2\theta_2,r}}
  B    ^{2}+  G_r
    L  K  \sin
  \psi  { b_{\phi_2\theta_2}}  \nonumber\\&&+G
  L    ^{2}B  a
  \sin  \psi    { b_{\phi_2\theta_2,r}}
    \sqrt {1-  B
  ^{2}}+G    L
^{2}  B_r    a
  \sin  \psi  { b_{\phi_2\theta_2}}  \sqrt {
1-  B    ^{2}}\nonumber\\&&+G
  L    ^{2}B
a_r    \sin  \psi  {
 b_{\phi_2\theta_2}}  \sqrt {1-  B
  ^{2}}+  L    ^{2}\sqrt {1-
  B    ^{2}}G  { b_{\psi r}
}  B  a  \sin \psi  \nonumber\\&&+L  G  { \Delta_2}
  \cos  { \theta_2}  { b_{\phi_2 r}}
  K  \sin  \psi  +L
  G  { b_{\psi r}}  K
  \cos  \psi  { b_{\phi_2\theta_2}}  +
  L    ^{2}\sqrt {1-  B
    ^{2}}G  { \Delta_2}
  \cos  { \theta_2}  { b_{\phi_2 r}}
B  a  \sin  \psi  \nonumber\\&&+
  L    ^{2}\sqrt {1-  B
    ^{2}}G  { b_{\psi r}}
  B  a  \cos  \psi
  { b_{\phi_2\theta_2}}  -L  \sqrt {1-
  B    ^{2}}\sqrt {{ F1}
  }K  \sin  \psi  { b_{\phi_2\theta_2}}
  \nonumber\\&&-L  G  { \Delta_2}
  \cos  { \theta_2}  { b_{\phi_2 r}}
  K  \sin  \psi    B
   ^{2}-L  G  {
b_{\psi r}}  K  \cos  \psi  {
 b_{\phi_2\theta_2}}    B    ^{2}\Big]\nonumber\\
\Omega_{r\psi\theta_2\phi_2}&=&-L    -  G_r
    B  L  -G
    B_r    L
  \nonumber\\&&-2\,G  B  L_r  +G  { \Delta_2}
  \cos  { \theta_2}  { b_{\phi_1 r}}
B  L  a  \sin \psi  +G  { \Delta_2}  \cos
  { \theta_2}  { b_{\phi_1 r}}  \sqrt {1-
  B    ^{2}}K  \sin
  \psi  \nonumber\\
\Omega_{r\psi\phi_1\phi_2}&=&\frac{-1}{ \sqrt {1-  B    ^{2}}}\Big[-
G_r
    L    ^{2}\sqrt {1-  B
    ^{2}}B  a
\sin  \psi  \nonumber\\&&-  G_r
  L  K  \sin  \psi
  +  G_r    L
 K  \sin  \psi    B
    ^{2}-G    L
    ^{2}  B_r
  a  \sin  \psi  \sqrt {1-
B    ^{2}}\nonumber\\&&-2\,G   L_r    \sqrt {1-  B
    ^{2}}B  L  a
 \sin  \psi  -G    L
    ^{2}B    a_r    \sin  \psi  \sqrt {1-
  B    ^{2}}\nonumber\\&&+G  L
 K  \sin  \psi  B
  B_r  -G  L
    K_r
\sin  \psi  +G  L
  K_r    \sin  \psi
    B    ^{2}-G
  L_r    K
\sin  \psi  \nonumber\\&&+G    L_r
    K  \sin  \psi
    B    ^{2}-  L
    ^{2}\sqrt {1-  B    ^{2
}}G  { b_{\psi r}}  B  a
  \cos  \psi  -G  L
  \cos  \psi  { b_{\psi r}}
  K  \nonumber\\&&+L  G  {
 b_{\psi r}}  K  \cos  \psi
    B    ^{2}+L
\sqrt {1-  B    ^{2}}\sqrt {{ F1}
  }K  \sin  \psi  \Big]\nonumber\\
\Omega_{r\theta_1\theta_2\phi_1}&=&\frac {1}{ \sqrt {1-  B
^{2}}}\Big[G B  { b_{\phi_2 r}}
  \sqrt {1-  B    ^{2}}{
 \Delta_2}  \sin  { \theta_2}
L    ^{2}  a    ^{ 2}+G  B  { b_{\phi_2 r}} \sqrt {1-  B    ^{2}}{
\Delta_2}
 \sin  { \theta_2}    K
  ^{2}\nonumber\\&&+G    { \Delta_{2,r}}
    \cos  { \theta_2}  { b_{\phi_2\theta_2}
}    K    ^{2}B
  \sqrt {1-  B    ^{2}}+G
  { \Delta_2}  \cos  { \theta_2}
    { b_{\phi_2\theta_2,r}}
  K    ^{2}B  \sqrt {1-
  B    ^{2}}\nonumber\\&&+2\,G  {
\Delta_2}  \cos  { \theta_2}  { b_{\phi_2\theta_2}}
  K  B    K_r    \sqrt {1-  B
    ^{2}}+G  { \Delta_2}
  \cos  { \theta_2}  { b_{\phi_2\theta_2}}
    K    ^{2}  B_r    \sqrt {1-  B
  ^{2}}\nonumber\\&&+G    { \Delta_{1,r}}
    \cos  { \theta_1}  B
    L    ^{2}\cos  \psi
  { b_{\phi_2\theta_2}}  a  \sqrt {1-
  B    ^{2}}+G  {
\Delta_1}  \cos  { \theta_1}    B_r      L
  ^{2}\cos  \psi  { b_{\phi_2\theta_2}}  a
  \sqrt {1-  B    ^{2}}\nonumber\\&&+2
\,G  { \Delta_1}  \cos  { \theta_1}  B  L  \cos  \psi
  { b_{\phi_2\theta_2}}  a    L_r    \sqrt {1-  B
    ^{2}}+  G_r
  { \Delta_1}  \cos  { \theta_1}
  K  L  \cos  \psi
  { b_{\phi_2\theta_2}}    B
  ^{2}\nonumber\\&&+G  { \Delta_1}  \cos
  { \theta_1}  K    L_r    \sin  \psi    B
    ^{2}-G  { \Delta_1}
  \cos  { \theta_1}  K
  L_r    \cos  \psi
  { b_{\phi_2\theta_2}}  +G  { \Delta_1}
  \cos  { \theta_1}  K
  L_r    \cos  \psi
  { b_{\phi_2\theta_2}}    B
  ^{2}\nonumber\\&&-G  { \Delta_1}  \cos
  { \theta_1}  K  L
\cos  \psi  { b_{\phi_2\theta_2,r}}  + G  { \Delta_1}  \cos  {
\theta_1}  K  L  \cos  \psi
    { b_{\phi_2\theta_2,r}}
  B    ^{2}\nonumber\\&&-G  {
\Delta_1}  \cos  { \theta_1}   K_r    L  \cos
  \psi  { b_{\phi_2\theta_2}}  +G
{ \Delta_1}  \cos  { \theta_1}
  K_r    L
\cos  \psi  { b_{\phi_2\theta_2}}    B
    ^{2}\nonumber\\&&-G    { \Delta_{1,r}}    \cos  { \theta_1}
  K  L  \sin  \psi
  +G    { \Delta_{1,r}}
    \cos  { \theta_1}  K
  L  \sin  \psi    B
   ^{2}\nonumber\\&&-G  { \Delta_1}
  \cos  { \theta_1}    K_r
    L  \sin  \psi
  +G  { \Delta_1}  \cos
  { \theta_1}    K_r
    L  \sin  \psi
B    ^{2}\nonumber\\&&-G    { \Delta_{1,r}}    \cos  { \theta_1}
  K  L  \cos  \psi
  { b_{\phi_2\theta_2}}  +G    { \Delta_{1,r}}    \cos  {
\theta_1}  K  L  \cos  \psi
  { b_{\phi_2\theta_2}}    B
  ^{2}\nonumber\\&&-  G_r    {
\Delta_1}  \cos  { \theta_1}  K
  L  \sin  \psi  +  {\frac {
d}{dr}}G    { \Delta_1}  \cos
  { \theta_1}  K  L
\sin  \psi    B    ^{2}\nonumber\\&&-
  G_r    { \Delta_1}
  \cos  { \theta_1}  K
L  \cos  \psi  { b_{\phi_2\theta_2}}
  +G  { \Delta_1}  \cos
  { \theta_1}  B    L
    ^{2}\cos  \psi
{ b_{\phi_2\theta_2,r}}    a  \sqrt {1-
  B    ^{2}}\nonumber\\&&+G  {
\Delta_1}  \cos  { \theta_1}  B
    L    ^{2}\cos  \psi
  { b_{\phi_2\theta_2}}    a_r
   \sqrt {1-  B    ^{2}}+
2\,G  { \Delta_2}  \cos  { \theta_2}  { b_{\phi_2\theta_2}}  B  L
    a    ^{2} L_r    \sqrt {1-  B
    ^{2}}\nonumber\\&&+2\,G  { \Delta_2}
  \cos  { \theta_2}  { b_{\phi_2\theta_2}}
  B    L    ^{2}a
    a_r
\sqrt {1-  B    ^{2}}+G
  { \Delta_{1,r}}    \cos
  { \theta_1}  B    L
    ^{2}\sin  \psi  a
\sqrt {1-  B    ^{2}}\nonumber\\&&+G { \Delta_1}  \cos  { \theta_1}
  B_r      L
    ^{2}\sin  \psi  a
\sqrt {1-  B    ^{2}}+2\,G
  { \Delta_1}  \cos  { \theta_1}
  B  L  \sin  \psi
  a    L_r
  \sqrt {1-  B    ^{2}}\nonumber\\&&+G
  { \Delta_1}  \cos  { \theta_1}
  B    L    ^{2}
\sin  \psi    a_r
  \sqrt {1-  B    ^{2}}+G
    { \Delta_{2,r}}
  \cos  { \theta_2}  { b_{\phi_2\theta_2}}
  B    L    ^{2}
  a    ^{2}\sqrt {1-  B
    ^{2}}\nonumber\\&&+G  { \Delta_2}
  \cos  { \theta_2}    {
b_{\phi_2\theta_2,r}}    B    L
    ^{2}  a    ^{2}\sqrt {1
-  B    ^{2}}+G  { \Delta_2}  \cos  { \theta_2}  {
b_{\phi_2\theta_2}}
    B_r
  L    ^{2}  a
  ^{2}\sqrt {1-  B    ^{2}}\nonumber\\&&+G
  { \Delta_1}  \cos  {
\theta_1}  K  L  \sin  \psi
  B  B_r  +G
  { \Delta_1}  \cos  {
\theta_1}  K  L  \cos  \psi
  { b_{\phi_2\theta_2}}  B  B_r  \nonumber\\&&+  G_r
  \sqrt {1-  B    ^{2}}{ \Delta_1
}  \cos  { \theta_1}  B
    L    ^{2}\cos  \psi
  { b_{\phi_2\theta_2}}  a  + G_r    \sqrt {1-  B
    ^{2}}{ \Delta_2}  \cos  {
\theta_2}  { b_{\phi_2\theta_2}}  B
  L    ^{2}  a
  ^{2}\nonumber\\&&+  G_r
\sqrt {1-  B    ^{2}}{ \Delta_1}
 \cos  { \theta_1}  B
L    ^{2}\sin  \psi  a
  +  G_r    \sqrt {1-
  B    ^{2}}{ \Delta_2}
  \cos  { \theta_2}  { b_{\phi_2\theta_2}}
    K    ^{2}B  \nonumber\\&&-G
  { \Delta_1}  \cos  {
\theta_1}  K    L_r
    \sin  \psi  +G  B
  { b_{\phi_2 r}}  \sqrt {1-  B
    ^{2}}  L    ^{
2}{ \Delta_1}  \sin  { \theta_1}  \Big]\nonumber\\
\Omega_{r\theta_1\theta_2\phi_2}&=&\frac {1}{\sqrt {1-  B
^{2}}}\Big[- G_r
  L  { \Delta_2}  \cos
{ \theta_2}  K  \sin  \psi  +
  G_r    L
{ \Delta_2}  \cos  { \theta_2}  K
  \sin  \psi    B
  ^{2}\nonumber\\&&-G    L_r
    { \Delta_2}  \cos  {
\theta_2}  { b_{\phi_1\theta_1}}  K  \cos
  \psi  +G    L_r
    { \Delta_2}  \cos  {
 \theta_2}  { b_{\phi_1\theta_1}}  K
\cos  \psi    B    ^{2}\nonumber\\&&-G
    L_r    {
 \Delta_2}  \cos  { \theta_2}  K
  \sin  \psi  +G
  L_r    { \Delta_2}
  \cos  { \theta_2}  K
\sin  \psi    B    ^{2}\nonumber\\&&-B
  G  { b_{\phi_1 r}}
\sqrt {1-  B    ^{2}}  L
    ^{2}{ \Delta_1}  \sin  {
\theta_1}  -B  G  { b_{\phi_1 r}}
  \sqrt {1-  B    ^{2}}{
 \Delta_2}  \sin  { \theta_2}
K    ^{2}\nonumber\\&&-B  G
  { b_{\phi_1 r}}  \sqrt {1-  B
    ^{2}}{ \Delta_2}  \sin  {
\theta_2}    L    ^{2}  a
    ^{2}-G  L
{ \Delta_2}  \cos  { \theta_2}  { b_{\phi_1\theta_1}}    K_r
  \cos  \psi  \nonumber\\&&+G  L
  { \Delta_2}  \cos  { \theta_2}
  { b_{\phi_1\theta_1}}    K_r
   \cos  \psi    B
    ^{2}-G  L    { \Delta_{2,r}}    \cos  {
\theta_2}  K  \sin  \psi  \nonumber\\&&+G
 L    { \Delta_{2,r}}
    \cos  { \theta_2}  K
  \sin  \psi    B
^{2}+  G_r    L
  { \Delta_2}  \cos  { \theta_2}
  { b_{\phi_1\theta_1}}  K  \cos
\psi    B    ^{2}\nonumber\\&&-G
  L  { \Delta_2}  \cos
{ \theta_2}    K_r
  \sin  \psi  +G  L
  { \Delta_2}  \cos  { \theta_2}
    K_r    \sin
  \psi    B    ^{2}\nonumber\\&&-G
  L  { \Delta_2}
\cos  { \theta_2}    { b_{\phi_1\theta_1,r}}
    K  \cos  \psi
  +G  L  { \Delta_2}
  \cos  { \theta_2}    {
b_{\phi_1\theta_1,r}}    K  \cos  \psi
    B    ^{2}\nonumber\\&&-G  L
    { \Delta_{2,r}}
    \cos  { \theta_2}  { b_{\phi_1\theta_1}}
 K  \cos  \psi  +G
  L    { \Delta_{2,r}}
    \cos  { \theta_2}  { b_{\phi_1\theta_1}
}  K  \cos  \psi
  B    ^{2}\nonumber\\&&-  G_r
    L  { \Delta_2}
  \cos  { \theta_2}  { b_{\phi_1\theta_1}}
  K  \cos  \psi  -  {\frac {
d}{dr}}G      L
  ^{2}\sqrt {1-  B    ^{2}}{
\Delta_1}  \cos  { \theta_1}  { b_{\phi_1\theta_1}}
  B \nonumber\\&& +  G_r
     L    ^{2}\sqrt {1-
  B    ^{2}}{ \Delta_2}
  \cos  { \theta_2}  B  \sin
  \psi  a  -  G_r
      L    ^{2}
\sqrt {1-  B    ^{2}}{ \Delta_2}
 \cos  { \theta_2}  { b_{\phi_1\theta_1}}
  B  a  \cos  \psi
  \nonumber\\&&-2\,G    L_r
    \sqrt {1-  B    ^{2}}{
 \Delta_1}  \cos  { \theta_1}  {
b_{\phi_1\theta_1}}  B  L  +2\,G
    L_r
\sqrt {1-  B    ^{2}}{ \Delta_2}
 \cos  { \theta_2}  B  \sin
  \psi  L  a  \nonumber\\&&-2\,G
    L_r
\sqrt {1-  B    ^{2}}{ \Delta_2}
 \cos  { \theta_2}  { b_{\phi_1\theta_1}}
  B  L  a  \cos
  \psi  +G  L  {
\Delta_2}  \cos  { \theta_2}  K
  \sin  \psi  B  B_r
  \nonumber\\&&+G  L  { \Delta_2}
  \cos  { \theta_2}  { b_{\phi_1\theta_1}}
 K  \cos  \psi  B
  B_r  -G
  L    ^{2}  {
\Delta_{1,r}}    \cos  { \theta_1}  {
 b_{\phi_1\theta_1}}  B  \sqrt {1-  B
    ^{2}}\nonumber\\&&-G    L
    ^{2}{ \Delta_1}  \cos  {
\theta_1}    { b_{\phi_1\theta_1,r}}
  B  \sqrt {1-  B
  ^{2}}-G    L
^{2}{ \Delta_1}  \cos  { \theta_1}  {
 b_{\phi_1\theta_1}}    B_r
  \sqrt {1-  B    ^{2}}\nonumber\\&&-G
    L    ^{2}  { \Delta_{2,r}}    \cos  { \theta_2}
  { b_{\phi_1\theta_1}}  B  a
  \cos  \psi  \sqrt {1-  B
  ^{2}}-G    L
^{2}{ \Delta_2}  \cos  { \theta_2}
  { b_{\phi_1\theta_1,r}}    B
 a  \cos  \psi  \sqrt {1-
  B    ^{2}}\nonumber\\&&-G    L
    ^{2}{ \Delta_2}  \cos
  { \theta_2}  { b_{\phi_1\theta_1}}    B_r    a  \cos
  \psi  \sqrt {1-  B    ^{2}
}-G    L    ^{2}{ \Delta_2}  \cos  { \theta_2}  {
b_{\phi_1\theta_1}}
  B    a_r
    \cos  \psi  \sqrt {1-  B
    ^{2}}\nonumber\\&&+G    L
  ^{2}  { \Delta_{2,r}}
  \cos  { \theta_2}  B  \sin
  \psi  a  \sqrt {1-  B
    ^{2}}+G    L
  ^{2}{ \Delta_2}  \cos  { \theta_2}
    B_r    \sin
  \psi  a  \sqrt {1- B
    ^{2}}\nonumber\\&&+G    L
  ^{2}{ \Delta_2}  \cos  { \theta_2}
  B  \sin  \psi    {\frac {d
}{dr}}a    \sqrt {1-  B
  ^{2}}\Big]\nonumber\\
\Omega_{r\theta_1\phi_1\phi_2}&=&\frac {1}{ \sqrt {1-  B ^{2}}}\Big[
G_r
  \sqrt {1-  B    ^{2}}{ \Delta_2
}  \cos  { \theta_2}  B
    L    ^{2}  a
    ^{2}+  G_r
  \sqrt {1-  B    ^{2}}{ \Delta_1
}  \cos  { \theta_1}  B
    L    ^{2}\cos  \psi
  a  \nonumber\\&&+  G_r
    \sqrt {1-  B    ^{2}}{
 \Delta_2}  \cos  { \theta_2}
K    ^{2}B  -  G_r    { \Delta_1}  \cos
  { \theta_1}  K  L
\cos  \psi  +  G_r
  { \Delta_1}  \cos  { \theta_1}
  K  L  \cos  \psi
    B    ^{2}\nonumber\\&&+G
  { \Delta_{2,r}}    \cos
  { \theta_2}  B    L
    ^{2}  a    ^{2}\sqrt {1
-  B    ^{2}}+G  { \Delta_2}  \cos  { \theta_2}    B_r      L
  ^{2}  a    ^{2}\sqrt {1-
B    ^{2}}\nonumber\\&&+2\,G  { \Delta_2}
  \cos  { \theta_2}  B
L    a    ^{2}  L_r    \sqrt {1-  B
    ^{2}}+2\,G  { \Delta_2}
  \cos  { \theta_2}  B    L
    ^{2}a    a_r    \sqrt {1-  B
  ^{2}}\nonumber\\&&+G    { \Delta_{1,r}}
    \cos  { \theta_1}  B
    L    ^{2}\cos  \psi
  a  \sqrt {1-  B
  ^{2}}+G  { \Delta_1}  \cos
  { \theta_1}    B_r
      L    ^{2}\cos
\psi  a  \sqrt {1-  B
  ^{2}}\nonumber\\&&+2\,G  { \Delta_1}
\cos  { \theta_1}  B  L
  \cos  \psi  a    {\frac {d
}{dr}}L    \sqrt {1-  B
  ^{2}}+G  { \Delta_1}  \cos
  { \theta_1}  B    L
    ^{2}\cos  \psi
a_r    \sqrt {1-  B
  ^{2}}\nonumber\\&&+G    { \Delta_{2,r}}
    \cos  { \theta_2}    K
    ^{2}B  \sqrt {1-  B
    ^{2}}+2\,G  { \Delta_2}
  \cos  { \theta_2}  K
B    K_r \sqrt {1-  B    ^{2}}\nonumber\\&&+G { \Delta_2}  \cos  {
\theta_2}
  K    ^{2}  B_r
    \sqrt {1-  B
  ^{2}}-G    { \Delta_{1,r}}
    \cos  { \theta_1}  K
  L  \cos  \psi  +G
    { \Delta_{1,r}}
  \cos  { \theta_1}  K  L
 \cos  \psi    B
  ^{2}\nonumber\\&&-G  { \Delta_1}  \cos
  { \theta_1}    K_r
    L  \cos  \psi  +G
  { \Delta_1}  \cos  {
\theta_1}    K_r    L
  \cos  \psi    B
  ^{2}\nonumber\\&&-G  { \Delta_1}  \cos
  { \theta_1}  K    L_r    \cos  \psi  +G
  { \Delta_1}  \cos  { \theta_1}
  K    L_r
  \cos  \psi    B
^{2}+G  { \Delta_1}  \cos  {
 \theta_1}  K  L  \cos
\psi  B  B_r  \Big]\nonumber\\
\Omega_{r\theta_2\phi_1\phi_2}&=&\frac {1}{\sqrt {1-  B ^{2}}}\Big[-
G_r
    L    ^{2}\sqrt {1-  B
    ^{2}}{ \Delta_2}  \cos
  { \theta_2}  B  a
\cos  \psi  -  G_r
  L  { \Delta_2}  \cos
{ \theta_2}  K  \cos  \psi  \nonumber\\&&+
  G_r    L
{ \Delta_2}  \cos  { \theta_2}  K
  \cos  \psi    B
  ^{2}-  G_r
  L    ^{2}\sqrt {1-  B
    ^{2}}{ \Delta_1}  \cos  {
\theta_1}  B  \nonumber\\&&-2\,G    L_r    \sqrt {1-  B
    ^{2}}{ \Delta_2}  \cos  {
\theta_2}  B  L  a
  \cos  \psi  -G    {\frac {
d}{dr}}L    { \Delta_2}  \cos
  { \theta_2}  K  \cos  \psi
  \nonumber\\&&+G    L_r
    { \Delta_2}  \cos  {
\theta_2}  K  \cos  \psi    B
    ^{2}-2\,G    {\frac {
d}{dr}}L    \sqrt {1-  B
    ^{2}}{ \Delta_1}  \cos  {
\theta_1}  B  L  \nonumber\\&&-G
    L    ^{2}  { \Delta_{2,r}}    \cos  { \theta_2}
  B  a  \cos  \psi
  \sqrt {1-  B    ^{2}}-G
    L    ^{2}{ \Delta_2}
  \cos  { \theta_2}    B_r
    a  \cos  \psi
  \sqrt {1-  B    ^{2}}\nonumber\\&&-G
    L    ^{2}{ \Delta_2}
  \cos  { \theta_2}  B    a_r    \cos  \psi
\sqrt {1-  B    ^{2}}-G L    { \Delta_{2,r}}
    \cos  { \theta_2}  K
\cos  \psi  \nonumber\\&&+G  L
  { \Delta_{2,r}}    \cos
  { \theta_2}  K  \cos  \psi
    B    ^{2}+G  L
  { \Delta_2}  \cos  {
\theta_2}  K  \cos  \psi  B
  B_r  -G  L
  { \Delta_2}  \cos  {
\theta_2}    K_r \cos  \psi \nonumber\\&& +G  L  { \Delta_2}  \cos
{ \theta_2} K_r    \cos  \psi
  B    ^{2}-G    L
    ^{2}  { \Delta_{1,r}}
    \cos  { \theta_1}  B
  \sqrt {1-  B    ^{2}}\nonumber\\&&-G
    L    ^{2}{ \Delta_1}
  \cos  { \theta_1}    B_r
    \sqrt {1-  B
  ^{2}}\Big]\nonumber\\
\Omega_{\psi\theta_1\theta_2\phi_1}&=&G  { \Delta_1}  \cos
  { \theta_1}  L    B
  \cos  \psi  L  a
  -B  L  a
\sin  \psi  { b_{\phi_2\theta_2}}  -K
  \sqrt {1-  B    ^{2}}\cos
  \psi  \nonumber\\&&+\sqrt {1-  B    ^{2
}}K  \sin  \psi  { b_{\phi_2\theta_2}} \nonumber\\
\Omega_{\psi\theta_1\theta_2\phi_2}&=&G  L  { \Delta_2}
 \cos  { \theta_2}    B
\cos  \psi  L  a  +{ b_{\phi_1\theta_1}}  B  L  a
  \sin  \psi  +{ b_{\phi_1\theta_1}}
  \sqrt {1-  B    ^{2}}K
  \sin  \psi  \nonumber\\&&-K  \sqrt {1-
  B    ^{2}}\cos  \psi  \nonumber\\
\Omega_{\psi\theta_1\phi_1\phi_2}&=&-G  { \Delta_1}  \cos
  { \theta_1}  L  \sin  \psi
    B  L  a
  -\sqrt {1-  B    ^{2}}K \nonumber\\
\Omega_{\psi\theta_2\phi_1\phi_2}&=&G  L  { \Delta_2}
 \cos  { \theta_2}  \sin  \psi
  B  L  a  +
\sqrt {1-  B    ^{2}}K \nonumber\\
\Omega_{\theta_1\theta_2\phi_1\phi_2}&=&G  B      L
    ^{2}{ \Delta_1}  \sin
  { \theta_1}  +{ \Delta_2}  \sin
  { \theta_2}    L    ^{2}
  a    ^{2}+{ \Delta_2}
  \sin  { \theta_2}    K
  ^{2}
\nd
\noindent Once we have all the components, we can plug this in the susy constraint equations \eqref{susconstr} and
get the additional relations between the parameters introduced in \cite{chen}. Together with \eqref{ghontida} we can
finally write the precise susy backgrounds in the chain of geometric transitions.

\end{document}